\documentclass[twocolumn,prb,showpacs,superscriptaddress,groupedaddress]{revtex4-1}

\usepackage{epsfig,color}
\usepackage{amssymb}
\usepackage{amsmath}
\usepackage{dcolumn}
\usepackage{graphicx}
\usepackage{color}
\usepackage[pdfborder={0 0 0}, colorlinks=true, linkcolor=blue, citecolor=blue, urlcolor=blue]{hyperref}

\def\beq{\begin{equation}}
\def\eeq{\end{equation}}
\def\bea{\begin{eqnarray}}
\def\eea{\end{eqnarray}}

\def\q{\mathbf{q}}
\def\k{\mathbf{k}}

\def\p{\mathbf{p}}

\def\Q{\mathbf{Q}}

\newcommand{\eps}{\varepsilon}

\newcommand{\su}{\uparrow}
\newcommand{\sd}{\downarrow}

\newcommand{\sign}{\mathrm{sgn}}

\newcommand{\ii}{{\mathrm{i}}}

\newcommand{\nn}{\nonumber}

\newcommand{\DLS}[1]{\textbf{#1}}
\newcommand{\DLL}[1]{\textit{#1}}

\begin{document}

\title{Effect of gap anisotropy on the spin resonance peak in the superconducting state of iron-based materials}
\author{M.M.~Korshunov}
\email{mkor@iph.krasn.ru}
\affiliation{Kirensky Institute of Physics, Federal Research Center KSC SB RAS, 660036 Krasnoyarsk, Russia}
%

\date{\today}

\begin{abstract}
Spin resonance in the superconducting state of Fe-based materials within the multiorbital model with unequal anisotropic gaps on different Fermi surface sheets is studied. On the basis of the model gap function and the one calculated within the spin fluctuation theory of pairing, I show that the resonance peak shifts to higher frequencies with increasing the zero-amplitude gap magnitude. On the contrary, with increasing the gap anisotropy, it shifts to lower frequencies and lose some intensity.
\end{abstract}

\maketitle

\section{Introduction \label{sec:intro}}

Many of iron-based superconductors, which include pnictides and chalcogenides, have high critical temperatures $T_c>50$~K allowing to refer to them as high-$T_c$ superconductors. The basic element is always a square lattice of Fe, though in some cases with orthorhombic distortions, surrounded by As or P in pnictides and by Se, Te, or S in chalcogenides~\cite{y_kamihara_08,SadovskiiReview2008,IzyumovReview2008,MazinReview,PaglioneReview,JohnstonReview,WenReview,StewartReview}. Weakly doped pnictides are antiferromagnetic metals. Though there is no ultimately accepted microscopic mechanism of superconductivity, the most promising candidate is the spin fluctuation mechanism of Cooper pairing~\cite{HirschfeldKorshunov2011,ChubukovReview2012,Korshunov2014eng,Hirschfeld2016}. It is tightly connected with the topology of the Fermi surface comprised of several sheets, namely, with the existence of hole and electron Fermi pockets for a wide range of doping concentrations $x$. Fermi surface, as well as states near the Fermi level, are formed by the iron $d$-orbitals and consists of two hole pockets near the $\Gamma=(0,0)$ point and two electron pockets centered at $(\pi,0)$ and $(0,\pi)$ points of the two-dimensional Brillouin zone corresponding to one Fe per unit cell. Proximity of the wave vector related to the scattering between particles at electron and hole sheets to the nesting wave vector $\Q$ results in strong antiferromagnetic fluctuations with the maximum of the spin susceptibility near $\Q$ that equal to $(\pi,0)$ or $(0,\pi)$. There is a qualitative and sometimes even quantitative agreement between the Fermi surface calculated within the density functional theory (DFT) and the one measured via quantum oscillations and by the angle-resolved photoemission spectroscopy (ARPES)~\cite{Kordyuk}. Absence of the insulating state in the undoped case points toward the moderate nature of the electronic correlations in such a multiorbital system~\cite{Anisimov2008eng,Kroll2008}. Iron magnetic moment differs from one family of Fe-based materials to another with the smallest value of $\sim 0.3\mu_B$ in LaFeAsO~\cite{KlaussKorshunov2008} to $\sim 3.3\mu_B$ in K$_2$Fe$_4$Se$_5$~\cite{Gretarsson2011}. This issue was discussed as originating from the effect of correlations~\cite{Haule2009,Hansmann2010,Toschi2012}. The concept of Hund's metal was put forward~\cite{Haule2009} to emphasize the role of Hund's exchange $J$ in the physics of Fe-based materials. In particular, the irreducible vertex corrections beyond the random phase approximation (RPA) for the magnetic susceptibility were calculated~\cite{Hansmann2010,Toschi2012} and compared to the neutron scattering experiments~\cite{Liu2012}. However, the RPA approach also gives reasonable results when compared to various experiments in the normal and superconducting states~\cite{HirschfeldKorshunov2011,Korshunov2014eng} thus providing the natural starting point for studying the low-energy physics of itinerant electrons in iron-based superconductors.

Different mechanisms of superconductivity result in specific symmetries and structures of the gap in iron-based materials~\cite{HirschfeldKorshunov2011}. In the spin fluctuation theory of pairing within the RPA and in the functional renormalization group (fRG) approach, the leading superconducting instability in a wide range of dopings is characterized by the extended $s$-wave gap having the opposite signs on hole and electron Fermi surface pockets~\cite{HirschfeldKorshunov2011,Korshunov2014eng,Mazin2008,Graser2009,Kuroki2008,Chubukov2008,MaitiKorshunovPRL2011,MaitiKorshunovPRB2011,Classen2017}. The corresponding gap structure belongs to the $A_{1g}$ representation of the tetragonal symmetry group and is called $s_{\pm}$ state. On the other hand, orbital fluctuations results in the $s_{++}$ state with the gap having the same sign on all Fermi surface sheets~\cite{Kontani}. Therefore, by determining the gap structure, one can deduce the microscopic mechanism of superconductivity. In this respect, inelastic neutron scattering plays a special role since the imaginary part of the dynamical spin susceptibility $\chi(\q,\omega)$ measured there carries information about the gap structure in the superconducting state. That is, the sign-changing $s_\pm$ gap leads to the formation of the spin resonance peak at or near the commensurate antiferromagnetic wave vector $\q = \Q$ connecting Fermi surface sheets with different signs of gaps on them~\cite{KorshunovEreminResonance2008,Maier2008,Maier2009}. In simple models, the peak appears at frequencies $\omega_R < 2\Delta$, where $\Delta$ is the gap magnitude. At present, the well defined peak was observed in neutron scattering on all iron-based superconductors for $T < T_c$ near the wave vector $\Q$, see, e.g., Refs.~\onlinecite{ChristiansonBKFA,ChristiansonBFCA,QiuFeSeTe,Park,Babkevich,Inosov2010,ArgyriouKorshunov2010,LumsdenReview,Castellan2011,Dai2015}.

However, by introducing an additional damping of quasiparticles and by adjusting parameters, one can attain the appearance of a peak in the spin susceptibility in the $s_{++}$ state at frequencies above $2\Delta$~\cite{Onari2010,Onari2011}. Therefore, to determine whether the observed peak is the true spin resonance one has to explore the effect of different details of the superconducting state on it and deduce some criterion. Previously, the characteristic feature of the spin resonance in the case of unequal gaps on hole and electron pockets were established~\cite{KorshunovPRB2016,KorshunovJMMM2017} -- in the presence of larger and smaller gaps, $\Delta_L$ and $\Delta_S$, the criterion is the condition for the spin resonance frequency, $\omega_R \leq \Delta_L+\Delta_S$. Comparison of data from the neutron scattering on the peak frequency and data from various techniques on gap magnitudes leads to the conclusion that in most cases the observed peak fulfills the condition and, therefore, indicates the $s_\pm$ gap structure~\cite{KorshunovPRB2016,KorshunovJMMM2017}. However, the role of the gap anisotropy in the formation of the spin resonance peak is still an open question. For example, results of ARPES~\cite{Shimojima2011} and Andreev spectroscopy~\cite{Abdel-Hafiez2014,Kuzmicheva2016,Kuzmicheva2017} demonstrate anisotropy of the larger gap as large as 30\% in pnictides. On the qualitative level, the question was discussed in Ref.~\onlinecite{Maiti2011}; however, within the very simple four-band model and without a particular recipe for comparison to the experimental data. Here I consider the effect of the gap anisotropy on the dynamical spin susceptibility and the spin resonance within the realistic five-orbital model from Ref.~\onlinecite{Graser2009}. Two approaches to the gap structure are used. One is phenomenological with the model gap function that is parameterized to reflect the general form of the experimentally observed and the theoretically obtained gap. Due to some freedom in the choice of parameters and ability to vary them, this approach allows us  to analyze basic effects of the gap anisotropy on the spin resonance peak. The other approach employs self-consistent calculation of the gap function within the spin fluctuation theory of pairing. Spin resonance peak is then calculated and compared to the results with the model gap. Obtained results lead to the adjustment of the condition $\omega_R \leq \Delta_L+\Delta_S$ that would allow us to make a comparison of experimental data on the peak frequency and gaps to answer the question on whether the observed peak is the true spin resonance originating from the $s_\pm$ state.

The paper is organized as follows. In Section~\ref{sec:model}, the model and the approaches are presented. Results for the spin susceptibility for the model gap function are given in Section~\ref{sec:resultsmodelgap} and the magnetic response for the gap calculated within the spin fluctuation theory of pairing is shown in Section~\ref{sec:resultscalcgap}. Concluding remarks and the brief analysis of the experimental data are given in Section~\ref{sec:conclusion}.

\section{Model and approximations \label{sec:model}}

I use here a Hamiltonian $H = H_0 + H_{int}$ consisting of the tight-binding model $H_0$~\cite{Graser2009} and an on-site Coulomb (Hubbard) multiorbital interaction $H_{int}$. Hamiltonian $H_0$ is based on the DFT band structure for LaFeAsO~\cite{Cao2008} and it includes five iron $d$-orbitals ($d_{xz}$, $d_{yz}$, $d_{xy}$, $d_{x^2-y^2}$, $d_{3z^2-r^2}$),
\begin{equation}
 H_0 = \sum_{\k \sigma} \sum_{l l'} \left[ t_{l l'}(\k) + \epsilon_{l} \delta_{l l'} \right] d_{\k l \sigma}^\dagger d_{\k l' \sigma},
 \label{eq:H0}
\end{equation}
where $d_{\k l \sigma}^\dagger$ is the annihilation operator for an electron with momentum $\k$, spin $\sigma$, and orbital index $l$. Hopping matrix elements $t_{l l'}(\k)$ and one-electron energies $\epsilon_{l}$ are given in Ref.~\onlinecite{Graser2009}. Fermi surface consists of two hole pockets, $\alpha_1$ and $\alpha_2$, near the $\Gamma$ point and two electron pockets, $\beta_1$ and $\beta_2$, centered at $(\pi,0)$ and $(0,\pi)$ points of the one-Fe Brillouin zone. Here I consider the case of small electron doping with $x=0.05$.

Interaction part $H_{int}$ has the following form~\cite{Graser2009,Kuroki2008,Castallani1978,Oles1983},
\bea
H_{int} &=& U \sum_{f, m} n_{f m \su} n_{f m \sd} + U' \sum_{f, m < l} n_{f l} n_{f m} \nn\\
  && + J \sum_{f, m < l} \sum_{\sigma,\sigma'} d_{f l \sigma}^\dag d_{f m \sigma'}^\dag d_{f l \sigma'} d_{f m \sigma} \nn\\
  && + J' \sum_{f, m \neq l} d_{f l \su}^\dag d_{f l \sd}^\dag d_{f m \sd} d_{f m \su}.
\label{eq:Hint}
\eea
where $n_{f m} = n_{f m \su} + n_{f m \sd}$, $n_{f m \sigma} = d_{f m \sigma}^\dag d_{f m \sigma}$ is the number of particles operator at site $f$, $U$ and $U'$ are intra- and interorbital Hubbard repulsions, $J$ is the Hund's exchange, and $J'$ is the pair hopping.
To limit a number of free parameters in the theory, let us assume the spin-rotational invariance (SRI) that adds two constraints, $U'=U-2J$ and $J'=J$. There are still two parameters to be determined, $U$ and $J$. Their values crucially depend on the orbital basis of the model. For example, constrained DFT gives $U=3.5$~eV and $J=0.8$~eV for the full set of Fe-$d$ and As-$p$ orbital set ($p-d$ model for LaFeAsO), while for the model that includes only $d$-orbitals, it gives $U=0.75$~eV and $J=0.51$~eV~\cite{Anisimov2008eng}. Another approach, constrained RPA (cRPA), results in $U=2.69$~eV and $J=0.79$~eV~\cite{Aichhorn2009,Aichhorn2011} or in $U=1.97$~eV and $J=0.77$~eV~\cite{Roekeghem2016} for the full set of $d$- and $p$-orbitals with excluded Coulomb interaction at the $p$-orbitals ($d-dp$ model). The same cRPA for the $d$-only orbital set gives
$U=2.2-3.3$~eV and $J=0.3-0.6$~eV~\cite{Miyake2008,Nakamura2008}. Such a dependence on the number of orbitals is due to the spatial extent of Wannier functions that are used to construct the matrix elements of the Coulomb interactions. As a general trend, a limited number of orbitals results in the smaller values of Hubbard parameters. For the five-orbital model studied here, the large values of $U$, greater than $\approx 1.5$~eV, results in the divergence of the spin susceptibility, i.e. the magnetic instability. Since the undoped LaFeAsO exhibits stripe antiferromagnetic order at low temperatures, the choice of parameters that provide closeness to the magnetic instability is reasonable. Therefore in what follows, I set $U=1.4$~eV. As for the Hund's exchange, it is taken to be $J=0.1-0.2$~eV. The $J/U$ ratio for the lower boundary, $J/U \approx 0.07$, is comparable to the widely discussed Hund's metal proposal for Fe-based materials with $J/U=0.35/4 \approx 0.08$~\cite{Haule2009}, while for the upper boundary, $J/U \approx 0.14$ is comparable with the cRPA ratio for the $d$-only orbital set, $J/U=0.43/2.92 \approx 0.14$~\cite{Miyake2008}.

Matrix elements of the transverse component of the spin susceptibility are equal to~\cite{Korshunov2014eng}
\bea \label{eq:chipmmu}
 \chi^{ll',mm'}_{(0)+-}(\q,\Omega) &=& -T \sum_{\p,\omega_n, \mu,\nu} \left[ \varphi^{\mu}_{\p m} {\varphi^*}^{\mu}_{\p l} G_{\mu \su}(\p,\omega_n) \right.\nn\\
 &\times& G_{\nu \sd}(\p+\q,\Omega+\omega_n) \varphi^{\nu}_{\p+\q l'} {\varphi^*}^{\nu}_{\p+\q m'} \nn\\
 &+& {\varphi^*}^{\mu}_{\p l} {\varphi^*}^{\mu}_{-\p m'} F^\dag_{\mu \su}(\p,\omega_n) \nn\\
 &\times& \left. F_{\nu \sd}(\p+\q,\Omega-\omega_n) \varphi^{\nu}_{\p+\q l'} \varphi^{\nu}_{-\p-\q m} \right],
\eea
where $\Omega$ and $\omega_n$ are bosonic and fermionic Matsubara frequencies, $G$ and $F$ are normal and anomalous (Gor'kov) Green's functions, $\mu$ and $\nu$ are band indices, $\varphi^{\mu}_{\k m}$ are matrix elements of orbital-to-band transformation, so that $d_{\k m \sigma} = \sum\limits_{\mu} \varphi^{\mu}_{\k m} b_{\k \mu \sigma}$. Here, $b_{\k \mu \sigma}$ is the electron annihilation operator in the band representation, where Green's function is diagonal with respect to band indices, $G_{\mu \sigma}(\k,\omega_n) = 1 / \left( \ii\omega_n - \eps_{\k\mu\sigma} \right)$.

Here I use two approaches to the superconducting state. The first one is phenomenological -- the gap function is chosen to simulate results of calculations and experimental findings, both of which are generally similar. Parameters of the gap function are treated as free, so one can model various situations including ones with the different sets of interaction parameters. In this case, the gap function belonging to the $A_{1g}$ representation of the tetragonal symmetry group and entering the anomalous Green's function is defined as
\beq \label{eq:delta}
 \Delta_{\k \mu} = \Delta_{\mu}^{0} + \Delta_{\mu}^{1} \left(\cos k_x + \cos k_y \right)/2.
\eeq
Here, parameter $\Delta_{\mu}^{1}$ controls changes of gap amplitude in the band $\mu$, while $\Delta_{\mu}^{0}$ controls the gap magnitude for zero amplitude (we refer to it later as the `zero-amplitude gap magnitude').
The simplest possible $s_{++}$ state takes place for $\Delta_{\mu}^{1} = 0$ and $\Delta_{\mu}^{0} = \Delta_{\mu'}^{0}$, and the simplest state of the $s_\pm$-type can be obtained taking $\Delta_{\alpha_{1,2}}^{0} = -\Delta_{\beta_{1,2}}^{0}$. The specific feature of the FS topology in pnictides is that due to the shift of $k_x$ or $k_y$ by $\pi$ with respect to $(0,0)$ point, the gap on electron pockets will have a local, i.e., with respect to pocket's center, $d$-wave symmetry~\cite{MaitiKorshunovPRB2011}.

The other approach to the superconducting state is to perform the spin fluctuation calculation of the gap function. I follow the procedure from Refs.~\onlinecite{Graser2009,Kemper2010,Korshunov2014eng}: calculate spin and charge susceptibilities in the RPA and combine them into the Cooper vertex entering the linearized gap equation. The latter is solved to obtain the eigenfunction $g(\k)$, which is the gap function, and the eigenvalue $\lambda$; the leading instability corresponds to the largest $\lambda$.

Below, all parameters of gaps are in units of $\Delta_0$ taken to be 5~meV in our calculations. Since all gaps have $A_{1g}$ symmetry and should not change upon the $\pi/2$ rotation, gaps on electron pockets $\beta_1$ and $\beta_2$ should be the same. Thus, $\Delta_{\beta_1}^{0,1} = \Delta_{\beta_2}^{0,1}$, which we denote simply as $\Delta_\beta^{0,1}$.

To calculate the spin response, the RPA is used with the local Coulomb interaction $H_{int}$. Sum of the corresponding ladder diagrams that include electron-hole bubble in the matrix form, $\hat\chi_{(0)+-}(\q,\omega)$, result in the following expression for the matrix of the RPA spin susceptibility~\cite{Korshunov2014eng}:
\begin{eqnarray}
 \hat\chi_{+-}(\q,\Omega) = \left[\hat{I} - \hat{U}_s \hat\chi_{(0)+-}(\q,\Omega)\right]^{-1} \hat\chi_{(0)+-}(\q,\Omega),
\label{eq:chi_s_sol}
\end{eqnarray}
where $\hat{I}$ and $\hat{U}_s$ are the unit and interaction matrices, respectively, in the orbital basis. Explicit form of the latter is given in Ref.~\onlinecite{Graser2009}. In the next section I present results for the physical susceptibility $\chi_{+-}(\q,\Omega) = \frac{1}{2} \sum_{l,m} \chi^{ll,mm}_{+-}(\q,\Omega)$ that was analytically continued to the real frequency axis $\omega$ ($\ii\Omega \to \omega + \ii\delta$, $\delta \to 0+$).

The mechanism of the spin resonance peak formation in the superconducting state with the sign-changing gap is quite transparent~\cite{KorshunovEreminResonance2008}. Since $\chi_{(0)+-}(\q,\omega)$ describes particle-hole excitations and since all excitations at frequencies less than about twice the gap magnitude are absent in the superconducting state, $\mathrm{Im}\chi_{(0)+-}(\q,\omega)$ becomes finite only above this frequency value. The anomalous Green's functions entering Eq.~(\ref{eq:chipmmu}) give rise to the anomalous coherence factors, $\left[1 - \frac{\Delta_\k \Delta_{\k+\q}}{E_{\k} E_{\k+\q}}\right]$. If $\Delta_\k$ and $\Delta_{\k+\q}$ have the same sign, as it is for the $s_{++}$ state, the coherence factors vanish leading to a gradual increase of the spin susceptibility with increasing frequency for $\omega > \omega_c$ with $\omega_c = \min \left(|\Delta_\k| + |\Delta_{\k+\q}| \right)$. For the $s_\pm$ state, vector $\q = \Q$ connects Fermi surfaces with different signs of the gap, $\sign {\Delta_\k} \neq \sign {\Delta_{\k+\q}}$, resulting in the finite coherence factors that leads to a jump in the imaginary part of $\chi_{(0)+-}$ at $\omega_c$. 
For a certain set of interaction parameters entering the matrix $\hat{U}_s$, this results in a divergence of $\mathrm{Im}\chi_{+-}(\Q,\omega)$~(\ref{eq:chi_s_sol}). The corresponding peak at a frequency $\omega_R \leq \omega_c$ is the true spin resonance. Since gaps entering the expression for $\omega_c$ correspond to bands separated by the wave vector $\q$, we can call $\omega_c$ the indirect or effective gap. That's the reason why in the case of unequal gaps in different bands, $\Delta_L$ and $\Delta_S$, connected by the wave vector $\Q$, we have $\omega_c = \Delta_L + \Delta_S$~\cite{KorshunovPRB2016}.

\section{Results for the model gap function \label{sec:resultsmodelgap}}

In this Section, Coulomb parameters are chosen to be $U=1.4$~eV and $J=0.15$~eV; the rest are constrained by the SRI.

\begin{figure}
\begin{center}
 \includegraphics[width=1.0\columnwidth]{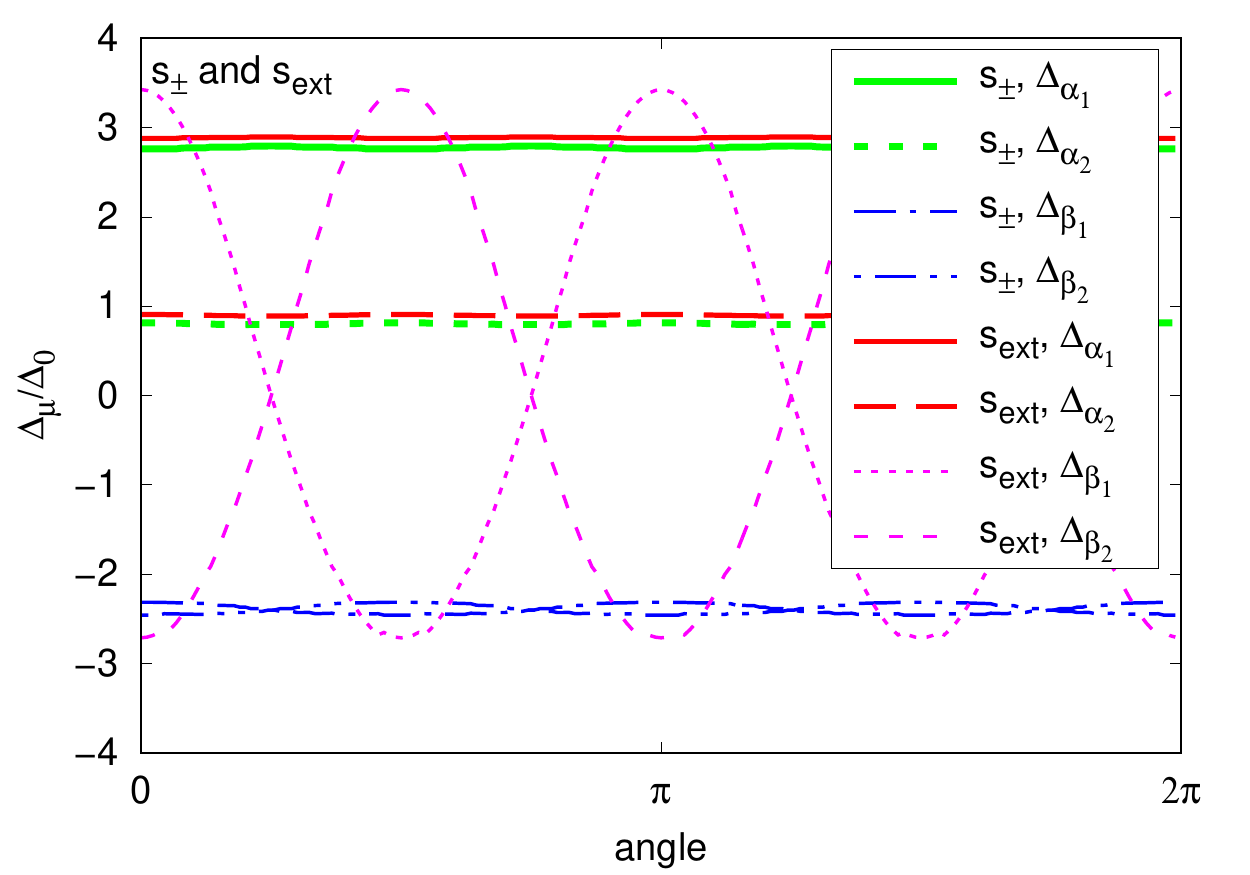}
 \includegraphics[width=1.0\columnwidth]{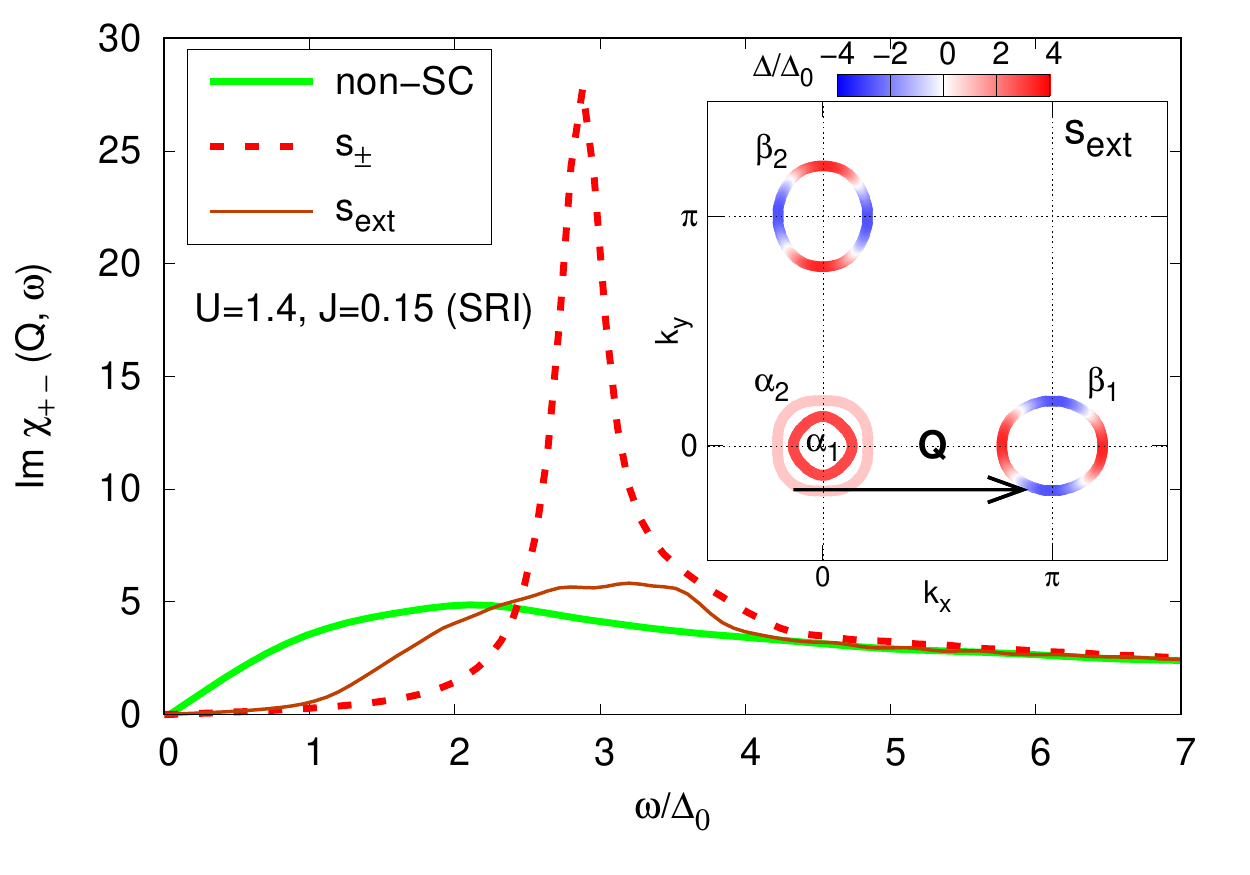}
 \caption{(Color online) Top: Angular dependencies of gaps on hole ($\alpha_{1,2}$) and electron ($\beta_{1,2}$) Fermi surface pockets for the $s_\pm$ and the $s_{ext}$ states. Bottom: Frequency dependencies of imaginary parts of the corresponding spin susceptibilities at the wave vector $\Q$, as well as the normal state (non-SC) spin response. Inset: Magnitudes of gaps on the Fermi surface for the $s_{ext}$ state and the wave vector $\Q$.}
\label{fig:spmsext}
\end{center}
\end{figure}

Gap angular dependencies on electron and hole sheets and the corresponding frequency dependencies of imaginary parts of spin susceptibilities at the wave vector $\Q$ for two extended $s$-wave symmetries, namely, $s_\pm$ and $s_{ext}$ states~\cite{Graser2009}, are shown in Fig.~\ref{fig:spmsext}. The former one is the widely discussed, fully gapped $s_\pm$ state with a small gap angular dependence on each Fermi surface pocket, $\Delta_{\k \mu} = \Delta_{\mu} \cos(k_x) \cos(k_y)$ with $\Delta_{\alpha_1,\beta}=3$ and $\Delta_{\alpha_2}=1$. In this state, the spin resonance peak is formed at frequencies lower than $\Delta_{\beta}+\Delta_{\alpha_2}$~\cite{KorshunovPRB2016}, see the lower panel of Fig.~\ref{fig:spmsext}. The $s_{ext}$ state corresponds to such a strong anisotropy on electron pockets, that the gap becomes sign-changing there and develops a nodal structure. The latter is clearly seen in the inset of Fig.~\ref{fig:spmsext}, where the gap magnitude on the Fermi surface is shown. Parameters in Eq.~(\ref{eq:delta}) were set to be: $\Delta_{\mu}^{0}=0$, $\Delta_{\alpha_1}^{1}=3$, $\Delta_{\alpha_2}^{1}=1$, $\Delta_{\beta}^{1}=30$. The spin resonance is absent in this case since only near-nodal states with a tiny gap on electron pockets $\beta_{1,2}$ contribute to the susceptibility at the wave vector $\Q$, as seen in the inset in Fig.~\ref{fig:spmsext}. Therefore, the discontinuous jump in $\mathrm{Im}\chi_{(0)+-}$ required for the formation of the spin resonance appears at vanishingly small frequencies and the RPA spin response gets only a small boost compared to the normal state, see Fig.~\ref{fig:spmsext}. This is similar to the case of $d_{x^2-y^2}$ gap symmetry where the spin resonance is absent at the commensurate wave vector~\cite{KorshunovEreminResonance2008}. Of course, the spin resonance may appear at the incommensurate wave vector different from $\Q$, see the discussion in Ref.~\onlinecite{Maier2009}.

\begin{figure}
\begin{center}
 \includegraphics[width=1.0\columnwidth]{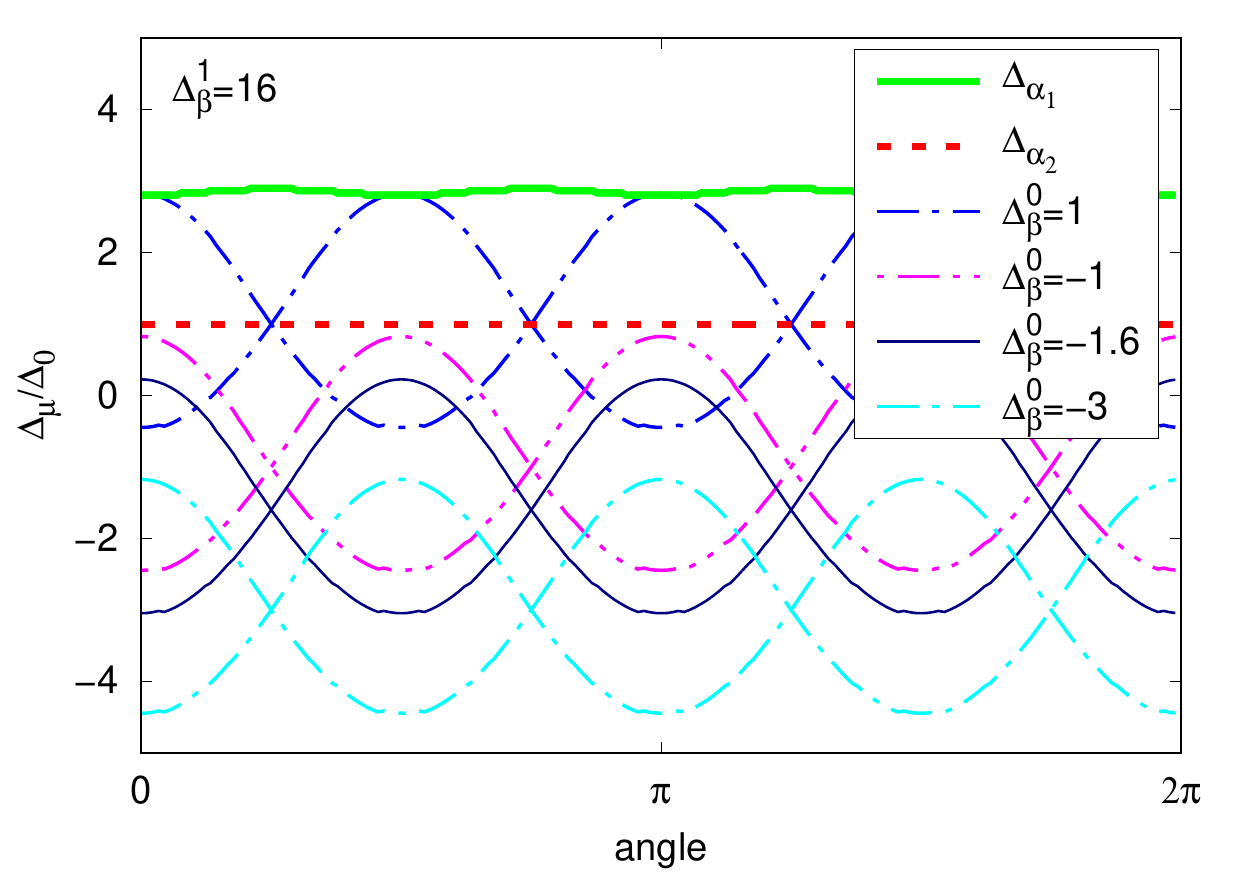}
 \includegraphics[width=1.0\columnwidth]{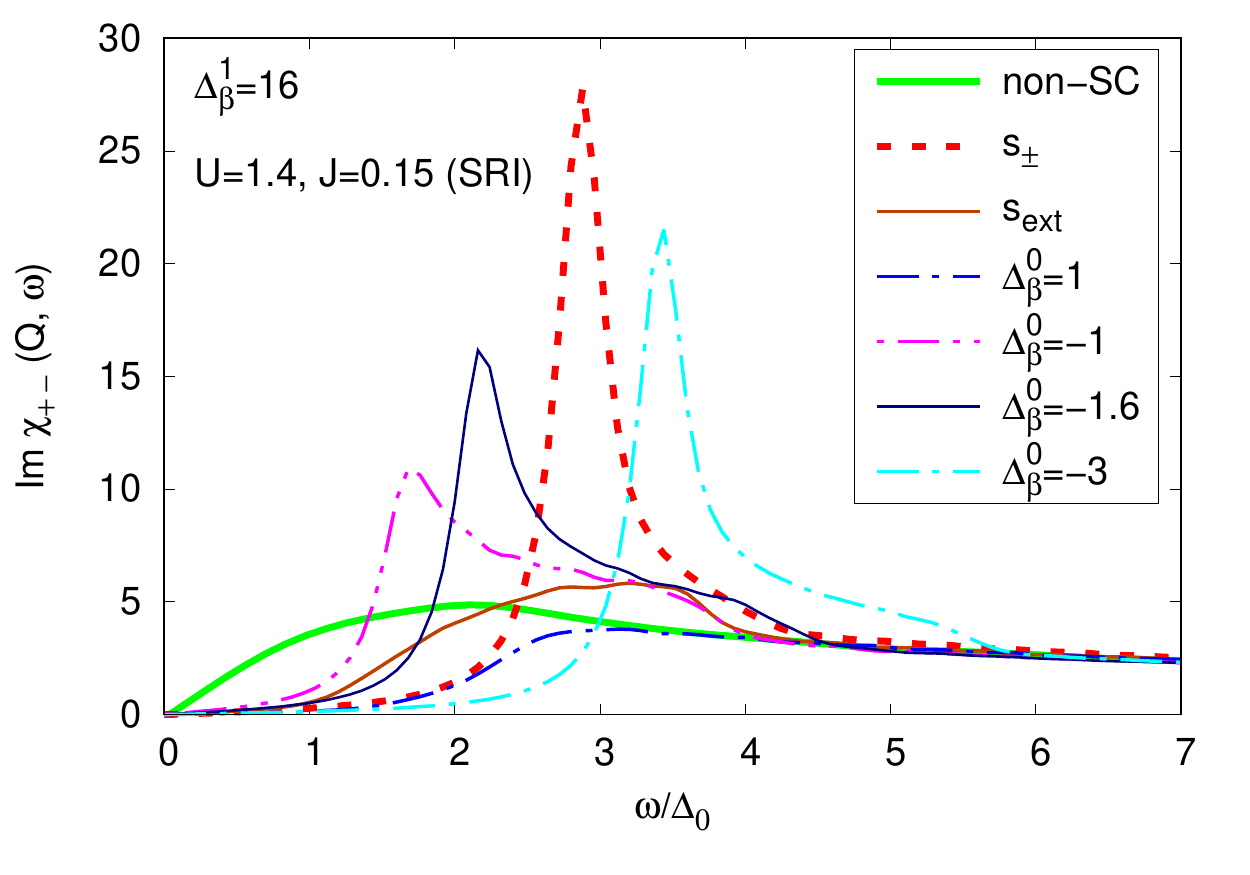}
 \caption{Top: Angular dependencies of gaps for several $A_{1g}$-type states with the fixed gap anisotropy on hole pockets and varying zero-amplitude gap magnitude on electron pockets. Bottom: Corresponding frequency dependencies of $\mathrm{Im}\chi_{+-}(\Q,\omega)$, as well as the spin response in the normal (non-SC), $s_\pm$, and $s_{ext}$ states.}
\label{fig:A1gSet1}
\end{center}
\end{figure}
\begin{figure}
\begin{center}
 \includegraphics[width=1.0\columnwidth]{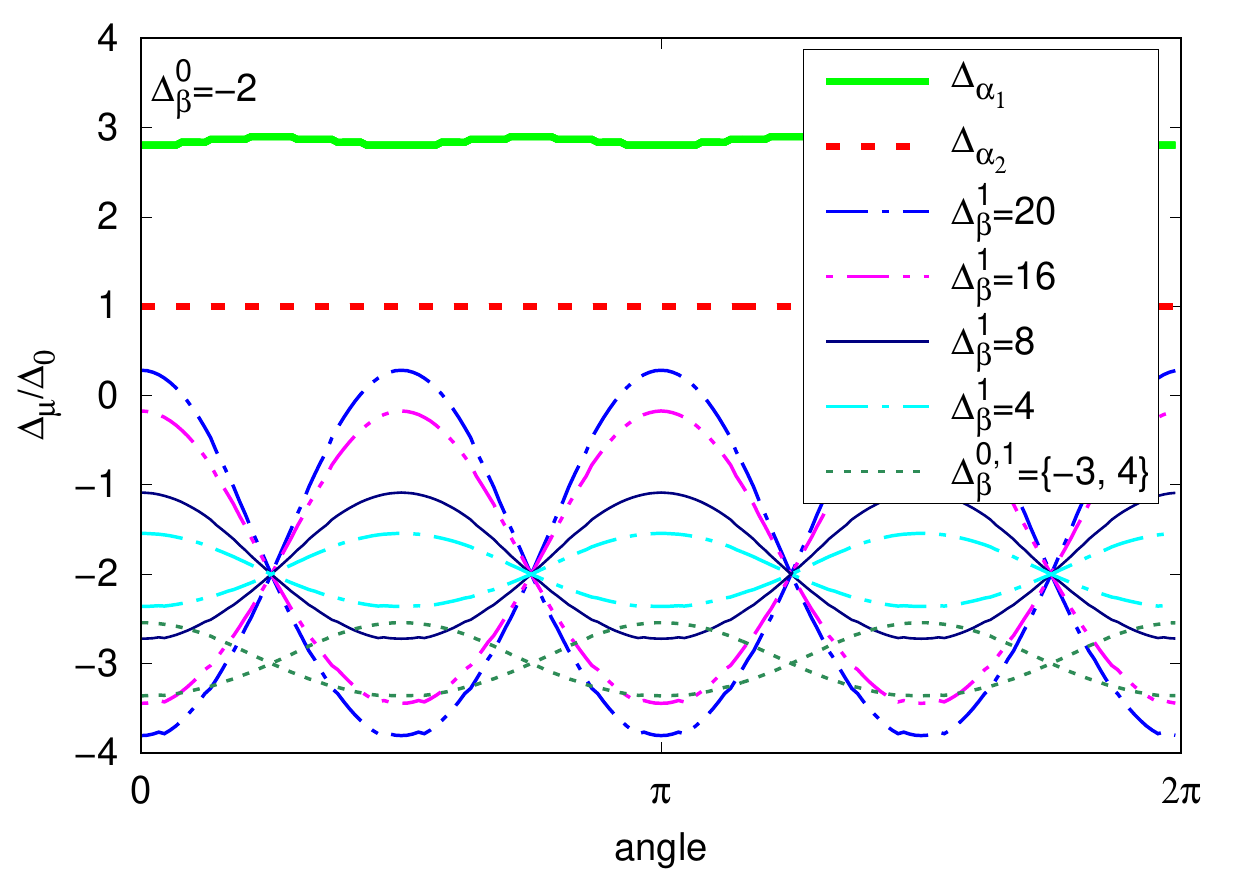}
 \includegraphics[width=1.0\columnwidth]{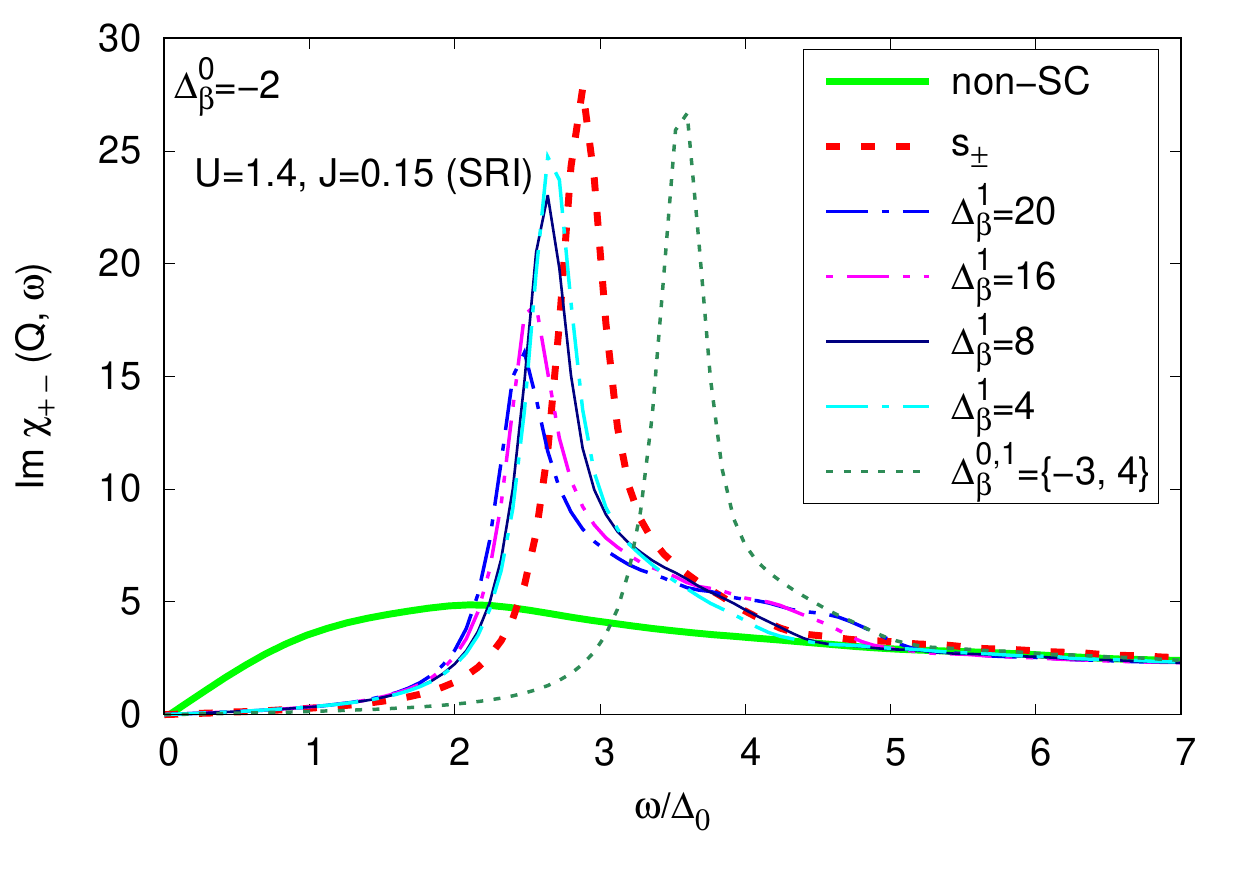}
 \caption{Top: Angular dependencies of gaps for several $A_{1g}$-type states with the fixed zero-amplitude gap magnitude and varying gap anisotropy on electron pockets. The case when $\Delta_{\beta}^{0}$ is shifted while $\Delta_{\beta}^{1}$ kept minimal is also shown. Bottom: Frequency dependence of $\mathrm{Im}\chi_{+-}(\Q,\omega)$ for these states, as well as for the normal and the $s_\pm$ states.}
\label{fig:A1gSet2}
\end{center}
\end{figure}

Most superconducting solutions in the spin fluctuation theory of pairing having the $A_{1g}$ symmetry are characterized by gaps with the weak angular dependence on hole pockets and a significant anisotropy on electron Fermi surface sheets~\cite{MaitiKorshunovPRB2011}. To model such a situation, I set the amplitude and the anisotropy of gaps on the hole pockets in Eq.~(\ref{eq:delta}) to be $\Delta_{\alpha_1}^{0}=1$, $\Delta_{\alpha_1}^{1}=0$, $\Delta_{\alpha_2}^{0}=-16.4$, $\Delta_{\alpha_2}^{1}=20$. This gives the constant gap on the inner hole pocket $\alpha_1$ and a weak anisotropy on the outer hole pocket $\alpha_2$. At the same time, the gap on $\alpha_1$ is approximately three times the gap on $\alpha_2$. This case is shown in Figs.~\ref{fig:A1gSet1}-\ref{fig:A1gSet2} and in~\onlinecite{PRBSuppl}.

First, I fix the gap anisotropy on electron pockets by setting $\Delta_{\beta}^{1}=16$ and vary the zero-amplitude magnitude, $\Delta_{\beta}^{0}$, of the gap there. The result is shown in Fig.~\ref{fig:A1gSet1}. Once the average gap on electron pockets have the same sign as on hole pockets (the case of $\Delta_{\beta}^{0}=1$), the resonance condition, $\Delta_{\k} = -\Delta_{\k+\Q}$, is not fulfilled and the spin resonance is absent. For the opposite signs of gaps at the wave vector $\Q$, the spin resonance forms and its frequency is as higher as larger the absolute value of the zero-amplitude gap magnitude on electron pockets.

Second, I change the gap anisotropy on electron pockets by varying $\Delta_{\beta}^{1}$ while the zero-amplitude gap magnitude is fixed, $\Delta_{\beta}^{0}=-2$. Results are shown in Fig.~\ref{fig:A1gSet2}. Experimentally observed gap anisotropy of 30\%~\cite{Shimojima2011,Abdel-Hafiez2014,Kuzmicheva2017} approximately corresponds to the case of $\Delta_{\beta}^{1}=4$ shown here. Evidently, decrease of the gap anisotropy leads to the increase of the spin resonance frequency. The same figure illustrates what happens when $\Delta_{\beta}^{0}$ is shifted to higher energies for the minimal amplitude shown. As expected, the spin resonance peak shifts to higher frequencies. Obviously, decrease of the spin resonance frequency with increasing the gap amplitude originates from the decrease of the effective gap at the wave vector $\Q$ entering the dynamical spin susceptibility. Decrease of the peak frequency is accompanied by the loss of its intensity due to the diminished spectral weight in agreement with the results of Ref.~\onlinecite{Maiti2011}.

\section{Results for the calculated gap function \label{sec:resultscalcgap}}

\begin{figure}
\begin{center}
 \includegraphics[width=1.0\columnwidth]{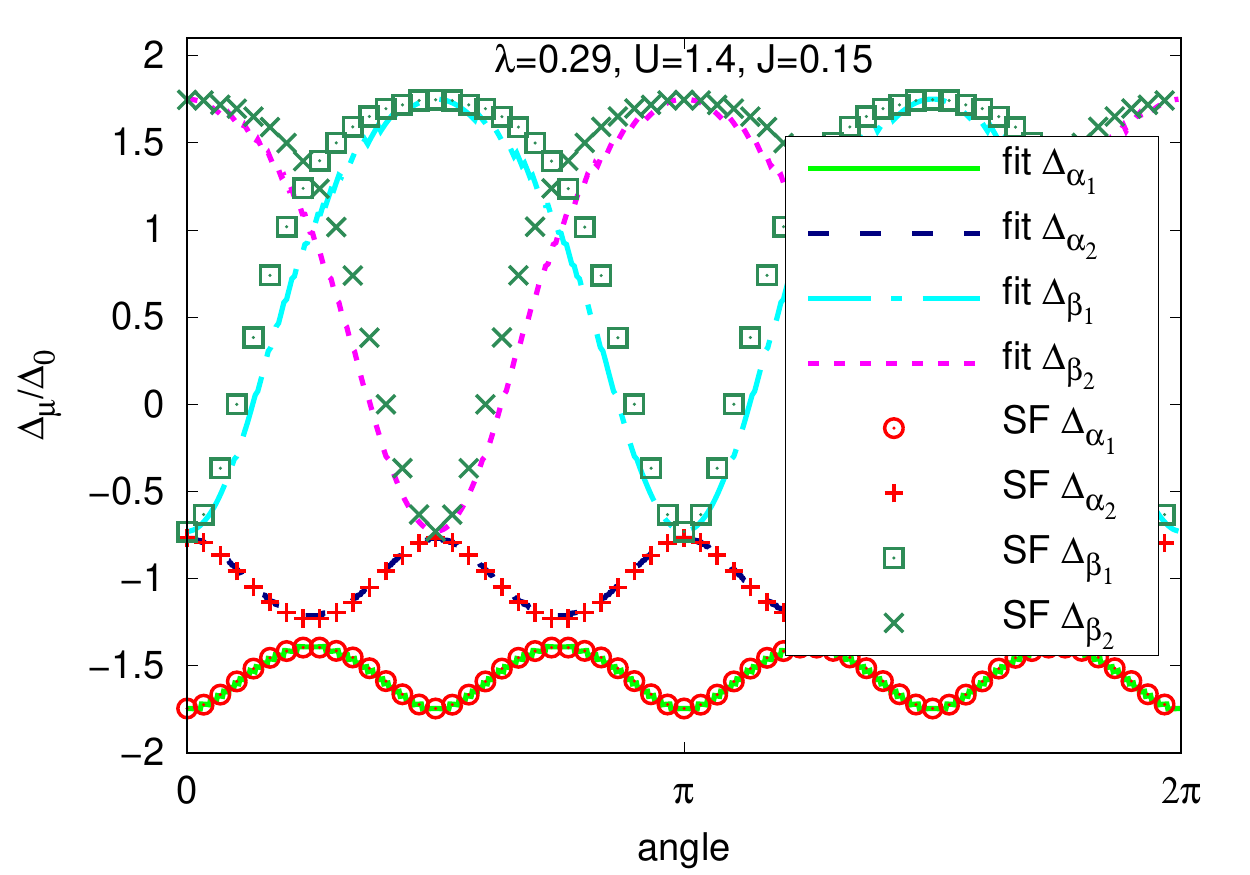}
 \includegraphics[width=1.0\columnwidth]{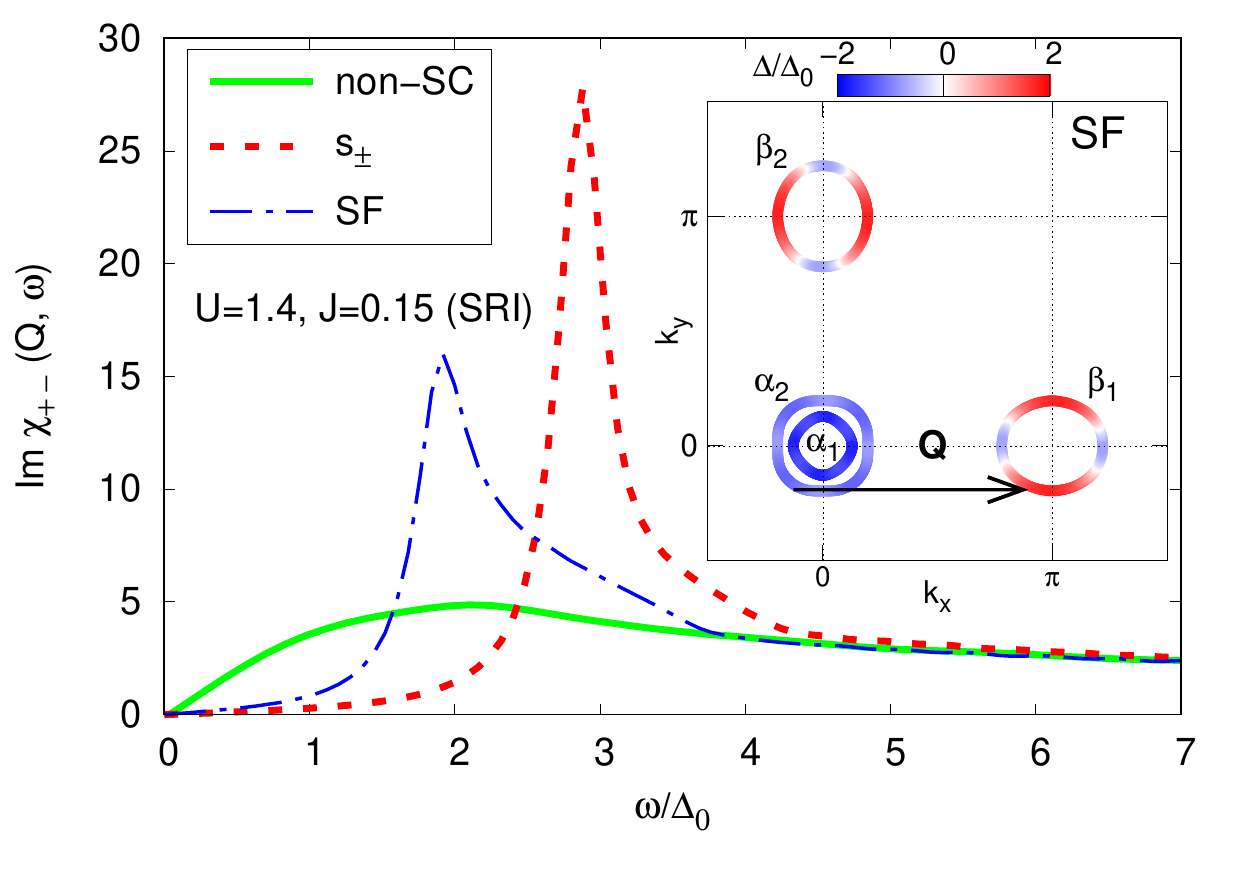}
 \caption{(Color online) Top: Angular dependencies of gaps on hole ($\alpha_{1,2}$) and electron ($\beta_{1,2}$) Fermi surface pockets calculated within the spin fluctuation pairing theory (SF gap) and obtained by fitting the parameters of Eq.~(\ref{eq:delta0123}). Bottom: Frequency dependencies of imaginary parts of the spin susceptibilities at the wave vector $\Q$ in the normal state (non-SC), for the model $s_\pm$ state, and for the SF gap. Magnitudes of the latter on the Fermi surface and the wave vector $\Q$ are shown in the inset.
 The SF gap was normalized by $\tilde\Delta_0=50$~meV to compare with our model results.}
\label{fig:SFset18}
\end{center}
\end{figure}

The linearized gap equation within the spin fluctuation theory of pairing was solved and the gap function $g(\k)$ and the corresponding eigenvalue $\lambda$ were obtained. For $U=1.4$~eV, $J=0.1$, $0.15$, and $0.2$~eV, the leading instability is the $A_{1g}$ gap that can be parameterized as
\bea \label{eq:delta0123}
 \Delta_{\k \mu} &=& \Delta_{\mu}^{0} + \Delta_{\mu}^{1} \left(\cos k_x + \cos k_y \right)/2 + \Delta_{\mu}^{2} \cos k_x \cos k_y \nn\\
 &+& \Delta_{\mu}^{3} \left(\cos 2k_x + \cos 2k_y \right)/2.
\eea
Two other instabilities has $d_{x^2-y^2}$ and $d_{xy}$ gap symmetries. For $U=1.4$~eV and $J=0.1$~eV as an example, $\lambda=0.24$, $0.19$, and $0.08$ corresponds to the $A_{1g}$ gap, $d_{x^2-y^2}$ gap, and $d_{xy}$ gap, respectively. Increase of $J$ doesn't change this hierarchy, see~\onlinecite{PRBSuppl} for details. In general, the observed situation is typical for iron-based superconductors and was extensively discussed within the leading angular harmonics approximation (LAHA)~\cite{MaitiKorshunovPRL2011,MaitiKorshunovPRB2011}.

In the following, we call the obtained $A_{1g}$ gap the SF gap. The resulting gap angular dependence for $U=1.4$~eV and $J=0.15$~eV is shown in Fig.~\ref{fig:SFset18}. It was fitted by the functional form~(\ref{eq:delta0123}) and the following parameters were obtained (only nonzero values in units of $\Delta_0$ are presented): $\Delta_{\alpha_1}^0=-23.76$, $\Delta_{\alpha_1}^3=26$, $\Delta_{\alpha_2}^0=-4.76$, $\Delta_{\alpha_1}^3=6$, $\Delta_{\beta}^0=6.99$, $\Delta_{\beta}^1=-15.5$, $\Delta_{\beta}^3=-10$.

Spin response for the gap function with the aforementioned parameters is shown in Fig.~\ref{fig:SFset18}. $\mathrm{Im}\chi_{+-}$ in the $s_\pm$ state is also shown there for comparison. The spin resonance peak appears in both cases, but at lower frequencies for the SF gap because of the smaller effective gap at the wave vector $\Q$. Note the similarity between the spin response for the SF gap and for the model gap with $\Delta_{\beta}^0=-1.6$ and $\Delta_{\beta}^1=16$ shown in Fig.~\ref{fig:A1gSet1}. The similarity stems again from the fact that the spin response at the wave vector $\Q$ is governed by the effective gap at the same wave vector. Therefore, even for the different functional forms of the gaps, (\ref{eq:delta}) and (\ref{eq:delta0123}), their comparable values at $\Q$ lead to the similarity in $\mathrm{Im}\chi_{+-}$.

\begin{figure}
\begin{center}
 \includegraphics[width=1.0\columnwidth]{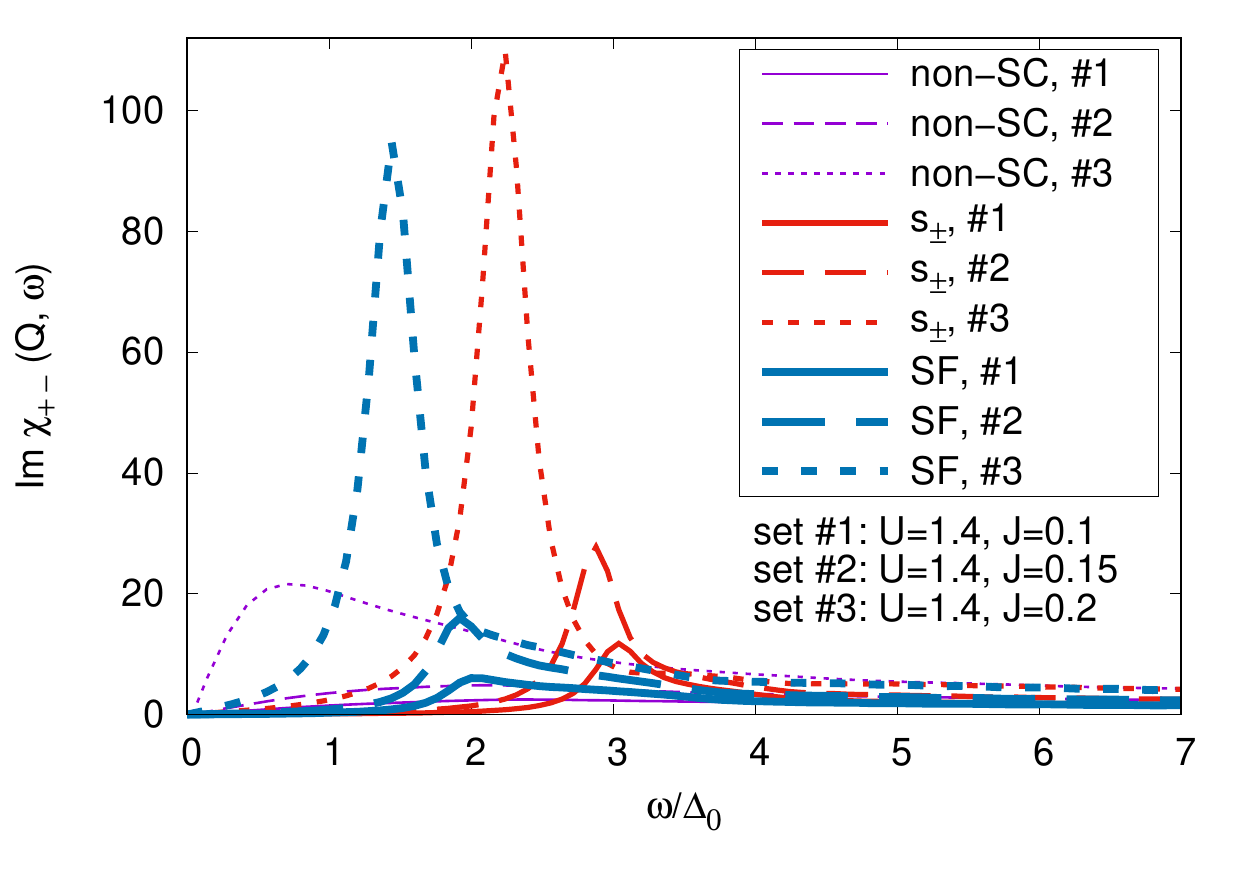}
 \caption{(Color online) Frequency dependence of imaginary part of the spin susceptibility at the wave vector $\Q$ in the normal state (non-SC), for the SF gap, and for the model $s_\pm$ state. Susceptibilities are shown for different sets of interaction parameters.}
\label{fig:SFsets}
\end{center}
\end{figure}

Now we discuss the interaction dependence of the spin resonance peak. Hubbard repulsion was chosen to be $U=1.4$~eV so that the system is on the verge of the magnetic instability; slight increase of it results in the spin susceptibility divergence. Therefore, the spin response in this case is very pronounced. To see what happens near the point with $J=0.15$~eV, $\mathrm{Im}\chi_{+-}$ was calculated for $J=0.1$~eV and $J=0.2$~eV. Since the SF gap structure doesn't change much for the mentioned values of Hund's exchange, the gap parameters are fixed to be the same as for $U=1.4$~eV and $J=0.15$~eV. The results for the SF gap and for the $s_\pm$ gap are shown in Fig.~\ref{fig:SFsets}. Apparently, the peak shifts to lower frequencies and becomes higher and sharper with the increase of $J$. This trend is similar for both the SF and the model $s_\pm$ gaps. Such a behavior is due to the structure of the RPA susceptibility denominator. As was discussed above, in accordance  with the Kramers-Kronig relations, the jump in $\mathrm{Im}\chi_{(0)+-}(\Q,\omega)$ at $\omega_c$ leads to a logarithmic singularity in the real part of the susceptibility. Sice the divergence condition determining the spin resonance peak is $\hat{U}_s \hat\chi_{(0)+-}(\q,\omega) = \hat{I}$, see Eq.~(\ref{eq:chi_s_sol}), the position of the peak is determined by the interaction matrix elements $\hat{U}_s$ and the behavior of $\mathrm{Re}\chi_{(0)+-}(\Q,\omega)$ near the logarithmic singularity. The relation between these two quantities determines $\omega_R$. Here I vary $J$ thus effectively changing $\omega_R$. Increase of interaction parameters decreases $\hat{U}_s^{-1}$ and the divergence can take place for smaller values of $\mathrm{Re}\chi_{(0)+-}(\Q,\omega)$. The latter appears at lower frequencies, thus $\omega_R$ shifts towards zero. That is exactly what is seen in Fig.~\ref{fig:SFsets}.

\section{Conclusions \label{sec:conclusion}}

\begin{table}
\caption{\label{tab} Comparison of peak energies in inelastic neutron scattering, $\omega_{INS}$, and larger and smaller gaps, $\Delta_L$ and $\Delta_S$, in various Fe-based superconductors. Values of frequencies and gaps are given in meV. Here $*$, $**$, and $\dag$ marks gaps extracted from Andreev experiments, BCS fit of $H_{c1}(T)$, and tunneling spectra, respectively; otherwise, gaps are from ARPES. Here, ``$?$'' marks the ``expected'' value of $\omega_{INS}$ (according to value for nearest doping) in a material for which the measurement is absent. If the peak frequency and gaps satisfy condition $\omega_{INS} \leq \min(\Delta_L)+\min(\Delta_S)$, frequency is written in \DLS{bold face}, and if they satisfy condition $\omega_{INS} \leq 2\min(\Delta_L)$, \DLL{italic} is used.}
\begin{tabular}{cccccc}
\hline
\hline
\centering{Material} & $T_c$ (K) & $\omega_{INS}$ & $\min(\Delta_L)$, $\min(\Delta_S)$\\
\hline
BaFe$_{1.9}$Co$_{0.1}$As$_2$ & 19 & \DLS{7.3}-\DLL{9.3}~\cite{Wang2016} & 5.0, 4.0~\cite{Wang2016} \\
BaFe$_{1.866}$Co$_{0.134}$As$_2$ & 25 & \DLS{7.0-8.0}~\cite{Wang2016} & 6.5, 4.6~\cite{Wang2016} \\
BaFe$_{1.81}$Co$_{0.19}$As$_2$ & 19 & \DLS{7.5-9.5}~\cite{Wang2016} & 5.6, 4.6~\cite{Wang2016} \\
BaFe$_{1.85}$Co$_{0.15}$As$_2$ & 25 & \DLS{7.7}-\DLL{10.0}~\cite{Inosov2010,Park} & 6.0, 3.8~\cite{Terashima2009} \\
BaFe$_{1.85}$Co$_{0.15}$As$_2$ & 25.5 & \DLS{9.5?} & 5.6, 4.0~\cite{Kawahara2010} \\
BaFe$_{1.8}$Co$_{0.2}$As$_2$ & 24.5 & \DLS{9.5?} & 8.2, 3.8$*$~\cite{Tortello2010} \\
\hline
Ba$_{0.6}$K$_{0.4}$Fe$_2$As$_2$ & 38 & \DLS{13-14}~\cite{ChristiansonBKFA,Castellan2011,Shan2012} & 10.0, 5.0~\cite{Ding2008} \\
Ba$_{0.6}$K$_{0.4}$Fe$_2$As$_2$ & 38 & \DLL{13-14}~\cite{ChristiansonBKFA,Castellan2011,Shan2012} & 8.0, 4.0~\cite{Wray2008} \\
Ba$_{0.6}$K$_{0.4}$Fe$_2$As$_2$ & 38 & \DLL{13-14}~\cite{ChristiansonBKFA,Castellan2011,Shan2012} & 8.0, 2.0~\cite{Zhang2010,Shimojima2011} \\
Ba$_{0.6}$K$_{0.4}$Fe$_2$As$_2$ & 38 & \DLL{13-14}~\cite{ChristiansonBKFA,Castellan2011,Shan2012} & 8.4, 3.2$\dag$~\cite{Shan2012} \\
Ba$_{0.6}$K$_{0.4}$Fe$_2$As$_2$ & 35 & \DLL{14.8}-15.2~\cite{Castellan2011} & 7.5, 5~\cite{Zhao2008ARPES} \\
Ba$_{0.6}$K$_{0.4}$Fe$_2$As$_2$ & 37.5 & \DLL{14.8-15.2}~\cite{Castellan2011} & 8.5, 1.7$**$~\cite{Ren2008} \\
Ba$_{0.65}$K$_{0.35}$Fe$_2$As$_2$ & 34 & 12.2-13.5~\cite{Castellan2011} & 5.7, 1.4$*$~\cite{Abdel-Hafiez2014} \\
Ba$_{1-x}$K$_{x}$Fe$_2$As$_2$ & 32 & \DLL{14.9-15.3}~\cite{Castellan2011} & 7.8, 1.1~\cite{Evtushinsky2009,Evtushinsky2009NJP} \\
\hline
FeSe & 8 & \DLS{4}~\cite{Wang.nmat4492} & 3.5, 2.5$\dag$~\cite{Kasahara.PNAS.111.16309} \\
FeSe & 8 & \DLL{4}~\cite{Wang.nmat4492} & 2.4, 0.6$*$~\cite{Ponomarev2013} \\
\hline
LiFeAs & 18 & \DLS{4}-12~\cite{Taylor.PhysRevB.83.220514} & 4.7, 2.5~\cite{Borisenko.PhysRevLett.105.067002,Borisenko.symmetry-04-00251,Umezawa.PhysRevLett.108.037002} \\
LiFeAs & 18 & \DLS{4}-12~\cite{Taylor.PhysRevB.83.220514} & 5.1, 0.9$*$~\cite{Kuzmichev2012,Kuzmichev2013} \\
LiFeAs & 18 & \DLS{4}-12~\cite{Taylor.PhysRevB.83.220514} & 5.2, 2.3$\dag$~\cite{Chi.PhysRevLett.109.087002,Hanaguri.PhysRevB.85.214505,Nag.srep27926} \\
\hline
NaFe$_{0.935}$Co$_{0.045}$As & 18 & \DLS{7}~\cite{Zhang.PhysRevLett.111.207002} & 4.5, 4.0~\cite{Zhang.PhysRevLett.111.207002,Ge.PhysRevX.3.011020} \\
NaFe$_{0.935}$Co$_{0.045}$As & 18 & \DLS{6.7-6.9}~\cite{Zhang.PhysRevB.88.064504} & 6.0, 5.0~\cite{Liu.PhysRevB.84.064519} \\
NaFe$_{0.95}$Co$_{0.05}$As & 18 & \DLS{7?} & 6.0, 5.0~\cite{Liu.PhysRevB.84.064519} \\
\hline
CaKFe$_4$As$_4$ & 18 & \DLS{12.5}~\cite{Iida2017} & 10.0, 6.0~\cite{Mou2016} \\
\hline
\hline
\end{tabular}
\end{table}

Within the five-orbital model for iron-based materials, I considered the question on what happens to the spin resonance when the anisotropy of the gap changes. By using both model gap function and the one calculated via the spin fluctuation theory of pairing, it is shown that the spin resonance peak forms for most of the superconducting solutions originating from the spin fluctuation approach to the pairing and having the $A_{1g}$ symmetry, including the $s_\pm$ state. The peak frequency is as higher as larger the zero-amplitude gap magnitude on electron pockets. On the contrary, the increase of the anisotropy leads to the decrease of the peak frequency that is connected with the decrease of the effective gap at the scattering wave vector $\Q$.

As for the experimental verification of the spin resonance appearance, the condition for the spin resonance frequency $\omega_R$ in the case of the anisotropic gaps $\Delta_{L,S}$ becomes $\omega_R \leq \min(\Delta_L)+\min(\Delta_S)$. If all values entering here fulfills this condition, then the observed peak is the true spin resonance. Otherwise, a calculation involving the details of the band structure and superconducting gap is required to make a definite conclusion. I collected available experimental data in Table~\ref{tab}. Note the data for Co-doped materials and a recently discovered CaKFe$_4$As$_4$ fall into the first category and, therefore, demonstrate presence of the $s_\pm$-type gap. Other materials require more efforts from both theoretical and experimental sides to (1) extract precise values of gaps and peak energies and (2) perform calculations for particular band and gap structures.

Additional information can be gained from the temperature dependence of the resonance peak. Since the peak frequency $\omega_R$ is determined by the amplitude of gaps, and the gaps decrease with temperature while approaching $T_c$, $\omega_R(T)$ should also scale with $\Delta_{L,S}(T)$. Simultaneous measurement of the temperature dependence of gaps and peak frequency is highly desirable for understanding of the spin resonance details.

\begin{acknowledgements}
I would like to thank I. Eremin, S.A. Kuzmichev, and T.E. Kuzmicheva for useful discussions. This work was supported in part by Presidium of RAS Program for the Fundamental Studies No. 12, Gosbudget program No. 0356-2017-0030, and Foundation for the advancement of theoretical physics and mathematics ``BASIS''.
\end{acknowledgements}

\bibliography{mmkbibl5}

\begin{thebibliography}{91}%
\makeatletter
\providecommand \@ifxundefined [1]{%
 \@ifx{#1\undefined}
}%
\providecommand \@ifnum [1]{%
 \ifnum #1\expandafter \@firstoftwo
 \else \expandafter \@secondoftwo
 \fi
}%
\providecommand \@ifx [1]{%
 \ifx #1\expandafter \@firstoftwo
 \else \expandafter \@secondoftwo
 \fi
}%
\providecommand \natexlab [1]{#1}%
\providecommand \enquote  [1]{``#1''}%
\providecommand \bibnamefont  [1]{#1}%
\providecommand \bibfnamefont [1]{#1}%
\providecommand \citenamefont [1]{#1}%
\providecommand \href@noop [0]{\@secondoftwo}%
\providecommand \href [0]{\begingroup \@sanitize@url \@href}%
\providecommand \@href[1]{\@@startlink{#1}\@@href}%
\providecommand \@@href[1]{\endgroup#1\@@endlink}%
\providecommand \@sanitize@url [0]{\catcode `\\12\catcode `\$12\catcode
  `\&12\catcode `\#12\catcode `\^12\catcode `\_12\catcode `\%12\relax}%
\providecommand \@@startlink[1]{}%
\providecommand \@@endlink[0]{}%
\providecommand \url  [0]{\begingroup\@sanitize@url \@url }%
\providecommand \@url [1]{\endgroup\@href {#1}{\urlprefix }}%
\providecommand \urlprefix  [0]{URL }%
\providecommand \Eprint [0]{\href }%
\providecommand \doibase [0]{http://dx.doi.org/}%
\providecommand \selectlanguage [0]{\@gobble}%
\providecommand \bibinfo  [0]{\@secondoftwo}%
\providecommand \bibfield  [0]{\@secondoftwo}%
\providecommand \translation [1]{[#1]}%
\providecommand \BibitemOpen [0]{}%
\providecommand \bibitemStop [0]{}%
\providecommand \bibitemNoStop [0]{.\EOS\space}%
\providecommand \EOS [0]{\spacefactor3000\relax}%
\providecommand \BibitemShut  [1]{\csname bibitem#1\endcsname}%
\let\auto@bib@innerbib\@empty
\bibitem [{\citenamefont {Kamihara}\ \emph {et~al.}(2008)\citenamefont
  {Kamihara}, \citenamefont {Watanabe}, \citenamefont {Hirano},\ and\
  \citenamefont {Hosono}}]{y_kamihara_08}%
  \BibitemOpen
  \bibfield  {author} {\bibinfo {author} {\bibfnamefont {Y.}~\bibnamefont
  {Kamihara}}, \bibinfo {author} {\bibfnamefont {T.}~\bibnamefont {Watanabe}},
  \bibinfo {author} {\bibfnamefont {M.}~\bibnamefont {Hirano}}, \ and\ \bibinfo
  {author} {\bibfnamefont {H.}~\bibnamefont {Hosono}},\ }\href {\doibase
  10.1021/ja800073m} {\bibfield  {journal} {\bibinfo  {journal} {Journal of the
  American Chemical Society}\ }\textbf {\bibinfo {volume} {130}},\ \bibinfo
  {pages} {3296} (\bibinfo {year} {2008})}\BibitemShut {NoStop}%
\bibitem [{\citenamefont {Sadovskii}(2008)}]{SadovskiiReview2008}%
  \BibitemOpen
  \bibfield  {author} {\bibinfo {author} {\bibfnamefont {M.~V.}\ \bibnamefont
  {Sadovskii}},\ }\href {\doibase 10.1070/PU2008v051n12ABEH006820} {\bibfield
  {journal} {\bibinfo  {journal} {Phys. Usp.}\ }\textbf {\bibinfo {volume}
  {51}},\ \bibinfo {pages} {1201} (\bibinfo {year} {2008})}\BibitemShut
  {NoStop}%
\bibitem [{\citenamefont {Izyumov}\ and\ \citenamefont
  {Kurmaev}(2008)}]{IzyumovReview2008}%
  \BibitemOpen
  \bibfield  {author} {\bibinfo {author} {\bibfnamefont {Y.~A.}\ \bibnamefont
  {Izyumov}}\ and\ \bibinfo {author} {\bibfnamefont {E.~Z.}\ \bibnamefont
  {Kurmaev}},\ }\href {\doibase 10.1070/PU2008v051n12ABEH006733} {\bibfield
  {journal} {\bibinfo  {journal} {Phys. Usp.}\ }\textbf {\bibinfo {volume}
  {51}},\ \bibinfo {pages} {1261} (\bibinfo {year} {2008})}\BibitemShut
  {NoStop}%
\bibitem [{\citenamefont {Mazin}(2010)}]{MazinReview}%
  \BibitemOpen
  \bibfield  {author} {\bibinfo {author} {\bibfnamefont {I.~I.}\ \bibnamefont
  {Mazin}},\ }\href {http://dx.doi.org/10.1038/nature08914} {\bibfield
  {journal} {\bibinfo  {journal} {Nature}\ }\textbf {\bibinfo {volume} {464}},\
  \bibinfo {pages} {183} (\bibinfo {year} {2010})}\BibitemShut {NoStop}%
\bibitem [{\citenamefont {Paglione}\ and\ \citenamefont
  {Greene}(2010)}]{PaglioneReview}%
  \BibitemOpen
  \bibfield  {author} {\bibinfo {author} {\bibfnamefont {J.}~\bibnamefont
  {Paglione}}\ and\ \bibinfo {author} {\bibfnamefont {R.~L.}\ \bibnamefont
  {Greene}},\ }\href {http://dx.doi.org/10.1038/nphys1759} {\bibfield
  {journal} {\bibinfo  {journal} {Nat. Phys.}\ }\textbf {\bibinfo {volume}
  {6}},\ \bibinfo {pages} {645} (\bibinfo {year} {2010})}\BibitemShut {NoStop}%
\bibitem [{\citenamefont {Johnston}(2010)}]{JohnstonReview}%
  \BibitemOpen
  \bibfield  {author} {\bibinfo {author} {\bibfnamefont {D.~C.}\ \bibnamefont
  {Johnston}},\ }\href {\doibase 10.1080/00018732.2010.513480} {\bibfield
  {journal} {\bibinfo  {journal} {Advances in Physics}\ }\textbf {\bibinfo
  {volume} {59}},\ \bibinfo {pages} {803} (\bibinfo {year} {2010})}\BibitemShut
  {NoStop}%
\bibitem [{\citenamefont {Wen}\ and\ \citenamefont {Li}(2011)}]{WenReview}%
  \BibitemOpen
  \bibfield  {author} {\bibinfo {author} {\bibfnamefont {H.-H.}\ \bibnamefont
  {Wen}}\ and\ \bibinfo {author} {\bibfnamefont {S.}~\bibnamefont {Li}},\
  }\href {\doibase 10.1146/annurev-conmatphys-062910-140518} {\bibfield
  {journal} {\bibinfo  {journal} {Annual Review of Condensed Matter Physics}\
  }\textbf {\bibinfo {volume} {2}},\ \bibinfo {pages} {121} (\bibinfo {year}
  {2011})}\BibitemShut {NoStop}%
\bibitem [{\citenamefont {Stewart}(2011)}]{StewartReview}%
  \BibitemOpen
  \bibfield  {author} {\bibinfo {author} {\bibfnamefont {G.~R.}\ \bibnamefont
  {Stewart}},\ }\href {\doibase 10.1103/RevModPhys.83.1589} {\bibfield
  {journal} {\bibinfo  {journal} {Rev. Mod. Phys.}\ }\textbf {\bibinfo {volume}
  {83}},\ \bibinfo {pages} {1589} (\bibinfo {year} {2011})}\BibitemShut
  {NoStop}%
\bibitem [{\citenamefont {Hirschfeld}\ \emph {et~al.}(2011)\citenamefont
  {Hirschfeld}, \citenamefont {Korshunov},\ and\ \citenamefont
  {Mazin}}]{HirschfeldKorshunov2011}%
  \BibitemOpen
  \bibfield  {author} {\bibinfo {author} {\bibfnamefont {P.~J.}\ \bibnamefont
  {Hirschfeld}}, \bibinfo {author} {\bibfnamefont {M.~M.}\ \bibnamefont
  {Korshunov}}, \ and\ \bibinfo {author} {\bibfnamefont {I.~I.}\ \bibnamefont
  {Mazin}},\ }\href {http://stacks.iop.org/0034-4885/74/i=12/a=124508}
  {\bibfield  {journal} {\bibinfo  {journal} {Reports on Progress in Physics}\
  }\textbf {\bibinfo {volume} {74}},\ \bibinfo {pages} {124508} (\bibinfo
  {year} {2011})}\BibitemShut {NoStop}%
\bibitem [{\citenamefont {Chubukov}(2012)}]{ChubukovReview2012}%
  \BibitemOpen
  \bibfield  {author} {\bibinfo {author} {\bibfnamefont {A.}~\bibnamefont
  {Chubukov}},\ }\href {\doibase 10.1146/annurev-conmatphys-020911-125055}
  {\bibfield  {journal} {\bibinfo  {journal} {Annual Review of Condensed Matter
  Physics}\ }\textbf {\bibinfo {volume} {3}},\ \bibinfo {pages} {57} (\bibinfo
  {year} {2012})}\BibitemShut {NoStop}%
\bibitem [{\citenamefont {Korshunov}(2014)}]{Korshunov2014eng}%
  \BibitemOpen
  \bibfield  {author} {\bibinfo {author} {\bibfnamefont {M.~M.}\ \bibnamefont
  {Korshunov}},\ }\href {\doibase 10.3367/UFNe.0184.201408h.0882} {\bibfield
  {journal} {\bibinfo  {journal} {Physics-Uspekhi}\ }\textbf {\bibinfo {volume}
  {57}},\ \bibinfo {pages} {813} (\bibinfo {year} {2014})}\BibitemShut
  {NoStop}%
\bibitem [{\citenamefont {Hirschfeld}(2016)}]{Hirschfeld2016}%
  \BibitemOpen
  \bibfield  {author} {\bibinfo {author} {\bibfnamefont {P.~J.}\ \bibnamefont
  {Hirschfeld}},\ }\href {\doibase 10.1016/j.crhy.2015.10.002} {\bibfield
  {journal} {\bibinfo  {journal} {Comptes Rendus Physique}\ }\textbf {\bibinfo
  {volume} {17}},\ \bibinfo {pages} {197 } (\bibinfo {year}
  {2016})}\BibitemShut {NoStop}%
\bibitem [{\citenamefont {Kordyuk}(2012)}]{Kordyuk}%
  \BibitemOpen
  \bibfield  {author} {\bibinfo {author} {\bibfnamefont {A.~A.}\ \bibnamefont
  {Kordyuk}},\ }\href {\doibase http://dx.doi.org/10.1063/1.4752092} {\bibfield
   {journal} {\bibinfo  {journal} {Low Temperature Physics}\ }\textbf {\bibinfo
  {volume} {38}},\ \bibinfo {pages} {888} (\bibinfo {year} {2012})}\BibitemShut
  {NoStop}%
\bibitem [{\citenamefont {Anisimov}\ \emph {et~al.}(2008)\citenamefont
  {Anisimov}, \citenamefont {Korotin}, \citenamefont {Streltsov}, \citenamefont
  {Kozhevnikov}, \citenamefont {Kune{\v{s}}}, \citenamefont {Shorikov},\ and\
  \citenamefont {Korotin}}]{Anisimov2008eng}%
  \BibitemOpen
  \bibfield  {author} {\bibinfo {author} {\bibfnamefont {V.~I.}\ \bibnamefont
  {Anisimov}}, \bibinfo {author} {\bibfnamefont {D.~M.}\ \bibnamefont
  {Korotin}}, \bibinfo {author} {\bibfnamefont {S.~V.}\ \bibnamefont
  {Streltsov}}, \bibinfo {author} {\bibfnamefont {A.~V.}\ \bibnamefont
  {Kozhevnikov}}, \bibinfo {author} {\bibfnamefont {J.}~\bibnamefont
  {Kune{\v{s}}}}, \bibinfo {author} {\bibfnamefont {A.~O.}\ \bibnamefont
  {Shorikov}}, \ and\ \bibinfo {author} {\bibfnamefont {M.~A.}\ \bibnamefont
  {Korotin}},\ }\href {\doibase 10.1134/S0021364008230069} {\bibfield
  {journal} {\bibinfo  {journal} {JETP Letters}\ }\textbf {\bibinfo {volume}
  {88}},\ \bibinfo {pages} {729} (\bibinfo {year} {2008})}\BibitemShut
  {NoStop}%
\bibitem [{\citenamefont {Kroll}\ \emph {et~al.}(2008)\citenamefont {Kroll},
  \citenamefont {Bonhommeau}, \citenamefont {Kachel}, \citenamefont {D\"urr},
  \citenamefont {Werner}, \citenamefont {Behr}, \citenamefont {Koitzsch},
  \citenamefont {H\"ubel}, \citenamefont {Leger}, \citenamefont
  {Sch\"onfelder}, \citenamefont {Ariffin}, \citenamefont {Manzke},
  \citenamefont {de~Groot}, \citenamefont {Fink}, \citenamefont {Eschrig},
  \citenamefont {B\"uchner},\ and\ \citenamefont {Knupfer}}]{Kroll2008}%
  \BibitemOpen
  \bibfield  {author} {\bibinfo {author} {\bibfnamefont {T.}~\bibnamefont
  {Kroll}}, \bibinfo {author} {\bibfnamefont {S.}~\bibnamefont {Bonhommeau}},
  \bibinfo {author} {\bibfnamefont {T.}~\bibnamefont {Kachel}}, \bibinfo
  {author} {\bibfnamefont {H.~A.}\ \bibnamefont {D\"urr}}, \bibinfo {author}
  {\bibfnamefont {J.}~\bibnamefont {Werner}}, \bibinfo {author} {\bibfnamefont
  {G.}~\bibnamefont {Behr}}, \bibinfo {author} {\bibfnamefont {A.}~\bibnamefont
  {Koitzsch}}, \bibinfo {author} {\bibfnamefont {R.}~\bibnamefont {H\"ubel}},
  \bibinfo {author} {\bibfnamefont {S.}~\bibnamefont {Leger}}, \bibinfo
  {author} {\bibfnamefont {R.}~\bibnamefont {Sch\"onfelder}}, \bibinfo {author}
  {\bibfnamefont {A.~K.}\ \bibnamefont {Ariffin}}, \bibinfo {author}
  {\bibfnamefont {R.}~\bibnamefont {Manzke}}, \bibinfo {author} {\bibfnamefont
  {F.~M.~F.}\ \bibnamefont {de~Groot}}, \bibinfo {author} {\bibfnamefont
  {J.}~\bibnamefont {Fink}}, \bibinfo {author} {\bibfnamefont {H.}~\bibnamefont
  {Eschrig}}, \bibinfo {author} {\bibfnamefont {B.}~\bibnamefont {B\"uchner}},
  \ and\ \bibinfo {author} {\bibfnamefont {M.}~\bibnamefont {Knupfer}},\ }\href
  {\doibase 10.1103/PhysRevB.78.220502} {\bibfield  {journal} {\bibinfo
  {journal} {Phys. Rev. B}\ }\textbf {\bibinfo {volume} {78}},\ \bibinfo
  {pages} {220502} (\bibinfo {year} {2008})}\BibitemShut {NoStop}%
\bibitem [{\citenamefont {Klauss}\ \emph {et~al.}(2008)\citenamefont {Klauss},
  \citenamefont {Luetkens}, \citenamefont {Klingeler}, \citenamefont {Hess},
  \citenamefont {Litterst}, \citenamefont {Kraken}, \citenamefont {Korshunov},
  \citenamefont {Eremin}, \citenamefont {Drechsler}, \citenamefont {Khasanov},
  \citenamefont {Amato}, \citenamefont {Hamann-Borrero}, \citenamefont {Leps},
  \citenamefont {Kondrat}, \citenamefont {Behr}, \citenamefont {Werner},\ and\
  \citenamefont {B\"uchner}}]{KlaussKorshunov2008}%
  \BibitemOpen
  \bibfield  {author} {\bibinfo {author} {\bibfnamefont {H.-H.}\ \bibnamefont
  {Klauss}}, \bibinfo {author} {\bibfnamefont {H.}~\bibnamefont {Luetkens}},
  \bibinfo {author} {\bibfnamefont {R.}~\bibnamefont {Klingeler}}, \bibinfo
  {author} {\bibfnamefont {C.}~\bibnamefont {Hess}}, \bibinfo {author}
  {\bibfnamefont {F.~J.}\ \bibnamefont {Litterst}}, \bibinfo {author}
  {\bibfnamefont {M.}~\bibnamefont {Kraken}}, \bibinfo {author} {\bibfnamefont
  {M.~M.}\ \bibnamefont {Korshunov}}, \bibinfo {author} {\bibfnamefont
  {I.}~\bibnamefont {Eremin}}, \bibinfo {author} {\bibfnamefont {S.-L.}\
  \bibnamefont {Drechsler}}, \bibinfo {author} {\bibfnamefont {R.}~\bibnamefont
  {Khasanov}}, \bibinfo {author} {\bibfnamefont {A.}~\bibnamefont {Amato}},
  \bibinfo {author} {\bibfnamefont {J.}~\bibnamefont {Hamann-Borrero}},
  \bibinfo {author} {\bibfnamefont {N.}~\bibnamefont {Leps}}, \bibinfo {author}
  {\bibfnamefont {A.}~\bibnamefont {Kondrat}}, \bibinfo {author} {\bibfnamefont
  {G.}~\bibnamefont {Behr}}, \bibinfo {author} {\bibfnamefont {J.}~\bibnamefont
  {Werner}}, \ and\ \bibinfo {author} {\bibfnamefont {B.}~\bibnamefont
  {B\"uchner}},\ }\href {\doibase 10.1103/PhysRevLett.101.077005} {\bibfield
  {journal} {\bibinfo  {journal} {Phys. Rev. Lett.}\ }\textbf {\bibinfo
  {volume} {101}},\ \bibinfo {pages} {077005} (\bibinfo {year}
  {2008})}\BibitemShut {NoStop}%
\bibitem [{\citenamefont {Gretarsson}\ \emph {et~al.}(2011)\citenamefont
  {Gretarsson}, \citenamefont {Lupascu}, \citenamefont {Kim}, \citenamefont
  {Casa}, \citenamefont {Gog}, \citenamefont {Wu}, \citenamefont {Julian},
  \citenamefont {Xu}, \citenamefont {Wen}, \citenamefont {Gu}, \citenamefont
  {Yuan}, \citenamefont {Chen}, \citenamefont {Wang}, \citenamefont {Khim},
  \citenamefont {Kim}, \citenamefont {Ishikado}, \citenamefont {Jarrige},
  \citenamefont {Shamoto}, \citenamefont {Chu}, \citenamefont {Fisher},\ and\
  \citenamefont {Kim}}]{Gretarsson2011}%
  \BibitemOpen
  \bibfield  {author} {\bibinfo {author} {\bibfnamefont {H.}~\bibnamefont
  {Gretarsson}}, \bibinfo {author} {\bibfnamefont {A.}~\bibnamefont {Lupascu}},
  \bibinfo {author} {\bibfnamefont {J.}~\bibnamefont {Kim}}, \bibinfo {author}
  {\bibfnamefont {D.}~\bibnamefont {Casa}}, \bibinfo {author} {\bibfnamefont
  {T.}~\bibnamefont {Gog}}, \bibinfo {author} {\bibfnamefont {W.}~\bibnamefont
  {Wu}}, \bibinfo {author} {\bibfnamefont {S.~R.}\ \bibnamefont {Julian}},
  \bibinfo {author} {\bibfnamefont {Z.~J.}\ \bibnamefont {Xu}}, \bibinfo
  {author} {\bibfnamefont {J.~S.}\ \bibnamefont {Wen}}, \bibinfo {author}
  {\bibfnamefont {G.~D.}\ \bibnamefont {Gu}}, \bibinfo {author} {\bibfnamefont
  {R.~H.}\ \bibnamefont {Yuan}}, \bibinfo {author} {\bibfnamefont {Z.~G.}\
  \bibnamefont {Chen}}, \bibinfo {author} {\bibfnamefont {N.-L.}\ \bibnamefont
  {Wang}}, \bibinfo {author} {\bibfnamefont {S.}~\bibnamefont {Khim}}, \bibinfo
  {author} {\bibfnamefont {K.~H.}\ \bibnamefont {Kim}}, \bibinfo {author}
  {\bibfnamefont {M.}~\bibnamefont {Ishikado}}, \bibinfo {author}
  {\bibfnamefont {I.}~\bibnamefont {Jarrige}}, \bibinfo {author} {\bibfnamefont
  {S.}~\bibnamefont {Shamoto}}, \bibinfo {author} {\bibfnamefont {J.-H.}\
  \bibnamefont {Chu}}, \bibinfo {author} {\bibfnamefont {I.~R.}\ \bibnamefont
  {Fisher}}, \ and\ \bibinfo {author} {\bibfnamefont {Y.-J.}\ \bibnamefont
  {Kim}},\ }\href {\doibase 10.1103/PhysRevB.84.100509} {\bibfield  {journal}
  {\bibinfo  {journal} {Phys. Rev. B}\ }\textbf {\bibinfo {volume} {84}},\
  \bibinfo {pages} {100509} (\bibinfo {year} {2011})}\BibitemShut {NoStop}%
\bibitem [{\citenamefont {Haule}\ and\ \citenamefont
  {Kotliar}(2009)}]{Haule2009}%
  \BibitemOpen
  \bibfield  {author} {\bibinfo {author} {\bibfnamefont {K.}~\bibnamefont
  {Haule}}\ and\ \bibinfo {author} {\bibfnamefont {G.}~\bibnamefont
  {Kotliar}},\ }\href {http://stacks.iop.org/1367-2630/11/i=2/a=025021}
  {\bibfield  {journal} {\bibinfo  {journal} {New Journal of Physics}\ }\textbf
  {\bibinfo {volume} {11}},\ \bibinfo {pages} {025021} (\bibinfo {year}
  {2009})}\BibitemShut {NoStop}%
\bibitem [{\citenamefont {Hansmann}\ \emph {et~al.}(2010)\citenamefont
  {Hansmann}, \citenamefont {Arita}, \citenamefont {Toschi}, \citenamefont
  {Sakai}, \citenamefont {Sangiovanni},\ and\ \citenamefont
  {Held}}]{Hansmann2010}%
  \BibitemOpen
  \bibfield  {author} {\bibinfo {author} {\bibfnamefont {P.}~\bibnamefont
  {Hansmann}}, \bibinfo {author} {\bibfnamefont {R.}~\bibnamefont {Arita}},
  \bibinfo {author} {\bibfnamefont {A.}~\bibnamefont {Toschi}}, \bibinfo
  {author} {\bibfnamefont {S.}~\bibnamefont {Sakai}}, \bibinfo {author}
  {\bibfnamefont {G.}~\bibnamefont {Sangiovanni}}, \ and\ \bibinfo {author}
  {\bibfnamefont {K.}~\bibnamefont {Held}},\ }\href {\doibase
  10.1103/PhysRevLett.104.197002} {\bibfield  {journal} {\bibinfo  {journal}
  {Phys. Rev. Lett.}\ }\textbf {\bibinfo {volume} {104}},\ \bibinfo {pages}
  {197002} (\bibinfo {year} {2010})}\BibitemShut {NoStop}%
\bibitem [{\citenamefont {Toschi}\ \emph {et~al.}(2012)\citenamefont {Toschi},
  \citenamefont {Arita}, \citenamefont {Hansmann}, \citenamefont
  {Sangiovanni},\ and\ \citenamefont {Held}}]{Toschi2012}%
  \BibitemOpen
  \bibfield  {author} {\bibinfo {author} {\bibfnamefont {A.}~\bibnamefont
  {Toschi}}, \bibinfo {author} {\bibfnamefont {R.}~\bibnamefont {Arita}},
  \bibinfo {author} {\bibfnamefont {P.}~\bibnamefont {Hansmann}}, \bibinfo
  {author} {\bibfnamefont {G.}~\bibnamefont {Sangiovanni}}, \ and\ \bibinfo
  {author} {\bibfnamefont {K.}~\bibnamefont {Held}},\ }\href {\doibase
  10.1103/PhysRevB.86.064411} {\bibfield  {journal} {\bibinfo  {journal} {Phys.
  Rev. B}\ }\textbf {\bibinfo {volume} {86}},\ \bibinfo {pages} {064411}
  (\bibinfo {year} {2012})}\BibitemShut {NoStop}%
\bibitem [{\citenamefont {Liu}\ \emph {et~al.}(2012)\citenamefont {Liu},
  \citenamefont {Harriger}, \citenamefont {Luo}, \citenamefont {Wang},
  \citenamefont {Ewings}, \citenamefont {Guidi}, \citenamefont {Park},
  \citenamefont {Haule}, \citenamefont {Kotliar}, \citenamefont {Hayden},\ and\
  \citenamefont {Dai}}]{Liu2012}%
  \BibitemOpen
  \bibfield  {author} {\bibinfo {author} {\bibfnamefont {M.}~\bibnamefont
  {Liu}}, \bibinfo {author} {\bibfnamefont {L.~W.}\ \bibnamefont {Harriger}},
  \bibinfo {author} {\bibfnamefont {H.}~\bibnamefont {Luo}}, \bibinfo {author}
  {\bibfnamefont {M.}~\bibnamefont {Wang}}, \bibinfo {author} {\bibfnamefont
  {R.~A.}\ \bibnamefont {Ewings}}, \bibinfo {author} {\bibfnamefont
  {T.}~\bibnamefont {Guidi}}, \bibinfo {author} {\bibfnamefont
  {H.}~\bibnamefont {Park}}, \bibinfo {author} {\bibfnamefont {K.}~\bibnamefont
  {Haule}}, \bibinfo {author} {\bibfnamefont {G.}~\bibnamefont {Kotliar}},
  \bibinfo {author} {\bibfnamefont {S.~M.}\ \bibnamefont {Hayden}}, \ and\
  \bibinfo {author} {\bibfnamefont {P.}~\bibnamefont {Dai}},\ }\href {\doibase
  10.1038/nphys2268} {\bibfield  {journal} {\bibinfo  {journal} {Nature
  Physics}\ }\textbf {\bibinfo {volume} {8}},\ \bibinfo {pages} {376} (\bibinfo
  {year} {2012})}\BibitemShut {NoStop}%
\bibitem [{\citenamefont {Mazin}\ \emph {et~al.}(2008)\citenamefont {Mazin},
  \citenamefont {Singh}, \citenamefont {Johannes},\ and\ \citenamefont
  {Du}}]{Mazin2008}%
  \BibitemOpen
  \bibfield  {author} {\bibinfo {author} {\bibfnamefont {I.~I.}\ \bibnamefont
  {Mazin}}, \bibinfo {author} {\bibfnamefont {D.~J.}\ \bibnamefont {Singh}},
  \bibinfo {author} {\bibfnamefont {M.~D.}\ \bibnamefont {Johannes}}, \ and\
  \bibinfo {author} {\bibfnamefont {M.~H.}\ \bibnamefont {Du}},\ }\href
  {\doibase 10.1103/PhysRevLett.101.057003} {\bibfield  {journal} {\bibinfo
  {journal} {Phys. Rev. Lett.}\ }\textbf {\bibinfo {volume} {101}},\ \bibinfo
  {pages} {057003} (\bibinfo {year} {2008})}\BibitemShut {NoStop}%
\bibitem [{\citenamefont {Graser}\ \emph {et~al.}(2009)\citenamefont {Graser},
  \citenamefont {Maier}, \citenamefont {Hirschfeld},\ and\ \citenamefont
  {Scalapino}}]{Graser2009}%
  \BibitemOpen
  \bibfield  {author} {\bibinfo {author} {\bibfnamefont {S.}~\bibnamefont
  {Graser}}, \bibinfo {author} {\bibfnamefont {T.}~\bibnamefont {Maier}},
  \bibinfo {author} {\bibfnamefont {P.}~\bibnamefont {Hirschfeld}}, \ and\
  \bibinfo {author} {\bibfnamefont {D.}~\bibnamefont {Scalapino}},\ }\href
  {http://stacks.iop.org/1367-2630/11/i=2/a=025016} {\bibfield  {journal}
  {\bibinfo  {journal} {New Journal of Physics}\ }\textbf {\bibinfo {volume}
  {11}},\ \bibinfo {pages} {025016} (\bibinfo {year} {2009})}\BibitemShut
  {NoStop}%
\bibitem [{\citenamefont {Kuroki}\ \emph {et~al.}(2008)\citenamefont {Kuroki},
  \citenamefont {Onari}, \citenamefont {Arita}, \citenamefont {Usui},
  \citenamefont {Tanaka}, \citenamefont {Kontani},\ and\ \citenamefont
  {Aoki}}]{Kuroki2008}%
  \BibitemOpen
  \bibfield  {author} {\bibinfo {author} {\bibfnamefont {K.}~\bibnamefont
  {Kuroki}}, \bibinfo {author} {\bibfnamefont {S.}~\bibnamefont {Onari}},
  \bibinfo {author} {\bibfnamefont {R.}~\bibnamefont {Arita}}, \bibinfo
  {author} {\bibfnamefont {H.}~\bibnamefont {Usui}}, \bibinfo {author}
  {\bibfnamefont {Y.}~\bibnamefont {Tanaka}}, \bibinfo {author} {\bibfnamefont
  {H.}~\bibnamefont {Kontani}}, \ and\ \bibinfo {author} {\bibfnamefont
  {H.}~\bibnamefont {Aoki}},\ }\href {\doibase 10.1103/PhysRevLett.101.087004}
  {\bibfield  {journal} {\bibinfo  {journal} {Phys. Rev. Lett.}\ }\textbf
  {\bibinfo {volume} {101}},\ \bibinfo {pages} {087004} (\bibinfo {year}
  {2008})}\BibitemShut {NoStop}%
\bibitem [{\citenamefont {Chubukov}\ \emph {et~al.}(2008)\citenamefont
  {Chubukov}, \citenamefont {Efremov},\ and\ \citenamefont
  {Eremin}}]{Chubukov2008}%
  \BibitemOpen
  \bibfield  {author} {\bibinfo {author} {\bibfnamefont {A.~V.}\ \bibnamefont
  {Chubukov}}, \bibinfo {author} {\bibfnamefont {D.~V.}\ \bibnamefont
  {Efremov}}, \ and\ \bibinfo {author} {\bibfnamefont {I.}~\bibnamefont
  {Eremin}},\ }\href {\doibase 10.1103/PhysRevB.78.134512} {\bibfield
  {journal} {\bibinfo  {journal} {Phys. Rev. B}\ }\textbf {\bibinfo {volume}
  {78}},\ \bibinfo {pages} {134512} (\bibinfo {year} {2008})}\BibitemShut
  {NoStop}%
\bibitem [{\citenamefont {Maiti}\ \emph
  {et~al.}(2011{\natexlab{a}})\citenamefont {Maiti}, \citenamefont {Korshunov},
  \citenamefont {Maier}, \citenamefont {Hirschfeld},\ and\ \citenamefont
  {Chubukov}}]{MaitiKorshunovPRL2011}%
  \BibitemOpen
  \bibfield  {author} {\bibinfo {author} {\bibfnamefont {S.}~\bibnamefont
  {Maiti}}, \bibinfo {author} {\bibfnamefont {M.~M.}\ \bibnamefont
  {Korshunov}}, \bibinfo {author} {\bibfnamefont {T.~A.}\ \bibnamefont
  {Maier}}, \bibinfo {author} {\bibfnamefont {P.~J.}\ \bibnamefont
  {Hirschfeld}}, \ and\ \bibinfo {author} {\bibfnamefont {A.~V.}\ \bibnamefont
  {Chubukov}},\ }\href {\doibase 10.1103/PhysRevLett.107.147002} {\bibfield
  {journal} {\bibinfo  {journal} {Phys. Rev. Lett.}\ }\textbf {\bibinfo
  {volume} {107}},\ \bibinfo {pages} {147002} (\bibinfo {year}
  {2011}{\natexlab{a}})}\BibitemShut {NoStop}%
\bibitem [{\citenamefont {Maiti}\ \emph
  {et~al.}(2011{\natexlab{b}})\citenamefont {Maiti}, \citenamefont {Korshunov},
  \citenamefont {Maier}, \citenamefont {Hirschfeld},\ and\ \citenamefont
  {Chubukov}}]{MaitiKorshunovPRB2011}%
  \BibitemOpen
  \bibfield  {author} {\bibinfo {author} {\bibfnamefont {S.}~\bibnamefont
  {Maiti}}, \bibinfo {author} {\bibfnamefont {M.~M.}\ \bibnamefont
  {Korshunov}}, \bibinfo {author} {\bibfnamefont {T.~A.}\ \bibnamefont
  {Maier}}, \bibinfo {author} {\bibfnamefont {P.~J.}\ \bibnamefont
  {Hirschfeld}}, \ and\ \bibinfo {author} {\bibfnamefont {A.~V.}\ \bibnamefont
  {Chubukov}},\ }\href {\doibase 10.1103/PhysRevB.84.224505} {\bibfield
  {journal} {\bibinfo  {journal} {Phys. Rev. B}\ }\textbf {\bibinfo {volume}
  {84}},\ \bibinfo {pages} {224505} (\bibinfo {year}
  {2011}{\natexlab{b}})}\BibitemShut {NoStop}%
\bibitem [{\citenamefont {Classen}\ \emph {et~al.}(2017)\citenamefont
  {Classen}, \citenamefont {Xing}, \citenamefont {Khodas},\ and\ \citenamefont
  {Chubukov}}]{Classen2017}%
  \BibitemOpen
  \bibfield  {author} {\bibinfo {author} {\bibfnamefont {L.}~\bibnamefont
  {Classen}}, \bibinfo {author} {\bibfnamefont {R.-Q.}\ \bibnamefont {Xing}},
  \bibinfo {author} {\bibfnamefont {M.}~\bibnamefont {Khodas}}, \ and\ \bibinfo
  {author} {\bibfnamefont {A.~V.}\ \bibnamefont {Chubukov}},\ }\href {\doibase
  10.1103/PhysRevLett.118.037001} {\bibfield  {journal} {\bibinfo  {journal}
  {Phys. Rev. Lett.}\ }\textbf {\bibinfo {volume} {118}},\ \bibinfo {pages}
  {037001} (\bibinfo {year} {2017})}\BibitemShut {NoStop}%
\bibitem [{\citenamefont {Kontani}\ and\ \citenamefont
  {Onari}(2010)}]{Kontani}%
  \BibitemOpen
  \bibfield  {author} {\bibinfo {author} {\bibfnamefont {H.}~\bibnamefont
  {Kontani}}\ and\ \bibinfo {author} {\bibfnamefont {S.}~\bibnamefont
  {Onari}},\ }\href {\doibase 10.1103/PhysRevLett.104.157001} {\bibfield
  {journal} {\bibinfo  {journal} {Phys. Rev. Lett.}\ }\textbf {\bibinfo
  {volume} {104}},\ \bibinfo {pages} {157001} (\bibinfo {year}
  {2010})}\BibitemShut {NoStop}%
\bibitem [{\citenamefont {Korshunov}\ and\ \citenamefont
  {Eremin}(2008)}]{KorshunovEreminResonance2008}%
  \BibitemOpen
  \bibfield  {author} {\bibinfo {author} {\bibfnamefont {M.~M.}\ \bibnamefont
  {Korshunov}}\ and\ \bibinfo {author} {\bibfnamefont {I.}~\bibnamefont
  {Eremin}},\ }\href {\doibase 10.1103/PhysRevB.78.140509} {\bibfield
  {journal} {\bibinfo  {journal} {Phys. Rev. B}\ }\textbf {\bibinfo {volume}
  {78}},\ \bibinfo {pages} {140509} (\bibinfo {year} {2008})}\BibitemShut
  {NoStop}%
\bibitem [{\citenamefont {Maier}\ and\ \citenamefont
  {Scalapino}(2008)}]{Maier2008}%
  \BibitemOpen
  \bibfield  {author} {\bibinfo {author} {\bibfnamefont {T.~A.}\ \bibnamefont
  {Maier}}\ and\ \bibinfo {author} {\bibfnamefont {D.~J.}\ \bibnamefont
  {Scalapino}},\ }\href {\doibase 10.1103/PhysRevB.78.020514} {\bibfield
  {journal} {\bibinfo  {journal} {Phys. Rev. B}\ }\textbf {\bibinfo {volume}
  {78}},\ \bibinfo {pages} {020514} (\bibinfo {year} {2008})}\BibitemShut
  {NoStop}%
\bibitem [{\citenamefont {Maier}\ \emph {et~al.}(2009)\citenamefont {Maier},
  \citenamefont {Graser}, \citenamefont {Scalapino},\ and\ \citenamefont
  {Hirschfeld}}]{Maier2009}%
  \BibitemOpen
  \bibfield  {author} {\bibinfo {author} {\bibfnamefont {T.~A.}\ \bibnamefont
  {Maier}}, \bibinfo {author} {\bibfnamefont {S.}~\bibnamefont {Graser}},
  \bibinfo {author} {\bibfnamefont {D.~J.}\ \bibnamefont {Scalapino}}, \ and\
  \bibinfo {author} {\bibfnamefont {P.}~\bibnamefont {Hirschfeld}},\ }\href
  {\doibase 10.1103/PhysRevB.79.134520} {\bibfield  {journal} {\bibinfo
  {journal} {Phys. Rev. B}\ }\textbf {\bibinfo {volume} {79}},\ \bibinfo
  {pages} {134520} (\bibinfo {year} {2009})}\BibitemShut {NoStop}%
\bibitem [{\citenamefont {Christianson}\ \emph {et~al.}(2008)\citenamefont
  {Christianson}, \citenamefont {Goremychkin}, \citenamefont {Osborn},
  \citenamefont {Rosenkranz}, \citenamefont {Lumsden}, \citenamefont
  {Malliakas}, \citenamefont {Todorov}, \citenamefont {Claus}, \citenamefont
  {Chung}, \citenamefont {Kanatzidis}, \citenamefont {Bewley},\ and\
  \citenamefont {Guidi}}]{ChristiansonBKFA}%
  \BibitemOpen
  \bibfield  {author} {\bibinfo {author} {\bibfnamefont {A.~D.}\ \bibnamefont
  {Christianson}}, \bibinfo {author} {\bibfnamefont {E.~A.}\ \bibnamefont
  {Goremychkin}}, \bibinfo {author} {\bibfnamefont {R.}~\bibnamefont {Osborn}},
  \bibinfo {author} {\bibfnamefont {S.}~\bibnamefont {Rosenkranz}}, \bibinfo
  {author} {\bibfnamefont {M.~D.}\ \bibnamefont {Lumsden}}, \bibinfo {author}
  {\bibfnamefont {C.~D.}\ \bibnamefont {Malliakas}}, \bibinfo {author}
  {\bibfnamefont {I.~S.}\ \bibnamefont {Todorov}}, \bibinfo {author}
  {\bibfnamefont {H.}~\bibnamefont {Claus}}, \bibinfo {author} {\bibfnamefont
  {D.~Y.}\ \bibnamefont {Chung}}, \bibinfo {author} {\bibfnamefont {M.~G.}\
  \bibnamefont {Kanatzidis}}, \bibinfo {author} {\bibfnamefont {R.~I.}\
  \bibnamefont {Bewley}}, \ and\ \bibinfo {author} {\bibfnamefont
  {T.}~\bibnamefont {Guidi}},\ }\href {\doibase 10.1038/nature07625} {\bibfield
   {journal} {\bibinfo  {journal} {Nature}\ }\textbf {\bibinfo {volume}
  {456}},\ \bibinfo {pages} {930} (\bibinfo {year} {2008})}\BibitemShut
  {NoStop}%
\bibitem [{\citenamefont {Christianson}\ \emph {et~al.}(2009)\citenamefont
  {Christianson}, \citenamefont {Lumsden}, \citenamefont {Nagler},
  \citenamefont {MacDougall}, \citenamefont {McGuire}, \citenamefont {Sefat},
  \citenamefont {Jin}, \citenamefont {Sales},\ and\ \citenamefont
  {Mandrus}}]{ChristiansonBFCA}%
  \BibitemOpen
  \bibfield  {author} {\bibinfo {author} {\bibfnamefont {A.~D.}\ \bibnamefont
  {Christianson}}, \bibinfo {author} {\bibfnamefont {M.~D.}\ \bibnamefont
  {Lumsden}}, \bibinfo {author} {\bibfnamefont {S.~E.}\ \bibnamefont {Nagler}},
  \bibinfo {author} {\bibfnamefont {G.~J.}\ \bibnamefont {MacDougall}},
  \bibinfo {author} {\bibfnamefont {M.~A.}\ \bibnamefont {McGuire}}, \bibinfo
  {author} {\bibfnamefont {A.~S.}\ \bibnamefont {Sefat}}, \bibinfo {author}
  {\bibfnamefont {R.}~\bibnamefont {Jin}}, \bibinfo {author} {\bibfnamefont
  {B.~C.}\ \bibnamefont {Sales}}, \ and\ \bibinfo {author} {\bibfnamefont
  {D.}~\bibnamefont {Mandrus}},\ }\href {\doibase
  10.1103/PhysRevLett.103.087002} {\bibfield  {journal} {\bibinfo  {journal}
  {Phys. Rev. Lett.}\ }\textbf {\bibinfo {volume} {103}},\ \bibinfo {pages}
  {087002} (\bibinfo {year} {2009})}\BibitemShut {NoStop}%
\bibitem [{\citenamefont {Qiu}\ \emph {et~al.}(2009)\citenamefont {Qiu},
  \citenamefont {Bao}, \citenamefont {Zhao}, \citenamefont {Broholm},
  \citenamefont {Stanev}, \citenamefont {Tesanovic}, \citenamefont
  {Gasparovic}, \citenamefont {Chang}, \citenamefont {Hu}, \citenamefont
  {Qian}, \citenamefont {Fang},\ and\ \citenamefont {Mao}}]{QiuFeSeTe}%
  \BibitemOpen
  \bibfield  {author} {\bibinfo {author} {\bibfnamefont {Y.}~\bibnamefont
  {Qiu}}, \bibinfo {author} {\bibfnamefont {W.}~\bibnamefont {Bao}}, \bibinfo
  {author} {\bibfnamefont {Y.}~\bibnamefont {Zhao}}, \bibinfo {author}
  {\bibfnamefont {C.}~\bibnamefont {Broholm}}, \bibinfo {author} {\bibfnamefont
  {V.}~\bibnamefont {Stanev}}, \bibinfo {author} {\bibfnamefont
  {Z.}~\bibnamefont {Tesanovic}}, \bibinfo {author} {\bibfnamefont {Y.~C.}\
  \bibnamefont {Gasparovic}}, \bibinfo {author} {\bibfnamefont
  {S.}~\bibnamefont {Chang}}, \bibinfo {author} {\bibfnamefont
  {J.}~\bibnamefont {Hu}}, \bibinfo {author} {\bibfnamefont {B.}~\bibnamefont
  {Qian}}, \bibinfo {author} {\bibfnamefont {M.}~\bibnamefont {Fang}}, \ and\
  \bibinfo {author} {\bibfnamefont {Z.}~\bibnamefont {Mao}},\ }\href {\doibase
  10.1103/PhysRevLett.103.067008} {\bibfield  {journal} {\bibinfo  {journal}
  {Phys. Rev. Lett.}\ }\textbf {\bibinfo {volume} {103}},\ \bibinfo {pages}
  {067008} (\bibinfo {year} {2009})}\BibitemShut {NoStop}%
\bibitem [{\citenamefont {Park}\ \emph {et~al.}(2010)\citenamefont {Park},
  \citenamefont {Inosov}, \citenamefont {Yaresko}, \citenamefont {Graser},
  \citenamefont {Sun}, \citenamefont {Bourges}, \citenamefont {Sidis},
  \citenamefont {Li}, \citenamefont {Kim}, \citenamefont {Haug}, \citenamefont
  {Ivanov}, \citenamefont {Hradil}, \citenamefont {Schneidewind}, \citenamefont
  {Link}, \citenamefont {Faulhaber}, \citenamefont {Glavatskyy}, \citenamefont
  {Lin}, \citenamefont {Keimer},\ and\ \citenamefont {Hinkov}}]{Park}%
  \BibitemOpen
  \bibfield  {author} {\bibinfo {author} {\bibfnamefont {J.~T.}\ \bibnamefont
  {Park}}, \bibinfo {author} {\bibfnamefont {D.~S.}\ \bibnamefont {Inosov}},
  \bibinfo {author} {\bibfnamefont {A.}~\bibnamefont {Yaresko}}, \bibinfo
  {author} {\bibfnamefont {S.}~\bibnamefont {Graser}}, \bibinfo {author}
  {\bibfnamefont {D.~L.}\ \bibnamefont {Sun}}, \bibinfo {author} {\bibfnamefont
  {P.}~\bibnamefont {Bourges}}, \bibinfo {author} {\bibfnamefont
  {Y.}~\bibnamefont {Sidis}}, \bibinfo {author} {\bibfnamefont
  {Y.}~\bibnamefont {Li}}, \bibinfo {author} {\bibfnamefont {J.-H.}\
  \bibnamefont {Kim}}, \bibinfo {author} {\bibfnamefont {D.}~\bibnamefont
  {Haug}}, \bibinfo {author} {\bibfnamefont {A.}~\bibnamefont {Ivanov}},
  \bibinfo {author} {\bibfnamefont {K.}~\bibnamefont {Hradil}}, \bibinfo
  {author} {\bibfnamefont {A.}~\bibnamefont {Schneidewind}}, \bibinfo {author}
  {\bibfnamefont {P.}~\bibnamefont {Link}}, \bibinfo {author} {\bibfnamefont
  {E.}~\bibnamefont {Faulhaber}}, \bibinfo {author} {\bibfnamefont
  {I.}~\bibnamefont {Glavatskyy}}, \bibinfo {author} {\bibfnamefont {C.~T.}\
  \bibnamefont {Lin}}, \bibinfo {author} {\bibfnamefont {B.}~\bibnamefont
  {Keimer}}, \ and\ \bibinfo {author} {\bibfnamefont {V.}~\bibnamefont
  {Hinkov}},\ }\href {\doibase 10.1103/PhysRevB.82.134503} {\bibfield
  {journal} {\bibinfo  {journal} {Phys. Rev. B}\ }\textbf {\bibinfo {volume}
  {82}},\ \bibinfo {pages} {134503} (\bibinfo {year} {2010})}\BibitemShut
  {NoStop}%
\bibitem [{\citenamefont {Babkevich}\ \emph {et~al.}(2010)\citenamefont
  {Babkevich}, \citenamefont {Bendele}, \citenamefont {Boothroyd},
  \citenamefont {Conder}, \citenamefont {Gvasaliya}, \citenamefont {Khasanov},
  \citenamefont {Pomjakushina},\ and\ \citenamefont {Roessli}}]{Babkevich}%
  \BibitemOpen
  \bibfield  {author} {\bibinfo {author} {\bibfnamefont {P.}~\bibnamefont
  {Babkevich}}, \bibinfo {author} {\bibfnamefont {M.}~\bibnamefont {Bendele}},
  \bibinfo {author} {\bibfnamefont {A.}~\bibnamefont {Boothroyd}}, \bibinfo
  {author} {\bibfnamefont {K.}~\bibnamefont {Conder}}, \bibinfo {author}
  {\bibfnamefont {S.}~\bibnamefont {Gvasaliya}}, \bibinfo {author}
  {\bibfnamefont {R.}~\bibnamefont {Khasanov}}, \bibinfo {author}
  {\bibfnamefont {E.}~\bibnamefont {Pomjakushina}}, \ and\ \bibinfo {author}
  {\bibfnamefont {B.}~\bibnamefont {Roessli}},\ }\href
  {http://stacks.iop.org/0953-8984/22/i=14/a=142202} {\bibfield  {journal}
  {\bibinfo  {journal} {Journal of Physics: Condensed Matter}\ }\textbf
  {\bibinfo {volume} {22}},\ \bibinfo {pages} {142202} (\bibinfo {year}
  {2010})}\BibitemShut {NoStop}%
\bibitem [{\citenamefont {Inosov}\ \emph {et~al.}(2010)\citenamefont {Inosov},
  \citenamefont {Park}, \citenamefont {Bourges}, \citenamefont {Sun},
  \citenamefont {Sidis}, \citenamefont {Schneidewind}, \citenamefont {Hradil},
  \citenamefont {Haug}, \citenamefont {Lin}, \citenamefont {Keimer},\ and\
  \citenamefont {Hinkov}}]{Inosov2010}%
  \BibitemOpen
  \bibfield  {author} {\bibinfo {author} {\bibfnamefont {D.~S.}\ \bibnamefont
  {Inosov}}, \bibinfo {author} {\bibfnamefont {J.~T.}\ \bibnamefont {Park}},
  \bibinfo {author} {\bibfnamefont {P.}~\bibnamefont {Bourges}}, \bibinfo
  {author} {\bibfnamefont {D.~L.}\ \bibnamefont {Sun}}, \bibinfo {author}
  {\bibfnamefont {Y.}~\bibnamefont {Sidis}}, \bibinfo {author} {\bibfnamefont
  {A.}~\bibnamefont {Schneidewind}}, \bibinfo {author} {\bibfnamefont
  {K.}~\bibnamefont {Hradil}}, \bibinfo {author} {\bibfnamefont
  {D.}~\bibnamefont {Haug}}, \bibinfo {author} {\bibfnamefont {C.~T.}\
  \bibnamefont {Lin}}, \bibinfo {author} {\bibfnamefont {B.}~\bibnamefont
  {Keimer}}, \ and\ \bibinfo {author} {\bibfnamefont {V.}~\bibnamefont
  {Hinkov}},\ }\href {\doibase 10.1038/nphys1483} {\bibfield  {journal}
  {\bibinfo  {journal} {Nat. Phys.}\ }\textbf {\bibinfo {volume} {6}},\
  \bibinfo {pages} {178} (\bibinfo {year} {2010})}\BibitemShut {NoStop}%
\bibitem [{\citenamefont {Argyriou}\ \emph {et~al.}(2010)\citenamefont
  {Argyriou}, \citenamefont {Hiess}, \citenamefont {Akbari}, \citenamefont
  {Eremin}, \citenamefont {Korshunov}, \citenamefont {Hu}, \citenamefont
  {Qian}, \citenamefont {Mao}, \citenamefont {Qiu}, \citenamefont {Broholm},\
  and\ \citenamefont {Bao}}]{ArgyriouKorshunov2010}%
  \BibitemOpen
  \bibfield  {author} {\bibinfo {author} {\bibfnamefont {D.~N.}\ \bibnamefont
  {Argyriou}}, \bibinfo {author} {\bibfnamefont {A.}~\bibnamefont {Hiess}},
  \bibinfo {author} {\bibfnamefont {A.}~\bibnamefont {Akbari}}, \bibinfo
  {author} {\bibfnamefont {I.}~\bibnamefont {Eremin}}, \bibinfo {author}
  {\bibfnamefont {M.~M.}\ \bibnamefont {Korshunov}}, \bibinfo {author}
  {\bibfnamefont {J.}~\bibnamefont {Hu}}, \bibinfo {author} {\bibfnamefont
  {B.}~\bibnamefont {Qian}}, \bibinfo {author} {\bibfnamefont {Z.}~\bibnamefont
  {Mao}}, \bibinfo {author} {\bibfnamefont {Y.}~\bibnamefont {Qiu}}, \bibinfo
  {author} {\bibfnamefont {C.}~\bibnamefont {Broholm}}, \ and\ \bibinfo
  {author} {\bibfnamefont {W.}~\bibnamefont {Bao}},\ }\href {\doibase
  10.1103/PhysRevB.81.220503} {\bibfield  {journal} {\bibinfo  {journal} {Phys.
  Rev. B}\ }\textbf {\bibinfo {volume} {81}},\ \bibinfo {pages} {220503}
  (\bibinfo {year} {2010})}\BibitemShut {NoStop}%
\bibitem [{\citenamefont {Lumsden}\ and\ \citenamefont
  {Christianson}(2010)}]{LumsdenReview}%
  \BibitemOpen
  \bibfield  {author} {\bibinfo {author} {\bibfnamefont {M.~D.}\ \bibnamefont
  {Lumsden}}\ and\ \bibinfo {author} {\bibfnamefont {A.~D.}\ \bibnamefont
  {Christianson}},\ }\href {http://stacks.iop.org/0953-8984/22/i=20/a=203203}
  {\bibfield  {journal} {\bibinfo  {journal} {Journal of Physics: Condensed
  Matter}\ }\textbf {\bibinfo {volume} {22}},\ \bibinfo {pages} {203203}
  (\bibinfo {year} {2010})}\BibitemShut {NoStop}%
\bibitem [{\citenamefont {Castellan}\ \emph {et~al.}(2011)\citenamefont
  {Castellan}, \citenamefont {Rosenkranz}, \citenamefont {Goremychkin},
  \citenamefont {Chung}, \citenamefont {Todorov}, \citenamefont {Kanatzidis},
  \citenamefont {Eremin}, \citenamefont {Knolle}, \citenamefont {Chubukov},
  \citenamefont {Maiti}, \citenamefont {Norman}, \citenamefont {Weber},
  \citenamefont {Claus}, \citenamefont {Guidi}, \citenamefont {Bewley},\ and\
  \citenamefont {Osborn}}]{Castellan2011}%
  \BibitemOpen
  \bibfield  {author} {\bibinfo {author} {\bibfnamefont {J.-P.}\ \bibnamefont
  {Castellan}}, \bibinfo {author} {\bibfnamefont {S.}~\bibnamefont
  {Rosenkranz}}, \bibinfo {author} {\bibfnamefont {E.~A.}\ \bibnamefont
  {Goremychkin}}, \bibinfo {author} {\bibfnamefont {D.~Y.}\ \bibnamefont
  {Chung}}, \bibinfo {author} {\bibfnamefont {I.~S.}\ \bibnamefont {Todorov}},
  \bibinfo {author} {\bibfnamefont {M.~G.}\ \bibnamefont {Kanatzidis}},
  \bibinfo {author} {\bibfnamefont {I.}~\bibnamefont {Eremin}}, \bibinfo
  {author} {\bibfnamefont {J.}~\bibnamefont {Knolle}}, \bibinfo {author}
  {\bibfnamefont {A.~V.}\ \bibnamefont {Chubukov}}, \bibinfo {author}
  {\bibfnamefont {S.}~\bibnamefont {Maiti}}, \bibinfo {author} {\bibfnamefont
  {M.~R.}\ \bibnamefont {Norman}}, \bibinfo {author} {\bibfnamefont
  {F.}~\bibnamefont {Weber}}, \bibinfo {author} {\bibfnamefont
  {H.}~\bibnamefont {Claus}}, \bibinfo {author} {\bibfnamefont
  {T.}~\bibnamefont {Guidi}}, \bibinfo {author} {\bibfnamefont {R.~I.}\
  \bibnamefont {Bewley}}, \ and\ \bibinfo {author} {\bibfnamefont
  {R.}~\bibnamefont {Osborn}},\ }\href {\doibase
  10.1103/PhysRevLett.107.177003} {\bibfield  {journal} {\bibinfo  {journal}
  {Phys. Rev. Lett.}\ }\textbf {\bibinfo {volume} {107}},\ \bibinfo {pages}
  {177003} (\bibinfo {year} {2011})}\BibitemShut {NoStop}%
\bibitem [{\citenamefont {Dai}(2015)}]{Dai2015}%
  \BibitemOpen
  \bibfield  {author} {\bibinfo {author} {\bibfnamefont {P.}~\bibnamefont
  {Dai}},\ }\href {\doibase 10.1103/RevModPhys.87.855} {\bibfield  {journal}
  {\bibinfo  {journal} {Rev. Mod. Phys.}\ }\textbf {\bibinfo {volume} {87}},\
  \bibinfo {pages} {855} (\bibinfo {year} {2015})}\BibitemShut {NoStop}%
\bibitem [{\citenamefont {Onari}\ \emph {et~al.}(2010)\citenamefont {Onari},
  \citenamefont {Kontani},\ and\ \citenamefont {Sato}}]{Onari2010}%
  \BibitemOpen
  \bibfield  {author} {\bibinfo {author} {\bibfnamefont {S.}~\bibnamefont
  {Onari}}, \bibinfo {author} {\bibfnamefont {H.}~\bibnamefont {Kontani}}, \
  and\ \bibinfo {author} {\bibfnamefont {M.}~\bibnamefont {Sato}},\ }\href
  {\doibase 10.1103/PhysRevB.81.060504} {\bibfield  {journal} {\bibinfo
  {journal} {Phys. Rev. B}\ }\textbf {\bibinfo {volume} {81}},\ \bibinfo
  {pages} {060504} (\bibinfo {year} {2010})}\BibitemShut {NoStop}%
\bibitem [{\citenamefont {Onari}\ and\ \citenamefont
  {Kontani}(2011)}]{Onari2011}%
  \BibitemOpen
  \bibfield  {author} {\bibinfo {author} {\bibfnamefont {S.}~\bibnamefont
  {Onari}}\ and\ \bibinfo {author} {\bibfnamefont {H.}~\bibnamefont
  {Kontani}},\ }\href {\doibase 10.1103/PhysRevB.84.144518} {\bibfield
  {journal} {\bibinfo  {journal} {Phys. Rev. B}\ }\textbf {\bibinfo {volume}
  {84}},\ \bibinfo {pages} {144518} (\bibinfo {year} {2011})}\BibitemShut
  {NoStop}%
\bibitem [{\citenamefont {Korshunov}\ \emph {et~al.}(2016)\citenamefont
  {Korshunov}, \citenamefont {Shestakov},\ and\ \citenamefont
  {Togushova}}]{KorshunovPRB2016}%
  \BibitemOpen
  \bibfield  {author} {\bibinfo {author} {\bibfnamefont {M.~M.}\ \bibnamefont
  {Korshunov}}, \bibinfo {author} {\bibfnamefont {V.~A.}\ \bibnamefont
  {Shestakov}}, \ and\ \bibinfo {author} {\bibfnamefont {Y.~N.}\ \bibnamefont
  {Togushova}},\ }\href {\doibase 10.1103/PhysRevB.94.094517} {\bibfield
  {journal} {\bibinfo  {journal} {Phys. Rev. B}\ }\textbf {\bibinfo {volume}
  {94}},\ \bibinfo {pages} {094517} (\bibinfo {year} {2016})}\BibitemShut
  {NoStop}%
\bibitem [{\citenamefont {Korshunov}\ \emph {et~al.}(2017)\citenamefont
  {Korshunov}, \citenamefont {Shestakov},\ and\ \citenamefont
  {Togushova}}]{KorshunovJMMM2017}%
  \BibitemOpen
  \bibfield  {author} {\bibinfo {author} {\bibfnamefont {M.}~\bibnamefont
  {Korshunov}}, \bibinfo {author} {\bibfnamefont {V.}~\bibnamefont
  {Shestakov}}, \ and\ \bibinfo {author} {\bibfnamefont {Y.}~\bibnamefont
  {Togushova}},\ }\href {\doibase 10.1016/j.jmmm.2016.12.082} {\bibfield
  {journal} {\bibinfo  {journal} {Journal of Magnetism and Magnetic Materials}\
  }\textbf {\bibinfo {volume} {440}},\ \bibinfo {pages} {133 } (\bibinfo {year}
  {2017})}\BibitemShut {NoStop}%
\bibitem [{\citenamefont {Shimojima}\ \emph {et~al.}(2011)\citenamefont
  {Shimojima}, \citenamefont {Sakaguchi}, \citenamefont {Ishizaka},
  \citenamefont {Ishida}, \citenamefont {Kiss}, \citenamefont {Okawa},
  \citenamefont {Togashi}, \citenamefont {Chen}, \citenamefont {Watanabe},
  \citenamefont {Arita}, \citenamefont {Shimada}, \citenamefont {Namatame},
  \citenamefont {Taniguchi}, \citenamefont {Ohgushi}, \citenamefont {Kasahara},
  \citenamefont {Terashima}, \citenamefont {Shibauchi}, \citenamefont
  {Matsuda}, \citenamefont {Chainani},\ and\ \citenamefont
  {Shin}}]{Shimojima2011}%
  \BibitemOpen
  \bibfield  {author} {\bibinfo {author} {\bibfnamefont {T.}~\bibnamefont
  {Shimojima}}, \bibinfo {author} {\bibfnamefont {F.}~\bibnamefont
  {Sakaguchi}}, \bibinfo {author} {\bibfnamefont {K.}~\bibnamefont {Ishizaka}},
  \bibinfo {author} {\bibfnamefont {Y.}~\bibnamefont {Ishida}}, \bibinfo
  {author} {\bibfnamefont {T.}~\bibnamefont {Kiss}}, \bibinfo {author}
  {\bibfnamefont {M.}~\bibnamefont {Okawa}}, \bibinfo {author} {\bibfnamefont
  {T.}~\bibnamefont {Togashi}}, \bibinfo {author} {\bibfnamefont {C.-T.}\
  \bibnamefont {Chen}}, \bibinfo {author} {\bibfnamefont {S.}~\bibnamefont
  {Watanabe}}, \bibinfo {author} {\bibfnamefont {M.}~\bibnamefont {Arita}},
  \bibinfo {author} {\bibfnamefont {K.}~\bibnamefont {Shimada}}, \bibinfo
  {author} {\bibfnamefont {H.}~\bibnamefont {Namatame}}, \bibinfo {author}
  {\bibfnamefont {M.}~\bibnamefont {Taniguchi}}, \bibinfo {author}
  {\bibfnamefont {K.}~\bibnamefont {Ohgushi}}, \bibinfo {author} {\bibfnamefont
  {S.}~\bibnamefont {Kasahara}}, \bibinfo {author} {\bibfnamefont
  {T.}~\bibnamefont {Terashima}}, \bibinfo {author} {\bibfnamefont
  {T.}~\bibnamefont {Shibauchi}}, \bibinfo {author} {\bibfnamefont
  {Y.}~\bibnamefont {Matsuda}}, \bibinfo {author} {\bibfnamefont
  {A.}~\bibnamefont {Chainani}}, \ and\ \bibinfo {author} {\bibfnamefont
  {S.}~\bibnamefont {Shin}},\ }\href {\doibase 10.1126/science.1202150}
  {\bibfield  {journal} {\bibinfo  {journal} {Science}\ }\textbf {\bibinfo
  {volume} {332}},\ \bibinfo {pages} {564} (\bibinfo {year}
  {2011})}\BibitemShut {NoStop}%
\bibitem [{\citenamefont {Abdel-Hafiez}\ \emph {et~al.}(2014)\citenamefont
  {Abdel-Hafiez}, \citenamefont {Pereira}, \citenamefont {Kuzmichev},
  \citenamefont {Kuzmicheva}, \citenamefont {Pudalov}, \citenamefont
  {Harnagea}, \citenamefont {Kordyuk}, \citenamefont {Silhanek}, \citenamefont
  {Moshchalkov}, \citenamefont {Shen}, \citenamefont {Wen}, \citenamefont
  {Vasiliev},\ and\ \citenamefont {Chen}}]{Abdel-Hafiez2014}%
  \BibitemOpen
  \bibfield  {author} {\bibinfo {author} {\bibfnamefont {M.}~\bibnamefont
  {Abdel-Hafiez}}, \bibinfo {author} {\bibfnamefont {P.~J.}\ \bibnamefont
  {Pereira}}, \bibinfo {author} {\bibfnamefont {S.~A.}\ \bibnamefont
  {Kuzmichev}}, \bibinfo {author} {\bibfnamefont {T.~E.}\ \bibnamefont
  {Kuzmicheva}}, \bibinfo {author} {\bibfnamefont {V.~M.}\ \bibnamefont
  {Pudalov}}, \bibinfo {author} {\bibfnamefont {L.}~\bibnamefont {Harnagea}},
  \bibinfo {author} {\bibfnamefont {A.~A.}\ \bibnamefont {Kordyuk}}, \bibinfo
  {author} {\bibfnamefont {A.~V.}\ \bibnamefont {Silhanek}}, \bibinfo {author}
  {\bibfnamefont {V.~V.}\ \bibnamefont {Moshchalkov}}, \bibinfo {author}
  {\bibfnamefont {B.}~\bibnamefont {Shen}}, \bibinfo {author} {\bibfnamefont
  {H.-H.}\ \bibnamefont {Wen}}, \bibinfo {author} {\bibfnamefont {A.~N.}\
  \bibnamefont {Vasiliev}}, \ and\ \bibinfo {author} {\bibfnamefont {X.-J.}\
  \bibnamefont {Chen}},\ }\href {\doibase 10.1103/PhysRevB.90.054524}
  {\bibfield  {journal} {\bibinfo  {journal} {Phys. Rev. B}\ }\textbf {\bibinfo
  {volume} {90}},\ \bibinfo {pages} {054524} (\bibinfo {year}
  {2014})}\BibitemShut {NoStop}%
\bibitem [{\citenamefont {Kuzmicheva}\ \emph {et~al.}(2016)\citenamefont
  {Kuzmicheva}, \citenamefont {Vlasenko}, \citenamefont {Gavrilkin},
  \citenamefont {Kuzmichev}, \citenamefont {Pervakov}, \citenamefont
  {Roshchina},\ and\ \citenamefont {Pudalov}}]{Kuzmicheva2016}%
  \BibitemOpen
  \bibfield  {author} {\bibinfo {author} {\bibfnamefont {T.~E.}\ \bibnamefont
  {Kuzmicheva}}, \bibinfo {author} {\bibfnamefont {V.~A.}\ \bibnamefont
  {Vlasenko}}, \bibinfo {author} {\bibfnamefont {S.~Y.}\ \bibnamefont
  {Gavrilkin}}, \bibinfo {author} {\bibfnamefont {S.~A.}\ \bibnamefont
  {Kuzmichev}}, \bibinfo {author} {\bibfnamefont {K.~S.}\ \bibnamefont
  {Pervakov}}, \bibinfo {author} {\bibfnamefont {I.~V.}\ \bibnamefont
  {Roshchina}}, \ and\ \bibinfo {author} {\bibfnamefont {V.~M.}\ \bibnamefont
  {Pudalov}},\ }\href {\doibase 10.1007/s10948-016-3816-4} {\bibfield
  {journal} {\bibinfo  {journal} {Journal of Superconductivity and Novel
  Magnetism}\ }\textbf {\bibinfo {volume} {29}},\ \bibinfo {pages} {3059}
  (\bibinfo {year} {2016})}\BibitemShut {NoStop}%
\bibitem [{\citenamefont {Kuzmicheva}\ \emph {et~al.}(2017)\citenamefont
  {Kuzmicheva}, \citenamefont {Pudalov}, \citenamefont {Kuzmichev},
  \citenamefont {Sadakov}, \citenamefont {Aleshenko}, \citenamefont {Vlasenko},
  \citenamefont {Martovitsky}, \citenamefont {Pervakov}, \citenamefont
  {Eltsev},\ and\ \citenamefont {Pudalov}}]{Kuzmicheva2017}%
  \BibitemOpen
  \bibfield  {author} {\bibinfo {author} {\bibfnamefont {T.~E.}\ \bibnamefont
  {Kuzmicheva}}, \bibinfo {author} {\bibfnamefont {V.~M.}\ \bibnamefont
  {Pudalov}}, \bibinfo {author} {\bibfnamefont {S.~A.}\ \bibnamefont
  {Kuzmichev}}, \bibinfo {author} {\bibfnamefont {A.~V.}\ \bibnamefont
  {Sadakov}}, \bibinfo {author} {\bibfnamefont {Y.~A.}\ \bibnamefont
  {Aleshenko}}, \bibinfo {author} {\bibfnamefont {V.~A.}\ \bibnamefont
  {Vlasenko}}, \bibinfo {author} {\bibfnamefont {V.~P.}\ \bibnamefont
  {Martovitsky}}, \bibinfo {author} {\bibfnamefont {K.~S.}\ \bibnamefont
  {Pervakov}}, \bibinfo {author} {\bibfnamefont {Y.~F.}\ \bibnamefont
  {Eltsev}}, \ and\ \bibinfo {author} {\bibfnamefont {V.~M.}\ \bibnamefont
  {Pudalov}},\ }\href {\doibase 10.3367/UFNe.2016.10.038002} {\bibfield
  {journal} {\bibinfo  {journal} {Phys. Usp.}\ }\textbf {\bibinfo {volume}
  {60}},\ \bibinfo {pages} {419} (\bibinfo {year} {2017})}\BibitemShut
  {NoStop}%
\bibitem [{\citenamefont {Maiti}\ \emph
  {et~al.}(2011{\natexlab{c}})\citenamefont {Maiti}, \citenamefont {Knolle},
  \citenamefont {Eremin},\ and\ \citenamefont {Chubukov}}]{Maiti2011}%
  \BibitemOpen
  \bibfield  {author} {\bibinfo {author} {\bibfnamefont {S.}~\bibnamefont
  {Maiti}}, \bibinfo {author} {\bibfnamefont {J.}~\bibnamefont {Knolle}},
  \bibinfo {author} {\bibfnamefont {I.}~\bibnamefont {Eremin}}, \ and\ \bibinfo
  {author} {\bibfnamefont {A.}~\bibnamefont {Chubukov}},\ }\href {\doibase
  10.1103/PhysRevB.84.144524} {\bibfield  {journal} {\bibinfo  {journal} {Phys.
  Rev. B}\ }\textbf {\bibinfo {volume} {84}},\ \bibinfo {pages} {144524}
  (\bibinfo {year} {2011}{\natexlab{c}})}\BibitemShut {NoStop}%
\bibitem [{\citenamefont {Cao}\ \emph {et~al.}(2008)\citenamefont {Cao},
  \citenamefont {Hirschfeld},\ and\ \citenamefont {Cheng}}]{Cao2008}%
  \BibitemOpen
  \bibfield  {author} {\bibinfo {author} {\bibfnamefont {C.}~\bibnamefont
  {Cao}}, \bibinfo {author} {\bibfnamefont {P.~J.}\ \bibnamefont {Hirschfeld}},
  \ and\ \bibinfo {author} {\bibfnamefont {H.-P.}\ \bibnamefont {Cheng}},\
  }\href {\doibase 10.1103/PhysRevB.77.220506} {\bibfield  {journal} {\bibinfo
  {journal} {Phys. Rev. B}\ }\textbf {\bibinfo {volume} {77}},\ \bibinfo
  {pages} {220506} (\bibinfo {year} {2008})}\BibitemShut {NoStop}%
\bibitem [{\citenamefont {Castellani}\ \emph {et~al.}(1978)\citenamefont
  {Castellani}, \citenamefont {Natoli},\ and\ \citenamefont
  {Ranninger}}]{Castallani1978}%
  \BibitemOpen
  \bibfield  {author} {\bibinfo {author} {\bibfnamefont {C.}~\bibnamefont
  {Castellani}}, \bibinfo {author} {\bibfnamefont {C.~R.}\ \bibnamefont
  {Natoli}}, \ and\ \bibinfo {author} {\bibfnamefont {J.}~\bibnamefont
  {Ranninger}},\ }\href {\doibase 10.1103/PhysRevB.18.4945} {\bibfield
  {journal} {\bibinfo  {journal} {Phys. Rev. B}\ }\textbf {\bibinfo {volume}
  {18}},\ \bibinfo {pages} {4945} (\bibinfo {year} {1978})}\BibitemShut
  {NoStop}%
\bibitem [{\citenamefont {Ole\'{s}}(1983)}]{Oles1983}%
  \BibitemOpen
  \bibfield  {author} {\bibinfo {author} {\bibfnamefont {A.~M.}\ \bibnamefont
  {Ole\'{s}}},\ }\href {\doibase 10.1103/PhysRevB.28.327} {\bibfield  {journal}
  {\bibinfo  {journal} {Phys. Rev. B}\ }\textbf {\bibinfo {volume} {28}},\
  \bibinfo {pages} {327} (\bibinfo {year} {1983})}\BibitemShut {NoStop}%
\bibitem [{\citenamefont {Aichhorn}\ \emph {et~al.}(2009)\citenamefont
  {Aichhorn}, \citenamefont {Pourovskii}, \citenamefont {Vildosola},
  \citenamefont {Ferrero}, \citenamefont {Parcollet}, \citenamefont {Miyake},
  \citenamefont {Georges},\ and\ \citenamefont {Biermann}}]{Aichhorn2009}%
  \BibitemOpen
  \bibfield  {author} {\bibinfo {author} {\bibfnamefont {M.}~\bibnamefont
  {Aichhorn}}, \bibinfo {author} {\bibfnamefont {L.}~\bibnamefont
  {Pourovskii}}, \bibinfo {author} {\bibfnamefont {V.}~\bibnamefont
  {Vildosola}}, \bibinfo {author} {\bibfnamefont {M.}~\bibnamefont {Ferrero}},
  \bibinfo {author} {\bibfnamefont {O.}~\bibnamefont {Parcollet}}, \bibinfo
  {author} {\bibfnamefont {T.}~\bibnamefont {Miyake}}, \bibinfo {author}
  {\bibfnamefont {A.}~\bibnamefont {Georges}}, \ and\ \bibinfo {author}
  {\bibfnamefont {S.}~\bibnamefont {Biermann}},\ }\href {\doibase
  10.1103/PhysRevB.80.085101} {\bibfield  {journal} {\bibinfo  {journal} {Phys.
  Rev. B}\ }\textbf {\bibinfo {volume} {80}},\ \bibinfo {pages} {085101}
  (\bibinfo {year} {2009})}\BibitemShut {NoStop}%
\bibitem [{\citenamefont {Aichhorn}\ \emph {et~al.}(2011)\citenamefont
  {Aichhorn}, \citenamefont {Pourovskii},\ and\ \citenamefont
  {Georges}}]{Aichhorn2011}%
  \BibitemOpen
  \bibfield  {author} {\bibinfo {author} {\bibfnamefont {M.}~\bibnamefont
  {Aichhorn}}, \bibinfo {author} {\bibfnamefont {L.}~\bibnamefont
  {Pourovskii}}, \ and\ \bibinfo {author} {\bibfnamefont {A.}~\bibnamefont
  {Georges}},\ }\href {\doibase 10.1103/PhysRevB.84.054529} {\bibfield
  {journal} {\bibinfo  {journal} {Phys. Rev. B}\ }\textbf {\bibinfo {volume}
  {84}},\ \bibinfo {pages} {054529} (\bibinfo {year} {2011})}\BibitemShut
  {NoStop}%
\bibitem [{\citenamefont {van Roekeghem}\ \emph {et~al.}(2016)\citenamefont
  {van Roekeghem}, \citenamefont {Vaugier}, \citenamefont {Jiang},\ and\
  \citenamefont {Biermann}}]{Roekeghem2016}%
  \BibitemOpen
  \bibfield  {author} {\bibinfo {author} {\bibfnamefont {A.}~\bibnamefont {van
  Roekeghem}}, \bibinfo {author} {\bibfnamefont {L.}~\bibnamefont {Vaugier}},
  \bibinfo {author} {\bibfnamefont {H.}~\bibnamefont {Jiang}}, \ and\ \bibinfo
  {author} {\bibfnamefont {S.}~\bibnamefont {Biermann}},\ }\href {\doibase
  10.1103/PhysRevB.94.125147} {\bibfield  {journal} {\bibinfo  {journal} {Phys.
  Rev. B}\ }\textbf {\bibinfo {volume} {94}},\ \bibinfo {pages} {125147}
  (\bibinfo {year} {2016})}\BibitemShut {NoStop}%
\bibitem [{\citenamefont {Miyake}\ \emph {et~al.}(2008)\citenamefont {Miyake},
  \citenamefont {Pourovskii}, \citenamefont {Vildosola}, \citenamefont
  {Biermann},\ and\ \citenamefont {Georges}}]{Miyake2008}%
  \BibitemOpen
  \bibfield  {author} {\bibinfo {author} {\bibfnamefont {T.}~\bibnamefont
  {Miyake}}, \bibinfo {author} {\bibfnamefont {L.}~\bibnamefont {Pourovskii}},
  \bibinfo {author} {\bibfnamefont {V.}~\bibnamefont {Vildosola}}, \bibinfo
  {author} {\bibfnamefont {S.}~\bibnamefont {Biermann}}, \ and\ \bibinfo
  {author} {\bibfnamefont {A.}~\bibnamefont {Georges}},\ }\href {\doibase
  10.1143/JPSJS.77SC.99} {\bibfield  {journal} {\bibinfo  {journal} {Journal of
  the Physical Society of Japan}\ }\textbf {\bibinfo {volume} {77}},\ \bibinfo
  {pages} {99} (\bibinfo {year} {2008})}\BibitemShut {NoStop}%
\bibitem [{\citenamefont {Nakamura}\ \emph {et~al.}(2008)\citenamefont
  {Nakamura}, \citenamefont {Arita},\ and\ \citenamefont
  {Imada}}]{Nakamura2008}%
  \BibitemOpen
  \bibfield  {author} {\bibinfo {author} {\bibfnamefont {K.}~\bibnamefont
  {Nakamura}}, \bibinfo {author} {\bibfnamefont {R.}~\bibnamefont {Arita}}, \
  and\ \bibinfo {author} {\bibfnamefont {M.}~\bibnamefont {Imada}},\ }\href
  {\doibase 10.1143/JPSJ.77.093711} {\bibfield  {journal} {\bibinfo  {journal}
  {Journal of the Physical Society of Japan}\ }\textbf {\bibinfo {volume}
  {77}},\ \bibinfo {pages} {093711} (\bibinfo {year} {2008})}\BibitemShut
  {NoStop}%
\bibitem [{\citenamefont {{Kemper}}\ \emph {et~al.}(2010)\citenamefont
  {{Kemper}}, \citenamefont {{Maier}}, \citenamefont {{Graser}}, \citenamefont
  {{Cheng}}, \citenamefont {{Hirschfeld}},\ and\ \citenamefont
  {{Scalapino}}}]{Kemper2010}%
  \BibitemOpen
  \bibfield  {author} {\bibinfo {author} {\bibfnamefont {A.~F.}\ \bibnamefont
  {{Kemper}}}, \bibinfo {author} {\bibfnamefont {T.~A.}\ \bibnamefont
  {{Maier}}}, \bibinfo {author} {\bibfnamefont {S.}~\bibnamefont {{Graser}}},
  \bibinfo {author} {\bibfnamefont {H.}~\bibnamefont {{Cheng}}}, \bibinfo
  {author} {\bibfnamefont {P.~J.}\ \bibnamefont {{Hirschfeld}}}, \ and\
  \bibinfo {author} {\bibfnamefont {D.~J.}\ \bibnamefont {{Scalapino}}},\
  }\href {\doibase 10.1088/1367-2630/12/7/073030} {\bibfield  {journal}
  {\bibinfo  {journal} {New Journal of Physics}\ }\textbf {\bibinfo {volume}
  {12}},\ \bibinfo {pages} {073030} (\bibinfo {year} {2010})}\BibitemShut
  {NoStop}%
\bibitem [{PRB()}]{PRBSuppl}%
  \BibitemOpen
  \href@noop {} {\bibinfo  {journal} {See Supplemental Material at [URL will be
  inserted by publisher] for intensity plots of gap structures within the
  Brillouin zone and results of gap calculations within the spin fluctuation
  theory of pairing}\ }\BibitemShut {NoStop}%
\bibitem [{\citenamefont {Wang}\ \emph
  {et~al.}(2016{\natexlab{a}})\citenamefont {Wang}, \citenamefont {Yi},
  \citenamefont {Sun}, \citenamefont {Valdivia}, \citenamefont {Kim},
  \citenamefont {Xu}, \citenamefont {Berlijn}, \citenamefont {Christianson},
  \citenamefont {Chi}, \citenamefont {Hashimoto}, \citenamefont {Lu},
  \citenamefont {Li}, \citenamefont {Bourret-Courchesne}, \citenamefont {Dai},
  \citenamefont {Lee}, \citenamefont {Maier},\ and\ \citenamefont
  {Birgeneau}}]{Wang2016}%
  \BibitemOpen
\bibfield  {journal} {  }\bibfield  {author} {\bibinfo {author} {\bibfnamefont
  {M.}~\bibnamefont {Wang}}, \bibinfo {author} {\bibfnamefont {M.}~\bibnamefont
  {Yi}}, \bibinfo {author} {\bibfnamefont {H.~L.}\ \bibnamefont {Sun}},
  \bibinfo {author} {\bibfnamefont {P.}~\bibnamefont {Valdivia}}, \bibinfo
  {author} {\bibfnamefont {M.~G.}\ \bibnamefont {Kim}}, \bibinfo {author}
  {\bibfnamefont {Z.~J.}\ \bibnamefont {Xu}}, \bibinfo {author} {\bibfnamefont
  {T.}~\bibnamefont {Berlijn}}, \bibinfo {author} {\bibfnamefont {A.~D.}\
  \bibnamefont {Christianson}}, \bibinfo {author} {\bibfnamefont
  {S.}~\bibnamefont {Chi}}, \bibinfo {author} {\bibfnamefont {M.}~\bibnamefont
  {Hashimoto}}, \bibinfo {author} {\bibfnamefont {D.~H.}\ \bibnamefont {Lu}},
  \bibinfo {author} {\bibfnamefont {X.~D.}\ \bibnamefont {Li}}, \bibinfo
  {author} {\bibfnamefont {E.}~\bibnamefont {Bourret-Courchesne}}, \bibinfo
  {author} {\bibfnamefont {P.}~\bibnamefont {Dai}}, \bibinfo {author}
  {\bibfnamefont {D.~H.}\ \bibnamefont {Lee}}, \bibinfo {author} {\bibfnamefont
  {T.~A.}\ \bibnamefont {Maier}}, \ and\ \bibinfo {author} {\bibfnamefont
  {R.~J.}\ \bibnamefont {Birgeneau}},\ }\href {\doibase
  10.1103/PhysRevB.93.205149} {\bibfield  {journal} {\bibinfo  {journal} {Phys.
  Rev. B}\ }\textbf {\bibinfo {volume} {93}},\ \bibinfo {pages} {205149}
  (\bibinfo {year} {2016}{\natexlab{a}})}\BibitemShut {NoStop}%
\bibitem [{\citenamefont {Terashima}\ \emph {et~al.}(2009)\citenamefont
  {Terashima}, \citenamefont {Sekiba}, \citenamefont {Bowen}, \citenamefont
  {Nakayama}, \citenamefont {Kawahara}, \citenamefont {Sato}, \citenamefont
  {Richard}, \citenamefont {Xu}, \citenamefont {Li}, \citenamefont {Cao},
  \citenamefont {Xu}, \citenamefont {Ding},\ and\ \citenamefont
  {Takahashi}}]{Terashima2009}%
  \BibitemOpen
  \bibfield  {author} {\bibinfo {author} {\bibfnamefont {K.}~\bibnamefont
  {Terashima}}, \bibinfo {author} {\bibfnamefont {Y.}~\bibnamefont {Sekiba}},
  \bibinfo {author} {\bibfnamefont {J.~H.}\ \bibnamefont {Bowen}}, \bibinfo
  {author} {\bibfnamefont {K.}~\bibnamefont {Nakayama}}, \bibinfo {author}
  {\bibfnamefont {T.}~\bibnamefont {Kawahara}}, \bibinfo {author}
  {\bibfnamefont {T.}~\bibnamefont {Sato}}, \bibinfo {author} {\bibfnamefont
  {P.}~\bibnamefont {Richard}}, \bibinfo {author} {\bibfnamefont {Y.-M.}\
  \bibnamefont {Xu}}, \bibinfo {author} {\bibfnamefont {L.~J.}\ \bibnamefont
  {Li}}, \bibinfo {author} {\bibfnamefont {G.~H.}\ \bibnamefont {Cao}},
  \bibinfo {author} {\bibfnamefont {Z.-A.}\ \bibnamefont {Xu}}, \bibinfo
  {author} {\bibfnamefont {H.}~\bibnamefont {Ding}}, \ and\ \bibinfo {author}
  {\bibfnamefont {T.}~\bibnamefont {Takahashi}},\ }\href {\doibase
  10.1073/pnas.0900469106} {\bibfield  {journal} {\bibinfo  {journal}
  {Proceedings of the National Academy of Sciences}\ }\textbf {\bibinfo
  {volume} {106}},\ \bibinfo {pages} {7330} (\bibinfo {year}
  {2009})}\BibitemShut {NoStop}%
\bibitem [{\citenamefont {Kawahara}\ \emph {et~al.}(2010)\citenamefont
  {Kawahara}, \citenamefont {Terashima}, \citenamefont {Sekiba}, \citenamefont
  {Bowen}, \citenamefont {Nakayama}, \citenamefont {Sato}, \citenamefont
  {Richard}, \citenamefont {Xu}, \citenamefont {Li}, \citenamefont {Cao},
  \citenamefont {Xu}, \citenamefont {Ding},\ and\ \citenamefont
  {Takahashi}}]{Kawahara2010}%
  \BibitemOpen
  \bibfield  {author} {\bibinfo {author} {\bibfnamefont {T.}~\bibnamefont
  {Kawahara}}, \bibinfo {author} {\bibfnamefont {K.}~\bibnamefont {Terashima}},
  \bibinfo {author} {\bibfnamefont {Y.}~\bibnamefont {Sekiba}}, \bibinfo
  {author} {\bibfnamefont {J.}~\bibnamefont {Bowen}}, \bibinfo {author}
  {\bibfnamefont {K.}~\bibnamefont {Nakayama}}, \bibinfo {author}
  {\bibfnamefont {T.}~\bibnamefont {Sato}}, \bibinfo {author} {\bibfnamefont
  {P.}~\bibnamefont {Richard}}, \bibinfo {author} {\bibfnamefont {Y.-M.}\
  \bibnamefont {Xu}}, \bibinfo {author} {\bibfnamefont {L.}~\bibnamefont {Li}},
  \bibinfo {author} {\bibfnamefont {G.}~\bibnamefont {Cao}}, \bibinfo {author}
  {\bibfnamefont {Z.-A.}\ \bibnamefont {Xu}}, \bibinfo {author} {\bibfnamefont
  {H.}~\bibnamefont {Ding}}, \ and\ \bibinfo {author} {\bibfnamefont
  {T.}~\bibnamefont {Takahashi}},\ }\href {\doibase
  http://dx.doi.org/10.1016/j.physc.2009.11.129} {\bibfield  {journal}
  {\bibinfo  {journal} {Physica C: Superconductivity and its Applications}\
  }\textbf {\bibinfo {volume} {470}},\ \bibinfo {pages} {S440 } (\bibinfo
  {year} {2010})}\BibitemShut {NoStop}%
\bibitem [{\citenamefont {Tortello}\ \emph {et~al.}(2010)\citenamefont
  {Tortello}, \citenamefont {Daghero}, \citenamefont {Ummarino}, \citenamefont
  {Stepanov}, \citenamefont {Jiang}, \citenamefont {Weiss}, \citenamefont
  {Hellstrom},\ and\ \citenamefont {Gonnelli}}]{Tortello2010}%
  \BibitemOpen
  \bibfield  {author} {\bibinfo {author} {\bibfnamefont {M.}~\bibnamefont
  {Tortello}}, \bibinfo {author} {\bibfnamefont {D.}~\bibnamefont {Daghero}},
  \bibinfo {author} {\bibfnamefont {G.~A.}\ \bibnamefont {Ummarino}}, \bibinfo
  {author} {\bibfnamefont {V.~A.}\ \bibnamefont {Stepanov}}, \bibinfo {author}
  {\bibfnamefont {J.}~\bibnamefont {Jiang}}, \bibinfo {author} {\bibfnamefont
  {J.~D.}\ \bibnamefont {Weiss}}, \bibinfo {author} {\bibfnamefont {E.~E.}\
  \bibnamefont {Hellstrom}}, \ and\ \bibinfo {author} {\bibfnamefont {R.~S.}\
  \bibnamefont {Gonnelli}},\ }\href {\doibase 10.1103/PhysRevLett.105.237002}
  {\bibfield  {journal} {\bibinfo  {journal} {Phys. Rev. Lett.}\ }\textbf
  {\bibinfo {volume} {105}},\ \bibinfo {pages} {237002} (\bibinfo {year}
  {2010})}\BibitemShut {NoStop}%
\bibitem [{\citenamefont {Shan}\ \emph {et~al.}(2012)\citenamefont {Shan},
  \citenamefont {Gong}, \citenamefont {Wang}, \citenamefont {Shen},
  \citenamefont {Hou}, \citenamefont {Ren}, \citenamefont {Li}, \citenamefont
  {Yang}, \citenamefont {Wen}, \citenamefont {Li},\ and\ \citenamefont
  {Dai}}]{Shan2012}%
  \BibitemOpen
  \bibfield  {author} {\bibinfo {author} {\bibfnamefont {L.}~\bibnamefont
  {Shan}}, \bibinfo {author} {\bibfnamefont {J.}~\bibnamefont {Gong}}, \bibinfo
  {author} {\bibfnamefont {Y.-L.}\ \bibnamefont {Wang}}, \bibinfo {author}
  {\bibfnamefont {B.}~\bibnamefont {Shen}}, \bibinfo {author} {\bibfnamefont
  {X.}~\bibnamefont {Hou}}, \bibinfo {author} {\bibfnamefont {C.}~\bibnamefont
  {Ren}}, \bibinfo {author} {\bibfnamefont {C.}~\bibnamefont {Li}}, \bibinfo
  {author} {\bibfnamefont {H.}~\bibnamefont {Yang}}, \bibinfo {author}
  {\bibfnamefont {H.-H.}\ \bibnamefont {Wen}}, \bibinfo {author} {\bibfnamefont
  {S.}~\bibnamefont {Li}}, \ and\ \bibinfo {author} {\bibfnamefont
  {P.}~\bibnamefont {Dai}},\ }\href {\doibase 10.1103/PhysRevLett.108.227002}
  {\bibfield  {journal} {\bibinfo  {journal} {Phys. Rev. Lett.}\ }\textbf
  {\bibinfo {volume} {108}},\ \bibinfo {pages} {227002} (\bibinfo {year}
  {2012})}\BibitemShut {NoStop}%
\bibitem [{\citenamefont {Ding}\ \emph {et~al.}(2008)\citenamefont {Ding},
  \citenamefont {Richard}, \citenamefont {Nakayama}, \citenamefont {Sugawara},
  \citenamefont {Arakane}, \citenamefont {Sekiba}, \citenamefont {Takayama},
  \citenamefont {Souma}, \citenamefont {Sato}, \citenamefont {Takahashi},
  \citenamefont {Wang}, \citenamefont {Dai}, \citenamefont {Fang},
  \citenamefont {Chen}, \citenamefont {Luo},\ and\ \citenamefont
  {Wang}}]{Ding2008}%
  \BibitemOpen
  \bibfield  {author} {\bibinfo {author} {\bibfnamefont {H.}~\bibnamefont
  {Ding}}, \bibinfo {author} {\bibfnamefont {P.}~\bibnamefont {Richard}},
  \bibinfo {author} {\bibfnamefont {K.}~\bibnamefont {Nakayama}}, \bibinfo
  {author} {\bibfnamefont {K.}~\bibnamefont {Sugawara}}, \bibinfo {author}
  {\bibfnamefont {T.}~\bibnamefont {Arakane}}, \bibinfo {author} {\bibfnamefont
  {Y.}~\bibnamefont {Sekiba}}, \bibinfo {author} {\bibfnamefont
  {A.}~\bibnamefont {Takayama}}, \bibinfo {author} {\bibfnamefont
  {S.}~\bibnamefont {Souma}}, \bibinfo {author} {\bibfnamefont
  {T.}~\bibnamefont {Sato}}, \bibinfo {author} {\bibfnamefont {T.}~\bibnamefont
  {Takahashi}}, \bibinfo {author} {\bibfnamefont {Z.}~\bibnamefont {Wang}},
  \bibinfo {author} {\bibfnamefont {X.}~\bibnamefont {Dai}}, \bibinfo {author}
  {\bibfnamefont {Z.}~\bibnamefont {Fang}}, \bibinfo {author} {\bibfnamefont
  {G.~F.}\ \bibnamefont {Chen}}, \bibinfo {author} {\bibfnamefont {J.~L.}\
  \bibnamefont {Luo}}, \ and\ \bibinfo {author} {\bibfnamefont {N.~L.}\
  \bibnamefont {Wang}},\ }\href
  {http://stacks.iop.org/0295-5075/83/i=4/a=47001} {\bibfield  {journal}
  {\bibinfo  {journal} {EPL (Europhysics Letters)}\ }\textbf {\bibinfo {volume}
  {83}},\ \bibinfo {pages} {47001} (\bibinfo {year} {2008})}\BibitemShut
  {NoStop}%
\bibitem [{\citenamefont {Wray}\ \emph {et~al.}(2008)\citenamefont {Wray},
  \citenamefont {Qian}, \citenamefont {Hsieh}, \citenamefont {Xia},
  \citenamefont {Li}, \citenamefont {Checkelsky}, \citenamefont {Pasupathy},
  \citenamefont {Gomes}, \citenamefont {Parker}, \citenamefont {Fedorov},
  \citenamefont {Chen}, \citenamefont {Luo}, \citenamefont {Yazdani},
  \citenamefont {Ong}, \citenamefont {Wang},\ and\ \citenamefont
  {Hasan}}]{Wray2008}%
  \BibitemOpen
  \bibfield  {author} {\bibinfo {author} {\bibfnamefont {L.}~\bibnamefont
  {Wray}}, \bibinfo {author} {\bibfnamefont {D.}~\bibnamefont {Qian}}, \bibinfo
  {author} {\bibfnamefont {D.}~\bibnamefont {Hsieh}}, \bibinfo {author}
  {\bibfnamefont {Y.}~\bibnamefont {Xia}}, \bibinfo {author} {\bibfnamefont
  {L.}~\bibnamefont {Li}}, \bibinfo {author} {\bibfnamefont {J.~G.}\
  \bibnamefont {Checkelsky}}, \bibinfo {author} {\bibfnamefont
  {A.}~\bibnamefont {Pasupathy}}, \bibinfo {author} {\bibfnamefont {K.~K.}\
  \bibnamefont {Gomes}}, \bibinfo {author} {\bibfnamefont {C.~V.}\ \bibnamefont
  {Parker}}, \bibinfo {author} {\bibfnamefont {A.~V.}\ \bibnamefont {Fedorov}},
  \bibinfo {author} {\bibfnamefont {G.~F.}\ \bibnamefont {Chen}}, \bibinfo
  {author} {\bibfnamefont {J.~L.}\ \bibnamefont {Luo}}, \bibinfo {author}
  {\bibfnamefont {A.}~\bibnamefont {Yazdani}}, \bibinfo {author} {\bibfnamefont
  {N.~P.}\ \bibnamefont {Ong}}, \bibinfo {author} {\bibfnamefont {N.~L.}\
  \bibnamefont {Wang}}, \ and\ \bibinfo {author} {\bibfnamefont {M.~Z.}\
  \bibnamefont {Hasan}},\ }\href {\doibase 10.1103/PhysRevB.78.184508}
  {\bibfield  {journal} {\bibinfo  {journal} {Phys. Rev. B}\ }\textbf {\bibinfo
  {volume} {78}},\ \bibinfo {pages} {184508} (\bibinfo {year}
  {2008})}\BibitemShut {NoStop}%
\bibitem [{\citenamefont {Zhang}\ \emph {et~al.}(2010)\citenamefont {Zhang},
  \citenamefont {Yang}, \citenamefont {Chen}, \citenamefont {Zhou},
  \citenamefont {Wang}, \citenamefont {Chen}, \citenamefont {Arita},
  \citenamefont {Shimada}, \citenamefont {Namatame}, \citenamefont {Taniguchi},
  \citenamefont {Hu}, \citenamefont {Xie},\ and\ \citenamefont
  {Feng}}]{Zhang2010}%
  \BibitemOpen
  \bibfield  {author} {\bibinfo {author} {\bibfnamefont {Y.}~\bibnamefont
  {Zhang}}, \bibinfo {author} {\bibfnamefont {L.~X.}\ \bibnamefont {Yang}},
  \bibinfo {author} {\bibfnamefont {F.}~\bibnamefont {Chen}}, \bibinfo {author}
  {\bibfnamefont {B.}~\bibnamefont {Zhou}}, \bibinfo {author} {\bibfnamefont
  {X.~F.}\ \bibnamefont {Wang}}, \bibinfo {author} {\bibfnamefont {X.~H.}\
  \bibnamefont {Chen}}, \bibinfo {author} {\bibfnamefont {M.}~\bibnamefont
  {Arita}}, \bibinfo {author} {\bibfnamefont {K.}~\bibnamefont {Shimada}},
  \bibinfo {author} {\bibfnamefont {H.}~\bibnamefont {Namatame}}, \bibinfo
  {author} {\bibfnamefont {M.}~\bibnamefont {Taniguchi}}, \bibinfo {author}
  {\bibfnamefont {J.~P.}\ \bibnamefont {Hu}}, \bibinfo {author} {\bibfnamefont
  {B.~P.}\ \bibnamefont {Xie}}, \ and\ \bibinfo {author} {\bibfnamefont
  {D.~L.}\ \bibnamefont {Feng}},\ }\href {\doibase
  10.1103/PhysRevLett.105.117003} {\bibfield  {journal} {\bibinfo  {journal}
  {Phys. Rev. Lett.}\ }\textbf {\bibinfo {volume} {105}},\ \bibinfo {pages}
  {117003} (\bibinfo {year} {2010})}\BibitemShut {NoStop}%
\bibitem [{\citenamefont {Zhao}\ \emph {et~al.}(2008)\citenamefont {Zhao},
  \citenamefont {Liu}, \citenamefont {Zhang}, \citenamefont {Meng},
  \citenamefont {Jia}, \citenamefont {Liu}, \citenamefont {Dong}, \citenamefont
  {Chen}, \citenamefont {Luo}, \citenamefont {Wang}, \citenamefont {Lu},
  \citenamefont {Wang}, \citenamefont {Zhou}, \citenamefont {Zhu},
  \citenamefont {Wang}, \citenamefont {Xu}, \citenamefont {Chen},\ and\
  \citenamefont {Zhou}}]{Zhao2008ARPES}%
  \BibitemOpen
  \bibfield  {author} {\bibinfo {author} {\bibfnamefont {L.}~\bibnamefont
  {Zhao}}, \bibinfo {author} {\bibfnamefont {H.-Y.}\ \bibnamefont {Liu}},
  \bibinfo {author} {\bibfnamefont {W.-T.}\ \bibnamefont {Zhang}}, \bibinfo
  {author} {\bibfnamefont {J.-Q.}\ \bibnamefont {Meng}}, \bibinfo {author}
  {\bibfnamefont {X.-W.}\ \bibnamefont {Jia}}, \bibinfo {author} {\bibfnamefont
  {G.-D.}\ \bibnamefont {Liu}}, \bibinfo {author} {\bibfnamefont {X.-L.}\
  \bibnamefont {Dong}}, \bibinfo {author} {\bibfnamefont {G.-F.}\ \bibnamefont
  {Chen}}, \bibinfo {author} {\bibfnamefont {J.-L.}\ \bibnamefont {Luo}},
  \bibinfo {author} {\bibfnamefont {N.-L.}\ \bibnamefont {Wang}}, \bibinfo
  {author} {\bibfnamefont {W.}~\bibnamefont {Lu}}, \bibinfo {author}
  {\bibfnamefont {G.-L.}\ \bibnamefont {Wang}}, \bibinfo {author}
  {\bibfnamefont {Y.}~\bibnamefont {Zhou}}, \bibinfo {author} {\bibfnamefont
  {Y.}~\bibnamefont {Zhu}}, \bibinfo {author} {\bibfnamefont {X.-Y.}\
  \bibnamefont {Wang}}, \bibinfo {author} {\bibfnamefont {Z.-Y.}\ \bibnamefont
  {Xu}}, \bibinfo {author} {\bibfnamefont {C.-T.}\ \bibnamefont {Chen}}, \ and\
  \bibinfo {author} {\bibfnamefont {X.-J.}\ \bibnamefont {Zhou}},\ }\href
  {http://stacks.iop.org/0256-307X/25/i=12/a=061} {\bibfield  {journal}
  {\bibinfo  {journal} {Chinese Physics Letters}\ }\textbf {\bibinfo {volume}
  {25}},\ \bibinfo {pages} {4402} (\bibinfo {year} {2008})}\BibitemShut
  {NoStop}%
\bibitem [{\citenamefont {Ren}\ \emph {et~al.}(2008)\citenamefont {Ren},
  \citenamefont {Wang}, \citenamefont {Luo}, \citenamefont {Yang},
  \citenamefont {Shan},\ and\ \citenamefont {Wen}}]{Ren2008}%
  \BibitemOpen
  \bibfield  {author} {\bibinfo {author} {\bibfnamefont {C.}~\bibnamefont
  {Ren}}, \bibinfo {author} {\bibfnamefont {Z.-S.}\ \bibnamefont {Wang}},
  \bibinfo {author} {\bibfnamefont {H.-Q.}\ \bibnamefont {Luo}}, \bibinfo
  {author} {\bibfnamefont {H.}~\bibnamefont {Yang}}, \bibinfo {author}
  {\bibfnamefont {L.}~\bibnamefont {Shan}}, \ and\ \bibinfo {author}
  {\bibfnamefont {H.-H.}\ \bibnamefont {Wen}},\ }\href {\doibase
  10.1103/PhysRevLett.101.257006} {\bibfield  {journal} {\bibinfo  {journal}
  {Phys. Rev. Lett.}\ }\textbf {\bibinfo {volume} {101}},\ \bibinfo {pages}
  {257006} (\bibinfo {year} {2008})}\BibitemShut {NoStop}%
\bibitem [{\citenamefont {Evtushinsky}\ \emph
  {et~al.}(2009{\natexlab{a}})\citenamefont {Evtushinsky}, \citenamefont
  {Inosov}, \citenamefont {Zabolotnyy}, \citenamefont {Koitzsch}, \citenamefont
  {Knupfer}, \citenamefont {B\"uchner}, \citenamefont {Viazovska},
  \citenamefont {Sun}, \citenamefont {Hinkov}, \citenamefont {Boris},
  \citenamefont {Lin}, \citenamefont {Keimer}, \citenamefont {Varykhalov},
  \citenamefont {Kordyuk},\ and\ \citenamefont {Borisenko}}]{Evtushinsky2009}%
  \BibitemOpen
  \bibfield  {author} {\bibinfo {author} {\bibfnamefont {D.~V.}\ \bibnamefont
  {Evtushinsky}}, \bibinfo {author} {\bibfnamefont {D.~S.}\ \bibnamefont
  {Inosov}}, \bibinfo {author} {\bibfnamefont {V.~B.}\ \bibnamefont
  {Zabolotnyy}}, \bibinfo {author} {\bibfnamefont {A.}~\bibnamefont
  {Koitzsch}}, \bibinfo {author} {\bibfnamefont {M.}~\bibnamefont {Knupfer}},
  \bibinfo {author} {\bibfnamefont {B.}~\bibnamefont {B\"uchner}}, \bibinfo
  {author} {\bibfnamefont {M.~S.}\ \bibnamefont {Viazovska}}, \bibinfo {author}
  {\bibfnamefont {G.~L.}\ \bibnamefont {Sun}}, \bibinfo {author} {\bibfnamefont
  {V.}~\bibnamefont {Hinkov}}, \bibinfo {author} {\bibfnamefont {A.~V.}\
  \bibnamefont {Boris}}, \bibinfo {author} {\bibfnamefont {C.~T.}\ \bibnamefont
  {Lin}}, \bibinfo {author} {\bibfnamefont {B.}~\bibnamefont {Keimer}},
  \bibinfo {author} {\bibfnamefont {A.}~\bibnamefont {Varykhalov}}, \bibinfo
  {author} {\bibfnamefont {A.~A.}\ \bibnamefont {Kordyuk}}, \ and\ \bibinfo
  {author} {\bibfnamefont {S.~V.}\ \bibnamefont {Borisenko}},\ }\href {\doibase
  10.1103/PhysRevB.79.054517} {\bibfield  {journal} {\bibinfo  {journal} {Phys.
  Rev. B}\ }\textbf {\bibinfo {volume} {79}},\ \bibinfo {pages} {054517}
  (\bibinfo {year} {2009}{\natexlab{a}})}\BibitemShut {NoStop}%
\bibitem [{\citenamefont {Evtushinsky}\ \emph
  {et~al.}(2009{\natexlab{b}})\citenamefont {Evtushinsky}, \citenamefont
  {Inosov}, \citenamefont {Zabolotnyy}, \citenamefont {Viazovska},
  \citenamefont {Khasanov}, \citenamefont {Amato}, \citenamefont {Klauss},
  \citenamefont {Luetkens}, \citenamefont {Niedermayer}, \citenamefont {Sun},
  \citenamefont {Hinkov}, \citenamefont {Lin}, \citenamefont {Varykhalov},
  \citenamefont {Koitzsch}, \citenamefont {Knupfer}, \citenamefont {B\"uchner},
  \citenamefont {Kordyuk},\ and\ \citenamefont
  {Borisenko}}]{Evtushinsky2009NJP}%
  \BibitemOpen
  \bibfield  {author} {\bibinfo {author} {\bibfnamefont {D.~V.}\ \bibnamefont
  {Evtushinsky}}, \bibinfo {author} {\bibfnamefont {D.~S.}\ \bibnamefont
  {Inosov}}, \bibinfo {author} {\bibfnamefont {V.~B.}\ \bibnamefont
  {Zabolotnyy}}, \bibinfo {author} {\bibfnamefont {M.~S.}\ \bibnamefont
  {Viazovska}}, \bibinfo {author} {\bibfnamefont {R.}~\bibnamefont {Khasanov}},
  \bibinfo {author} {\bibfnamefont {A.}~\bibnamefont {Amato}}, \bibinfo
  {author} {\bibfnamefont {H.-H.}\ \bibnamefont {Klauss}}, \bibinfo {author}
  {\bibfnamefont {H.}~\bibnamefont {Luetkens}}, \bibinfo {author}
  {\bibfnamefont {C.}~\bibnamefont {Niedermayer}}, \bibinfo {author}
  {\bibfnamefont {G.~L.}\ \bibnamefont {Sun}}, \bibinfo {author} {\bibfnamefont
  {V.}~\bibnamefont {Hinkov}}, \bibinfo {author} {\bibfnamefont {C.~T.}\
  \bibnamefont {Lin}}, \bibinfo {author} {\bibfnamefont {A.}~\bibnamefont
  {Varykhalov}}, \bibinfo {author} {\bibfnamefont {A.}~\bibnamefont
  {Koitzsch}}, \bibinfo {author} {\bibfnamefont {M.}~\bibnamefont {Knupfer}},
  \bibinfo {author} {\bibfnamefont {B.}~\bibnamefont {B\"uchner}}, \bibinfo
  {author} {\bibfnamefont {A.~A.}\ \bibnamefont {Kordyuk}}, \ and\ \bibinfo
  {author} {\bibfnamefont {S.~V.}\ \bibnamefont {Borisenko}},\ }\href
  {http://stacks.iop.org/1367-2630/11/i=5/a=055069} {\bibfield  {journal}
  {\bibinfo  {journal} {New Journal of Physics}\ }\textbf {\bibinfo {volume}
  {11}},\ \bibinfo {pages} {055069} (\bibinfo {year}
  {2009}{\natexlab{b}})}\BibitemShut {NoStop}%
\bibitem [{\citenamefont {Wang}\ \emph
  {et~al.}(2016{\natexlab{b}})\citenamefont {Wang}, \citenamefont {Shen},
  \citenamefont {Pan}, \citenamefont {Hao}, \citenamefont {Ma}, \citenamefont
  {Zhou}, \citenamefont {Steffens}, \citenamefont {Schmalzl}, \citenamefont
  {Forrest}, \citenamefont {Abdel-Hafiez}, \citenamefont {Chen}, \citenamefont
  {Chareev}, \citenamefont {Vasiliev}, \citenamefont {Bourges}, \citenamefont
  {Sidis}, \citenamefont {Cao},\ and\ \citenamefont {Zhao}}]{Wang.nmat4492}%
  \BibitemOpen
  \bibfield  {author} {\bibinfo {author} {\bibfnamefont {Q.}~\bibnamefont
  {Wang}}, \bibinfo {author} {\bibfnamefont {Y.}~\bibnamefont {Shen}}, \bibinfo
  {author} {\bibfnamefont {B.}~\bibnamefont {Pan}}, \bibinfo {author}
  {\bibfnamefont {Y.}~\bibnamefont {Hao}}, \bibinfo {author} {\bibfnamefont
  {M.}~\bibnamefont {Ma}}, \bibinfo {author} {\bibfnamefont {F.}~\bibnamefont
  {Zhou}}, \bibinfo {author} {\bibfnamefont {P.}~\bibnamefont {Steffens}},
  \bibinfo {author} {\bibfnamefont {K.}~\bibnamefont {Schmalzl}}, \bibinfo
  {author} {\bibfnamefont {T.~R.}\ \bibnamefont {Forrest}}, \bibinfo {author}
  {\bibfnamefont {M.}~\bibnamefont {Abdel-Hafiez}}, \bibinfo {author}
  {\bibfnamefont {X.}~\bibnamefont {Chen}}, \bibinfo {author} {\bibfnamefont
  {D.~A.}\ \bibnamefont {Chareev}}, \bibinfo {author} {\bibfnamefont {A.~N.}\
  \bibnamefont {Vasiliev}}, \bibinfo {author} {\bibfnamefont {P.}~\bibnamefont
  {Bourges}}, \bibinfo {author} {\bibfnamefont {Y.}~\bibnamefont {Sidis}},
  \bibinfo {author} {\bibfnamefont {H.}~\bibnamefont {Cao}}, \ and\ \bibinfo
  {author} {\bibfnamefont {J.}~\bibnamefont {Zhao}},\ }\href
  {http://dx.doi.org/10.1038/nmat4492} {\bibfield  {journal} {\bibinfo
  {journal} {Nat Mater}\ }\textbf {\bibinfo {volume} {15}},\ \bibinfo {pages}
  {159} (\bibinfo {year} {2016}{\natexlab{b}})},\ \bibinfo {note}
  {letter}\BibitemShut {NoStop}%
\bibitem [{\citenamefont {Kasahara}\ \emph {et~al.}(2014)\citenamefont
  {Kasahara}, \citenamefont {Watashige}, \citenamefont {Hanaguri},
  \citenamefont {Kohsaka}, \citenamefont {Yamashita}, \citenamefont
  {Shimoyama}, \citenamefont {Mizukami}, \citenamefont {Endo}, \citenamefont
  {Ikeda}, \citenamefont {Aoyama}, \citenamefont {Terashima}, \citenamefont
  {Uji}, \citenamefont {Wolf}, \citenamefont {von L\"ohneysen}, \citenamefont
  {Shibauchi},\ and\ \citenamefont {Matsuda}}]{Kasahara.PNAS.111.16309}%
  \BibitemOpen
  \bibfield  {author} {\bibinfo {author} {\bibfnamefont {S.}~\bibnamefont
  {Kasahara}}, \bibinfo {author} {\bibfnamefont {T.}~\bibnamefont {Watashige}},
  \bibinfo {author} {\bibfnamefont {T.}~\bibnamefont {Hanaguri}}, \bibinfo
  {author} {\bibfnamefont {Y.}~\bibnamefont {Kohsaka}}, \bibinfo {author}
  {\bibfnamefont {T.}~\bibnamefont {Yamashita}}, \bibinfo {author}
  {\bibfnamefont {Y.}~\bibnamefont {Shimoyama}}, \bibinfo {author}
  {\bibfnamefont {Y.}~\bibnamefont {Mizukami}}, \bibinfo {author}
  {\bibfnamefont {R.}~\bibnamefont {Endo}}, \bibinfo {author} {\bibfnamefont
  {H.}~\bibnamefont {Ikeda}}, \bibinfo {author} {\bibfnamefont
  {K.}~\bibnamefont {Aoyama}}, \bibinfo {author} {\bibfnamefont
  {T.}~\bibnamefont {Terashima}}, \bibinfo {author} {\bibfnamefont
  {S.}~\bibnamefont {Uji}}, \bibinfo {author} {\bibfnamefont {T.}~\bibnamefont
  {Wolf}}, \bibinfo {author} {\bibfnamefont {H.}~\bibnamefont {von
  L\"ohneysen}}, \bibinfo {author} {\bibfnamefont {T.}~\bibnamefont
  {Shibauchi}}, \ and\ \bibinfo {author} {\bibfnamefont {Y.}~\bibnamefont
  {Matsuda}},\ }\href {\doibase 10.1073/pnas.1413477111} {\bibfield  {journal}
  {\bibinfo  {journal} {Proceedings of the National Academy of Sciences}\
  }\textbf {\bibinfo {volume} {111}},\ \bibinfo {pages} {16309} (\bibinfo
  {year} {2014})}\BibitemShut {NoStop}%
\bibitem [{\citenamefont {Ponomarev}\ \emph {et~al.}(2013)\citenamefont
  {Ponomarev}, \citenamefont {Kuzmichev}, \citenamefont {Kuzmicheva},
  \citenamefont {Mikheev}, \citenamefont {Sudakova}, \citenamefont
  {Tchesnokov}, \citenamefont {Volkova}, \citenamefont {Vasiliev},
  \citenamefont {Pudalov}, \citenamefont {Sadakov}, \citenamefont {Usol'tsev},
  \citenamefont {Wolf}, \citenamefont {Khlybov},\ and\ \citenamefont
  {Kulikova}}]{Ponomarev2013}%
  \BibitemOpen
  \bibfield  {author} {\bibinfo {author} {\bibfnamefont {Y.~G.}\ \bibnamefont
  {Ponomarev}}, \bibinfo {author} {\bibfnamefont {S.~A.}\ \bibnamefont
  {Kuzmichev}}, \bibinfo {author} {\bibfnamefont {T.~E.}\ \bibnamefont
  {Kuzmicheva}}, \bibinfo {author} {\bibfnamefont {M.~G.}\ \bibnamefont
  {Mikheev}}, \bibinfo {author} {\bibfnamefont {M.~V.}\ \bibnamefont
  {Sudakova}}, \bibinfo {author} {\bibfnamefont {S.~N.}\ \bibnamefont
  {Tchesnokov}}, \bibinfo {author} {\bibfnamefont {O.~S.}\ \bibnamefont
  {Volkova}}, \bibinfo {author} {\bibfnamefont {A.~N.}\ \bibnamefont
  {Vasiliev}}, \bibinfo {author} {\bibfnamefont {V.~M.}\ \bibnamefont
  {Pudalov}}, \bibinfo {author} {\bibfnamefont {A.~V.}\ \bibnamefont
  {Sadakov}}, \bibinfo {author} {\bibfnamefont {A.~S.}\ \bibnamefont
  {Usol'tsev}}, \bibinfo {author} {\bibfnamefont {T.}~\bibnamefont {Wolf}},
  \bibinfo {author} {\bibfnamefont {E.~P.}\ \bibnamefont {Khlybov}}, \ and\
  \bibinfo {author} {\bibfnamefont {L.~F.}\ \bibnamefont {Kulikova}},\ }\href
  {\doibase 10.1007/s10948-013-2264-7} {\bibfield  {journal} {\bibinfo
  {journal} {Journal of Superconductivity and Novel Magnetism}\ }\textbf
  {\bibinfo {volume} {26}},\ \bibinfo {pages} {2867} (\bibinfo {year}
  {2013})}\BibitemShut {NoStop}%
\bibitem [{\citenamefont {Taylor}\ \emph {et~al.}(2011)\citenamefont {Taylor},
  \citenamefont {Pitcher}, \citenamefont {Ewings}, \citenamefont {Perring},
  \citenamefont {Clarke},\ and\ \citenamefont
  {Boothroyd}}]{Taylor.PhysRevB.83.220514}%
  \BibitemOpen
  \bibfield  {author} {\bibinfo {author} {\bibfnamefont {A.~E.}\ \bibnamefont
  {Taylor}}, \bibinfo {author} {\bibfnamefont {M.~J.}\ \bibnamefont {Pitcher}},
  \bibinfo {author} {\bibfnamefont {R.~A.}\ \bibnamefont {Ewings}}, \bibinfo
  {author} {\bibfnamefont {T.~G.}\ \bibnamefont {Perring}}, \bibinfo {author}
  {\bibfnamefont {S.~J.}\ \bibnamefont {Clarke}}, \ and\ \bibinfo {author}
  {\bibfnamefont {A.~T.}\ \bibnamefont {Boothroyd}},\ }\href {\doibase
  10.1103/PhysRevB.83.220514} {\bibfield  {journal} {\bibinfo  {journal} {Phys.
  Rev. B}\ }\textbf {\bibinfo {volume} {83}},\ \bibinfo {pages} {220514}
  (\bibinfo {year} {2011})}\BibitemShut {NoStop}%
\bibitem [{\citenamefont {Borisenko}\ \emph {et~al.}(2010)\citenamefont
  {Borisenko}, \citenamefont {Zabolotnyy}, \citenamefont {Evtushinsky},
  \citenamefont {Kim}, \citenamefont {Morozov}, \citenamefont {Yaresko},
  \citenamefont {Kordyuk}, \citenamefont {Behr}, \citenamefont {Vasiliev},
  \citenamefont {Follath},\ and\ \citenamefont
  {B\"uchner}}]{Borisenko.PhysRevLett.105.067002}%
  \BibitemOpen
  \bibfield  {author} {\bibinfo {author} {\bibfnamefont {S.~V.}\ \bibnamefont
  {Borisenko}}, \bibinfo {author} {\bibfnamefont {V.~B.}\ \bibnamefont
  {Zabolotnyy}}, \bibinfo {author} {\bibfnamefont {D.~V.}\ \bibnamefont
  {Evtushinsky}}, \bibinfo {author} {\bibfnamefont {T.~K.}\ \bibnamefont
  {Kim}}, \bibinfo {author} {\bibfnamefont {I.~V.}\ \bibnamefont {Morozov}},
  \bibinfo {author} {\bibfnamefont {A.~N.}\ \bibnamefont {Yaresko}}, \bibinfo
  {author} {\bibfnamefont {A.~A.}\ \bibnamefont {Kordyuk}}, \bibinfo {author}
  {\bibfnamefont {G.}~\bibnamefont {Behr}}, \bibinfo {author} {\bibfnamefont
  {A.}~\bibnamefont {Vasiliev}}, \bibinfo {author} {\bibfnamefont
  {R.}~\bibnamefont {Follath}}, \ and\ \bibinfo {author} {\bibfnamefont
  {B.}~\bibnamefont {B\"uchner}},\ }\href {\doibase
  10.1103/PhysRevLett.105.067002} {\bibfield  {journal} {\bibinfo  {journal}
  {Phys. Rev. Lett.}\ }\textbf {\bibinfo {volume} {105}},\ \bibinfo {pages}
  {067002} (\bibinfo {year} {2010})}\BibitemShut {NoStop}%
\bibitem [{\citenamefont {Borisenko}\ \emph {et~al.}(2012)\citenamefont
  {Borisenko}, \citenamefont {Zabolotnyy}, \citenamefont {Kordyuk},
  \citenamefont {Evtushinsky}, \citenamefont {Kim}, \citenamefont {Morozov},
  \citenamefont {Follath},\ and\ \citenamefont
  {B\"uchner}}]{Borisenko.symmetry-04-00251}%
  \BibitemOpen
  \bibfield  {author} {\bibinfo {author} {\bibfnamefont {S.~V.}\ \bibnamefont
  {Borisenko}}, \bibinfo {author} {\bibfnamefont {V.~B.}\ \bibnamefont
  {Zabolotnyy}}, \bibinfo {author} {\bibfnamefont {A.~A.}\ \bibnamefont
  {Kordyuk}}, \bibinfo {author} {\bibfnamefont {D.~V.}\ \bibnamefont
  {Evtushinsky}}, \bibinfo {author} {\bibfnamefont {T.~K.}\ \bibnamefont
  {Kim}}, \bibinfo {author} {\bibfnamefont {I.~V.}\ \bibnamefont {Morozov}},
  \bibinfo {author} {\bibfnamefont {R.}~\bibnamefont {Follath}}, \ and\
  \bibinfo {author} {\bibfnamefont {B.}~\bibnamefont {B\"uchner}},\ }\href
  {\doibase 10.3390/sym4010251} {\bibfield  {journal} {\bibinfo  {journal}
  {Symmetry}\ }\textbf {\bibinfo {volume} {4}},\ \bibinfo {pages} {251}
  (\bibinfo {year} {2012})}\BibitemShut {NoStop}%
\bibitem [{\citenamefont {Umezawa}\ \emph {et~al.}(2012)\citenamefont
  {Umezawa}, \citenamefont {Li}, \citenamefont {Miao}, \citenamefont
  {Nakayama}, \citenamefont {Liu}, \citenamefont {Richard}, \citenamefont
  {Sato}, \citenamefont {He}, \citenamefont {Wang}, \citenamefont {Chen},
  \citenamefont {Ding}, \citenamefont {Takahashi},\ and\ \citenamefont
  {Wang}}]{Umezawa.PhysRevLett.108.037002}%
  \BibitemOpen
  \bibfield  {author} {\bibinfo {author} {\bibfnamefont {K.}~\bibnamefont
  {Umezawa}}, \bibinfo {author} {\bibfnamefont {Y.}~\bibnamefont {Li}},
  \bibinfo {author} {\bibfnamefont {H.}~\bibnamefont {Miao}}, \bibinfo {author}
  {\bibfnamefont {K.}~\bibnamefont {Nakayama}}, \bibinfo {author}
  {\bibfnamefont {Z.-H.}\ \bibnamefont {Liu}}, \bibinfo {author} {\bibfnamefont
  {P.}~\bibnamefont {Richard}}, \bibinfo {author} {\bibfnamefont
  {T.}~\bibnamefont {Sato}}, \bibinfo {author} {\bibfnamefont {J.~B.}\
  \bibnamefont {He}}, \bibinfo {author} {\bibfnamefont {D.-M.}\ \bibnamefont
  {Wang}}, \bibinfo {author} {\bibfnamefont {G.~F.}\ \bibnamefont {Chen}},
  \bibinfo {author} {\bibfnamefont {H.}~\bibnamefont {Ding}}, \bibinfo {author}
  {\bibfnamefont {T.}~\bibnamefont {Takahashi}}, \ and\ \bibinfo {author}
  {\bibfnamefont {S.-C.}\ \bibnamefont {Wang}},\ }\href {\doibase
  10.1103/PhysRevLett.108.037002} {\bibfield  {journal} {\bibinfo  {journal}
  {Phys. Rev. Lett.}\ }\textbf {\bibinfo {volume} {108}},\ \bibinfo {pages}
  {037002} (\bibinfo {year} {2012})}\BibitemShut {NoStop}%
\bibitem [{\citenamefont {Kuzmichev}\ \emph {et~al.}(2012)\citenamefont
  {Kuzmichev}, \citenamefont {Shanygina}, \citenamefont {Morozov},
  \citenamefont {Boltalin}, \citenamefont {Roslova}, \citenamefont {Wurmehl},\
  and\ \citenamefont {B{\"u}chner}}]{Kuzmichev2012}%
  \BibitemOpen
  \bibfield  {author} {\bibinfo {author} {\bibfnamefont {S.~A.}\ \bibnamefont
  {Kuzmichev}}, \bibinfo {author} {\bibfnamefont {T.~E.}\ \bibnamefont
  {Shanygina}}, \bibinfo {author} {\bibfnamefont {I.~V.}\ \bibnamefont
  {Morozov}}, \bibinfo {author} {\bibfnamefont {A.~I.}\ \bibnamefont
  {Boltalin}}, \bibinfo {author} {\bibfnamefont {M.~V.}\ \bibnamefont
  {Roslova}}, \bibinfo {author} {\bibfnamefont {S.}~\bibnamefont {Wurmehl}}, \
  and\ \bibinfo {author} {\bibfnamefont {B.}~\bibnamefont {B{\"u}chner}},\
  }\href {\doibase 10.1134/S0021364012100086} {\bibfield  {journal} {\bibinfo
  {journal} {JETP Letters}\ }\textbf {\bibinfo {volume} {95}},\ \bibinfo
  {pages} {537} (\bibinfo {year} {2012})}\BibitemShut {NoStop}%
\bibitem [{\citenamefont {Kuzmichev}\ \emph {et~al.}(2014)\citenamefont
  {Kuzmichev}, \citenamefont {Kuzmicheva}, \citenamefont {Boltalin},\ and\
  \citenamefont {Morozov}}]{Kuzmichev2013}%
  \BibitemOpen
  \bibfield  {author} {\bibinfo {author} {\bibfnamefont {S.~A.}\ \bibnamefont
  {Kuzmichev}}, \bibinfo {author} {\bibfnamefont {T.~E.}\ \bibnamefont
  {Kuzmicheva}}, \bibinfo {author} {\bibfnamefont {A.~I.}\ \bibnamefont
  {Boltalin}}, \ and\ \bibinfo {author} {\bibfnamefont {I.~V.}\ \bibnamefont
  {Morozov}},\ }\href {\doibase 10.1134/S0021364013240120} {\bibfield
  {journal} {\bibinfo  {journal} {JETP Letters}\ }\textbf {\bibinfo {volume}
  {98}},\ \bibinfo {pages} {722} (\bibinfo {year} {2014})}\BibitemShut
  {NoStop}%
\bibitem [{\citenamefont {Chi}\ \emph {et~al.}(2012)\citenamefont {Chi},
  \citenamefont {Grothe}, \citenamefont {Liang}, \citenamefont {Dosanjh},
  \citenamefont {Hardy}, \citenamefont {Burke}, \citenamefont {Bonn},\ and\
  \citenamefont {Pennec}}]{Chi.PhysRevLett.109.087002}%
  \BibitemOpen
  \bibfield  {author} {\bibinfo {author} {\bibfnamefont {S.}~\bibnamefont
  {Chi}}, \bibinfo {author} {\bibfnamefont {S.}~\bibnamefont {Grothe}},
  \bibinfo {author} {\bibfnamefont {R.}~\bibnamefont {Liang}}, \bibinfo
  {author} {\bibfnamefont {P.}~\bibnamefont {Dosanjh}}, \bibinfo {author}
  {\bibfnamefont {W.~N.}\ \bibnamefont {Hardy}}, \bibinfo {author}
  {\bibfnamefont {S.~A.}\ \bibnamefont {Burke}}, \bibinfo {author}
  {\bibfnamefont {D.~A.}\ \bibnamefont {Bonn}}, \ and\ \bibinfo {author}
  {\bibfnamefont {Y.}~\bibnamefont {Pennec}},\ }\href {\doibase
  10.1103/PhysRevLett.109.087002} {\bibfield  {journal} {\bibinfo  {journal}
  {Phys. Rev. Lett.}\ }\textbf {\bibinfo {volume} {109}},\ \bibinfo {pages}
  {087002} (\bibinfo {year} {2012})}\BibitemShut {NoStop}%
\bibitem [{\citenamefont {Hanaguri}\ \emph {et~al.}(2012)\citenamefont
  {Hanaguri}, \citenamefont {Kitagawa}, \citenamefont {Matsubayashi},
  \citenamefont {Mazaki}, \citenamefont {Uwatoko},\ and\ \citenamefont
  {Takagi}}]{Hanaguri.PhysRevB.85.214505}%
  \BibitemOpen
  \bibfield  {author} {\bibinfo {author} {\bibfnamefont {T.}~\bibnamefont
  {Hanaguri}}, \bibinfo {author} {\bibfnamefont {K.}~\bibnamefont {Kitagawa}},
  \bibinfo {author} {\bibfnamefont {K.}~\bibnamefont {Matsubayashi}}, \bibinfo
  {author} {\bibfnamefont {Y.}~\bibnamefont {Mazaki}}, \bibinfo {author}
  {\bibfnamefont {Y.}~\bibnamefont {Uwatoko}}, \ and\ \bibinfo {author}
  {\bibfnamefont {H.}~\bibnamefont {Takagi}},\ }\href {\doibase
  10.1103/PhysRevB.85.214505} {\bibfield  {journal} {\bibinfo  {journal} {Phys.
  Rev. B}\ }\textbf {\bibinfo {volume} {85}},\ \bibinfo {pages} {214505}
  (\bibinfo {year} {2012})}\BibitemShut {NoStop}%
\bibitem [{\citenamefont {Nag}\ \emph {et~al.}(2016)\citenamefont {Nag},
  \citenamefont {Schlegel}, \citenamefont {Baumann}, \citenamefont {Grafe},
  \citenamefont {Beck}, \citenamefont {Wurmehl}, \citenamefont {B{\"u}chner},\
  and\ \citenamefont {Hess}}]{Nag.srep27926}%
  \BibitemOpen
  \bibfield  {author} {\bibinfo {author} {\bibfnamefont {P.~K.}\ \bibnamefont
  {Nag}}, \bibinfo {author} {\bibfnamefont {R.}~\bibnamefont {Schlegel}},
  \bibinfo {author} {\bibfnamefont {D.}~\bibnamefont {Baumann}}, \bibinfo
  {author} {\bibfnamefont {H.-J.}\ \bibnamefont {Grafe}}, \bibinfo {author}
  {\bibfnamefont {R.}~\bibnamefont {Beck}}, \bibinfo {author} {\bibfnamefont
  {S.}~\bibnamefont {Wurmehl}}, \bibinfo {author} {\bibfnamefont
  {B.}~\bibnamefont {B{\"u}chner}}, \ and\ \bibinfo {author} {\bibfnamefont
  {C.}~\bibnamefont {Hess}},\ }\href {http://dx.doi.org/10.1038/srep27926}
  {\bibfield  {journal} {\bibinfo  {journal} {Scientific Reports}\ }\textbf
  {\bibinfo {volume} {6}},\ \bibinfo {pages} {27926} (\bibinfo {year}
  {2016})}\BibitemShut {NoStop}%
\bibitem [{\citenamefont {Zhang}\ \emph
  {et~al.}(2013{\natexlab{a}})\citenamefont {Zhang}, \citenamefont {Yu},
  \citenamefont {Su}, \citenamefont {Song}, \citenamefont {Wang}, \citenamefont
  {Tan}, \citenamefont {Egami}, \citenamefont {Fernandez-Baca}, \citenamefont
  {Faulhaber}, \citenamefont {Si},\ and\ \citenamefont
  {Dai}}]{Zhang.PhysRevLett.111.207002}%
  \BibitemOpen
  \bibfield  {author} {\bibinfo {author} {\bibfnamefont {C.}~\bibnamefont
  {Zhang}}, \bibinfo {author} {\bibfnamefont {R.}~\bibnamefont {Yu}}, \bibinfo
  {author} {\bibfnamefont {Y.}~\bibnamefont {Su}}, \bibinfo {author}
  {\bibfnamefont {Y.}~\bibnamefont {Song}}, \bibinfo {author} {\bibfnamefont
  {M.}~\bibnamefont {Wang}}, \bibinfo {author} {\bibfnamefont {G.}~\bibnamefont
  {Tan}}, \bibinfo {author} {\bibfnamefont {T.}~\bibnamefont {Egami}}, \bibinfo
  {author} {\bibfnamefont {J.~A.}\ \bibnamefont {Fernandez-Baca}}, \bibinfo
  {author} {\bibfnamefont {E.}~\bibnamefont {Faulhaber}}, \bibinfo {author}
  {\bibfnamefont {Q.}~\bibnamefont {Si}}, \ and\ \bibinfo {author}
  {\bibfnamefont {P.}~\bibnamefont {Dai}},\ }\href {\doibase
  10.1103/PhysRevLett.111.207002} {\bibfield  {journal} {\bibinfo  {journal}
  {Phys. Rev. Lett.}\ }\textbf {\bibinfo {volume} {111}},\ \bibinfo {pages}
  {207002} (\bibinfo {year} {2013}{\natexlab{a}})}\BibitemShut {NoStop}%
\bibitem [{\citenamefont {Ge}\ \emph {et~al.}(2013)\citenamefont {Ge},
  \citenamefont {Ye}, \citenamefont {Xu}, \citenamefont {Zhang}, \citenamefont
  {Jiang}, \citenamefont {Xie}, \citenamefont {Song}, \citenamefont {Zhang},
  \citenamefont {Dai},\ and\ \citenamefont {Feng}}]{Ge.PhysRevX.3.011020}%
  \BibitemOpen
  \bibfield  {author} {\bibinfo {author} {\bibfnamefont {Q.~Q.}\ \bibnamefont
  {Ge}}, \bibinfo {author} {\bibfnamefont {Z.~R.}\ \bibnamefont {Ye}}, \bibinfo
  {author} {\bibfnamefont {M.}~\bibnamefont {Xu}}, \bibinfo {author}
  {\bibfnamefont {Y.}~\bibnamefont {Zhang}}, \bibinfo {author} {\bibfnamefont
  {J.}~\bibnamefont {Jiang}}, \bibinfo {author} {\bibfnamefont {B.~P.}\
  \bibnamefont {Xie}}, \bibinfo {author} {\bibfnamefont {Y.}~\bibnamefont
  {Song}}, \bibinfo {author} {\bibfnamefont {C.~L.}\ \bibnamefont {Zhang}},
  \bibinfo {author} {\bibfnamefont {P.}~\bibnamefont {Dai}}, \ and\ \bibinfo
  {author} {\bibfnamefont {D.~L.}\ \bibnamefont {Feng}},\ }\href {\doibase
  10.1103/PhysRevX.3.011020} {\bibfield  {journal} {\bibinfo  {journal} {Phys.
  Rev. X}\ }\textbf {\bibinfo {volume} {3}},\ \bibinfo {pages} {011020}
  (\bibinfo {year} {2013})}\BibitemShut {NoStop}%
\bibitem [{\citenamefont {Zhang}\ \emph
  {et~al.}(2013{\natexlab{b}})\citenamefont {Zhang}, \citenamefont {Li},
  \citenamefont {Song}, \citenamefont {Su}, \citenamefont {Tan}, \citenamefont
  {Netherton}, \citenamefont {Redding}, \citenamefont {Carr}, \citenamefont
  {Sobolev}, \citenamefont {Schneidewind}, \citenamefont {Faulhaber},
  \citenamefont {Harriger}, \citenamefont {Li}, \citenamefont {Lu},
  \citenamefont {Yao}, \citenamefont {Das}, \citenamefont {Balatsky},
  \citenamefont {Br\"uckel}, \citenamefont {Lynn},\ and\ \citenamefont
  {Dai}}]{Zhang.PhysRevB.88.064504}%
  \BibitemOpen
  \bibfield  {author} {\bibinfo {author} {\bibfnamefont {C.}~\bibnamefont
  {Zhang}}, \bibinfo {author} {\bibfnamefont {H.-F.}\ \bibnamefont {Li}},
  \bibinfo {author} {\bibfnamefont {Y.}~\bibnamefont {Song}}, \bibinfo {author}
  {\bibfnamefont {Y.}~\bibnamefont {Su}}, \bibinfo {author} {\bibfnamefont
  {G.}~\bibnamefont {Tan}}, \bibinfo {author} {\bibfnamefont {T.}~\bibnamefont
  {Netherton}}, \bibinfo {author} {\bibfnamefont {C.}~\bibnamefont {Redding}},
  \bibinfo {author} {\bibfnamefont {S.~V.}\ \bibnamefont {Carr}}, \bibinfo
  {author} {\bibfnamefont {O.}~\bibnamefont {Sobolev}}, \bibinfo {author}
  {\bibfnamefont {A.}~\bibnamefont {Schneidewind}}, \bibinfo {author}
  {\bibfnamefont {E.}~\bibnamefont {Faulhaber}}, \bibinfo {author}
  {\bibfnamefont {L.~W.}\ \bibnamefont {Harriger}}, \bibinfo {author}
  {\bibfnamefont {S.}~\bibnamefont {Li}}, \bibinfo {author} {\bibfnamefont
  {X.}~\bibnamefont {Lu}}, \bibinfo {author} {\bibfnamefont {D.-X.}\
  \bibnamefont {Yao}}, \bibinfo {author} {\bibfnamefont {T.}~\bibnamefont
  {Das}}, \bibinfo {author} {\bibfnamefont {A.~V.}\ \bibnamefont {Balatsky}},
  \bibinfo {author} {\bibfnamefont {T.}~\bibnamefont {Br\"uckel}}, \bibinfo
  {author} {\bibfnamefont {J.~W.}\ \bibnamefont {Lynn}}, \ and\ \bibinfo
  {author} {\bibfnamefont {P.}~\bibnamefont {Dai}},\ }\href {\doibase
  10.1103/PhysRevB.88.064504} {\bibfield  {journal} {\bibinfo  {journal} {Phys.
  Rev. B}\ }\textbf {\bibinfo {volume} {88}},\ \bibinfo {pages} {064504}
  (\bibinfo {year} {2013}{\natexlab{b}})}\BibitemShut {NoStop}%
\bibitem [{\citenamefont {Liu}\ \emph {et~al.}(2011)\citenamefont {Liu},
  \citenamefont {Richard}, \citenamefont {Nakayama}, \citenamefont {Chen},
  \citenamefont {Dong}, \citenamefont {He}, \citenamefont {Wang}, \citenamefont
  {Xia}, \citenamefont {Umezawa}, \citenamefont {Kawahara}, \citenamefont
  {Souma}, \citenamefont {Sato}, \citenamefont {Takahashi}, \citenamefont
  {Qian}, \citenamefont {Huang}, \citenamefont {Xu}, \citenamefont {Shi},
  \citenamefont {Ding},\ and\ \citenamefont {Wang}}]{Liu.PhysRevB.84.064519}%
  \BibitemOpen
  \bibfield  {author} {\bibinfo {author} {\bibfnamefont {Z.-H.}\ \bibnamefont
  {Liu}}, \bibinfo {author} {\bibfnamefont {P.}~\bibnamefont {Richard}},
  \bibinfo {author} {\bibfnamefont {K.}~\bibnamefont {Nakayama}}, \bibinfo
  {author} {\bibfnamefont {G.-F.}\ \bibnamefont {Chen}}, \bibinfo {author}
  {\bibfnamefont {S.}~\bibnamefont {Dong}}, \bibinfo {author} {\bibfnamefont
  {J.-B.}\ \bibnamefont {He}}, \bibinfo {author} {\bibfnamefont {D.-M.}\
  \bibnamefont {Wang}}, \bibinfo {author} {\bibfnamefont {T.-L.}\ \bibnamefont
  {Xia}}, \bibinfo {author} {\bibfnamefont {K.}~\bibnamefont {Umezawa}},
  \bibinfo {author} {\bibfnamefont {T.}~\bibnamefont {Kawahara}}, \bibinfo
  {author} {\bibfnamefont {S.}~\bibnamefont {Souma}}, \bibinfo {author}
  {\bibfnamefont {T.}~\bibnamefont {Sato}}, \bibinfo {author} {\bibfnamefont
  {T.}~\bibnamefont {Takahashi}}, \bibinfo {author} {\bibfnamefont
  {T.}~\bibnamefont {Qian}}, \bibinfo {author} {\bibfnamefont {Y.}~\bibnamefont
  {Huang}}, \bibinfo {author} {\bibfnamefont {N.}~\bibnamefont {Xu}}, \bibinfo
  {author} {\bibfnamefont {Y.}~\bibnamefont {Shi}}, \bibinfo {author}
  {\bibfnamefont {H.}~\bibnamefont {Ding}}, \ and\ \bibinfo {author}
  {\bibfnamefont {S.-C.}\ \bibnamefont {Wang}},\ }\href {\doibase
  10.1103/PhysRevB.84.064519} {\bibfield  {journal} {\bibinfo  {journal} {Phys.
  Rev. B}\ }\textbf {\bibinfo {volume} {84}},\ \bibinfo {pages} {064519}
  (\bibinfo {year} {2011})}\BibitemShut {NoStop}%
\bibitem [{\citenamefont {Iida}\ \emph {et~al.}(2017)\citenamefont {Iida},
  \citenamefont {Ishikado}, \citenamefont {Nagai}, \citenamefont {Yoshida},
  \citenamefont {Christianson}, \citenamefont {Murai}, \citenamefont
  {Kawashima}, \citenamefont {Yoshida}, \citenamefont {Eisaki},\ and\
  \citenamefont {Iyo}}]{Iida2017}%
  \BibitemOpen
  \bibfield  {author} {\bibinfo {author} {\bibfnamefont {K.}~\bibnamefont
  {Iida}}, \bibinfo {author} {\bibfnamefont {M.}~\bibnamefont {Ishikado}},
  \bibinfo {author} {\bibfnamefont {Y.}~\bibnamefont {Nagai}}, \bibinfo
  {author} {\bibfnamefont {H.}~\bibnamefont {Yoshida}}, \bibinfo {author}
  {\bibfnamefont {A.~D.}\ \bibnamefont {Christianson}}, \bibinfo {author}
  {\bibfnamefont {N.}~\bibnamefont {Murai}}, \bibinfo {author} {\bibfnamefont
  {K.}~\bibnamefont {Kawashima}}, \bibinfo {author} {\bibfnamefont
  {Y.}~\bibnamefont {Yoshida}}, \bibinfo {author} {\bibfnamefont
  {H.}~\bibnamefont {Eisaki}}, \ and\ \bibinfo {author} {\bibfnamefont
  {A.}~\bibnamefont {Iyo}},\ }\href {\doibase 10.7566/JPSJ.86.093703}
  {\bibfield  {journal} {\bibinfo  {journal} {Journal of the Physical Society
  of Japan}\ }\textbf {\bibinfo {volume} {86}},\ \bibinfo {pages} {093703}
  (\bibinfo {year} {2017})}\BibitemShut {NoStop}%
\bibitem [{\citenamefont {Mou}\ \emph {et~al.}(2016)\citenamefont {Mou},
  \citenamefont {Kong}, \citenamefont {Meier}, \citenamefont {Lochner},
  \citenamefont {Wang}, \citenamefont {Lin}, \citenamefont {Wu}, \citenamefont
  {Bud'ko}, \citenamefont {Eremin}, \citenamefont {Johnson}, \citenamefont
  {Canfield},\ and\ \citenamefont {Kaminski}}]{Mou2016}%
  \BibitemOpen
  \bibfield  {author} {\bibinfo {author} {\bibfnamefont {D.}~\bibnamefont
  {Mou}}, \bibinfo {author} {\bibfnamefont {T.}~\bibnamefont {Kong}}, \bibinfo
  {author} {\bibfnamefont {W.~R.}\ \bibnamefont {Meier}}, \bibinfo {author}
  {\bibfnamefont {F.}~\bibnamefont {Lochner}}, \bibinfo {author} {\bibfnamefont
  {L.-L.}\ \bibnamefont {Wang}}, \bibinfo {author} {\bibfnamefont
  {Q.}~\bibnamefont {Lin}}, \bibinfo {author} {\bibfnamefont {Y.}~\bibnamefont
  {Wu}}, \bibinfo {author} {\bibfnamefont {S.~L.}\ \bibnamefont {Bud'ko}},
  \bibinfo {author} {\bibfnamefont {I.}~\bibnamefont {Eremin}}, \bibinfo
  {author} {\bibfnamefont {D.~D.}\ \bibnamefont {Johnson}}, \bibinfo {author}
  {\bibfnamefont {P.~C.}\ \bibnamefont {Canfield}}, \ and\ \bibinfo {author}
  {\bibfnamefont {A.}~\bibnamefont {Kaminski}},\ }\href {\doibase
  10.1103/PhysRevLett.117.277001} {\bibfield  {journal} {\bibinfo  {journal}
  {Phys. Rev. Lett.}\ }\textbf {\bibinfo {volume} {117}},\ \bibinfo {pages}
  {277001} (\bibinfo {year} {2016})}\BibitemShut {NoStop}%
\end{thebibliography}%
\bibliographystyle{apsrev4-1}

\clearpage

\section{Supplemental Material for ``Effect of gap anisotropy on the spin resonance peak in the superconducting state of iron-based materials'' \label{suppl}}

\textit{Here I provide intensity plots of gap magnitudes that shown as the functions of angles in the main text. The purpose is to give an additional visual representation of the discussed gap structures and highlight nodal structures which become apparent in the intensity plots. Also, the gap functions resulting from the spin fluctuation calculation are shown for the three leading eigenvalues.}

\vspace{1cm}

\begin{figure}[t]
\begin{center}
 \includegraphics[width=1.0\columnwidth]{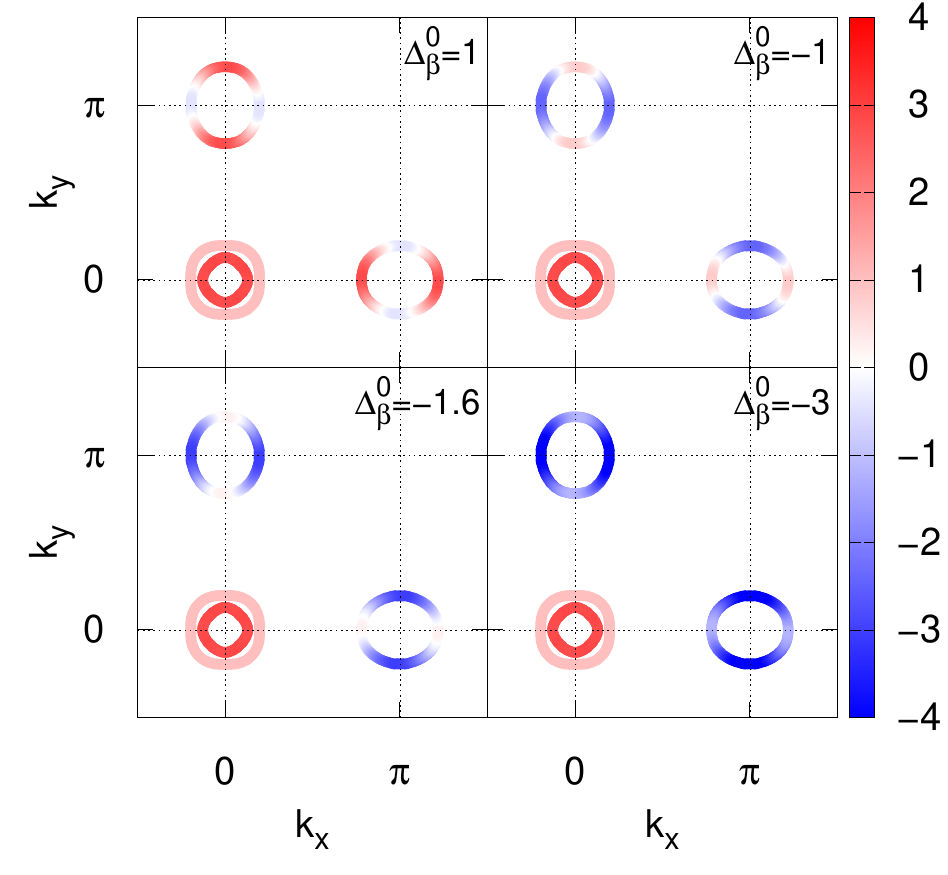}
 \caption{Gaps at the Fermi surface for $\Delta_\beta^{1}=16$ and various $\Delta_\beta^{0}$.}
\label{fig:set1}
\end{center}
\end{figure}
\begin{figure}[t]
\begin{center}
 \includegraphics[width=1.0\columnwidth]{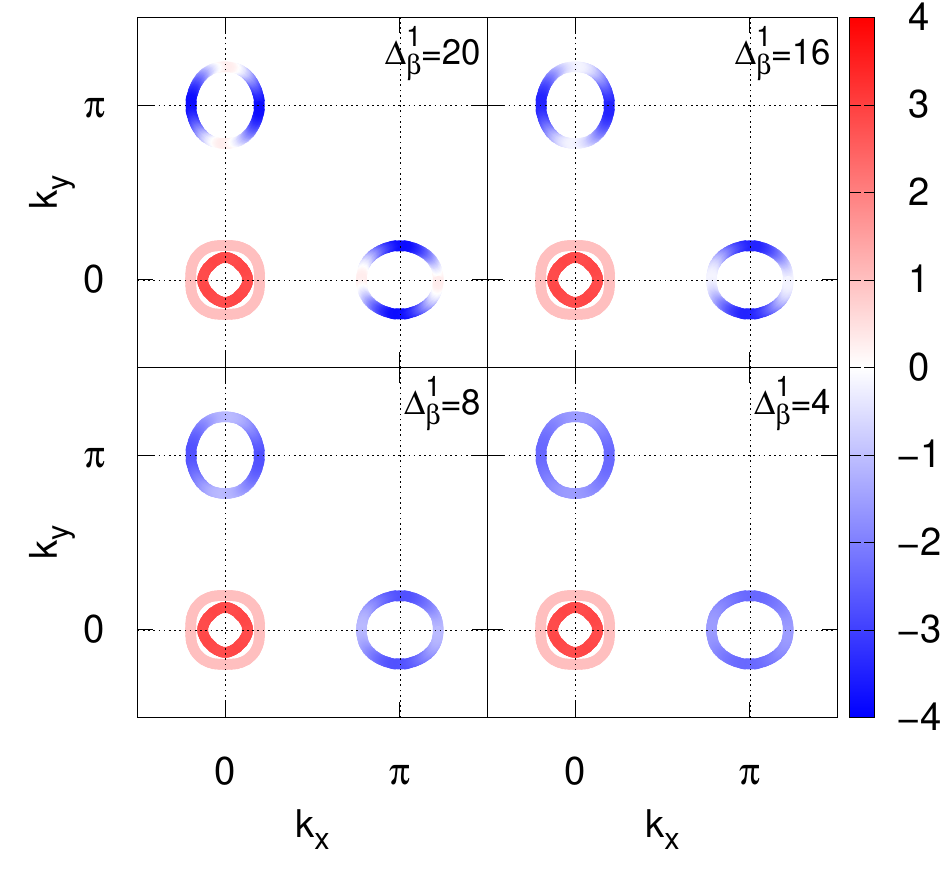}
 \caption{Gaps at the Fermi surface for $\Delta_\beta^{0}=-2$ and various $\Delta_\beta^{1}$.}
\label{fig:set2}
\end{center}
\end{figure}
\begin{figure}
\begin{center}
 \includegraphics[width=0.7\columnwidth]{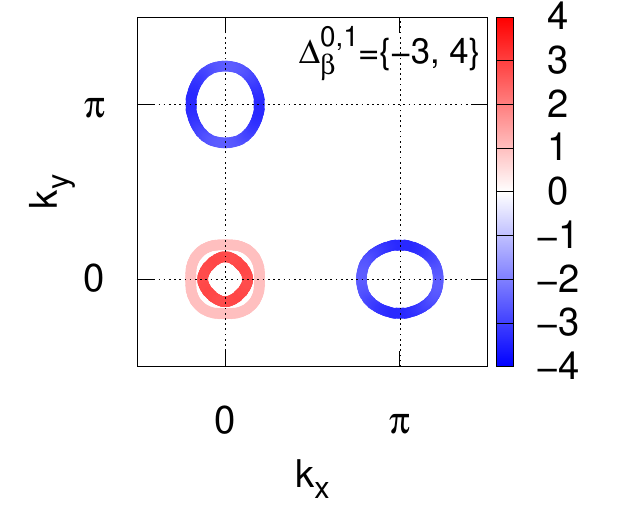}
 \caption{Gaps at the Fermi surface for $\Delta_\beta^{0}=-3$ and $\Delta_\beta^{1}=4$.}
\label{fig:set2_4}
\end{center}
\end{figure}

Figs.~\ref{fig:set1}-\ref{fig:set2_4} show superconducting gap amplitudes at the Fermi surface for different sets of $\Delta_\beta^0$ and $\Delta_\beta^1$ parameters. Using these parameters, the gap function is defined as
\beq
 \Delta_{\k \mu} = \Delta_{\mu}^{0} + \Delta_{\mu}^{1} \left(\cos k_x + \cos k_y \right)/2.
\eeq
All parameters are in units of $\Delta_0$ taken to be 5~meV. Since all gaps have $A_{1g}$ symmetry and should not change upon the $\pi/2$ rotation, gaps at electron pockets $\beta_1$ and $\beta_2$ should be the same. Thus $\Delta_{\beta_1}^{0,1} = \Delta_{\beta_2}^{0,1}$ that we denote simply as $\Delta_\beta^{0,1}$.

Gaps at hole pockets $\alpha_{1,2}$ are parameterized as $\Delta_{\alpha_1}^{0}=1$, $\Delta_{\alpha_1}^{1}=0$, $\Delta_{\alpha_2}^{0}=-16.4$, $\Delta_{\alpha_2}^{1}=20$, which gives constant gap at the inner hole pocket $\alpha_1$ and a weak anisotropy at the outer hole pocket $\alpha_2$. At the same time, gap at $\alpha_1$ approximately three times larger than the gap at $\alpha_2$.

Fig.~\ref{fig:set1} illustrates the decrease of $\Delta_\beta^{0}$ resulting in the shift of the zero-amplitude gap magnitude while the gap amplitude ($\Delta_\beta^{1}$) is constant. Lifting of nodes can be seen. Another situation is presented in Fig.~\ref{fig:set2} where the decrease of the gap amplitude $\Delta_\beta^{1}$ is shown for constant $\Delta_\beta^{0}$. Here we observe a gradual `isotropization' of the gap at the electron pockets. Effect of decreasing $\Delta_\beta^{0}$ for $\Delta_\beta^{1}=4$ is shown in Fig.~\ref{fig:set2_4}.

\begin{figure*}
\begin{center}
 (a)\includegraphics[width=0.3\linewidth]{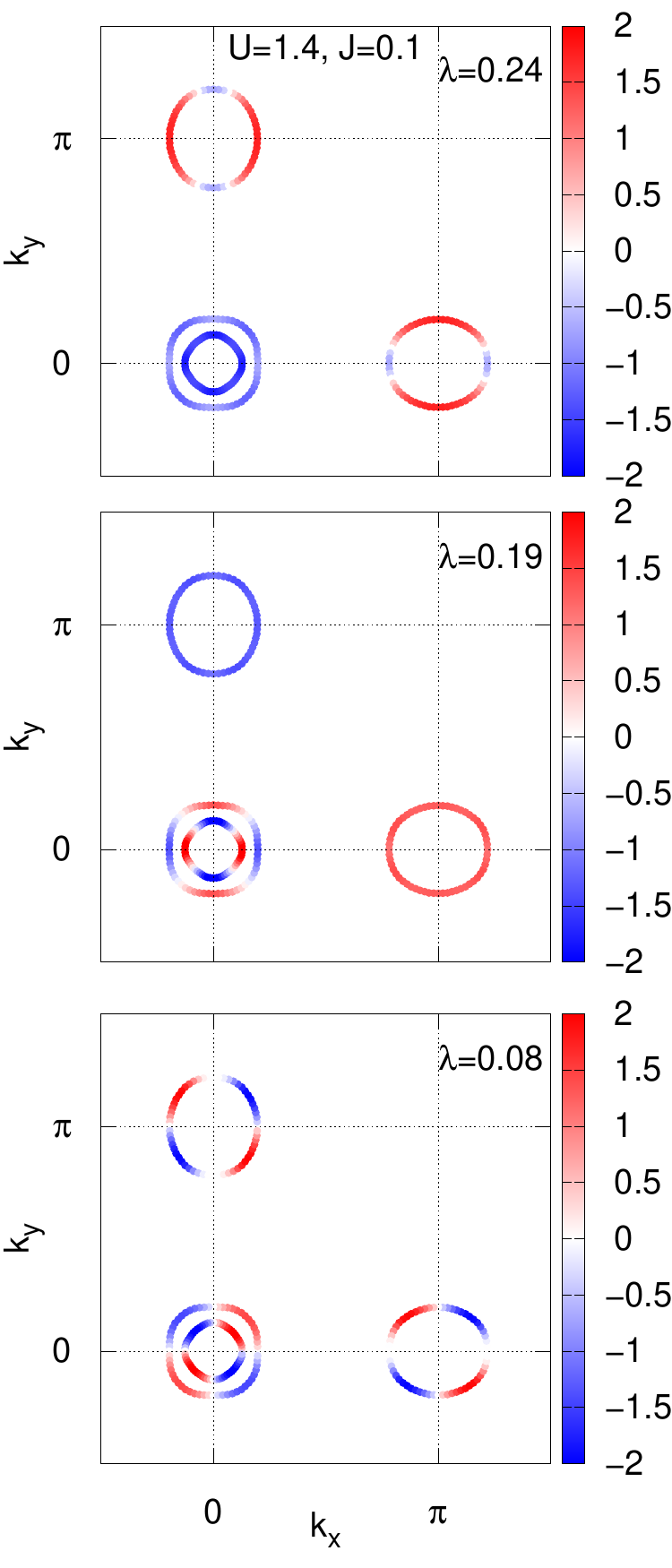}
 (b)\includegraphics[width=0.3\linewidth]{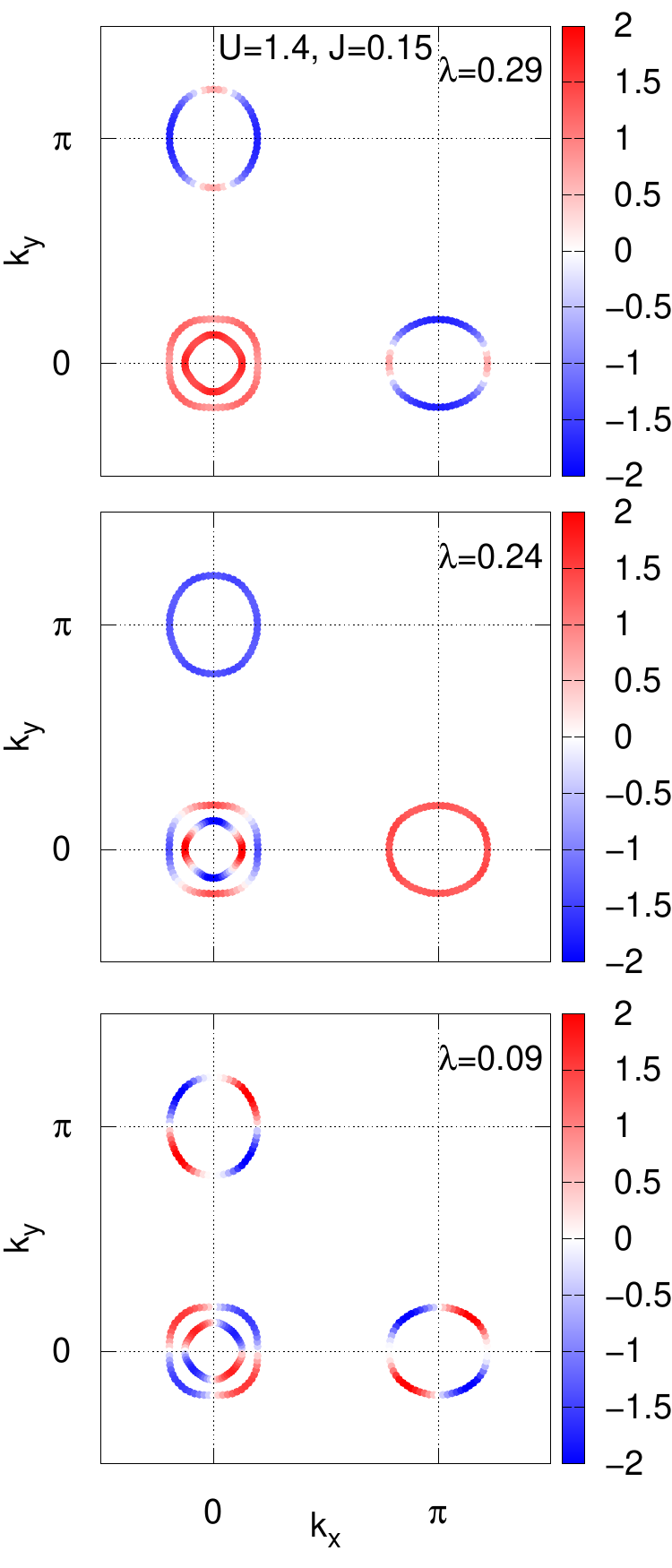}
 (c)\includegraphics[width=0.3\linewidth]{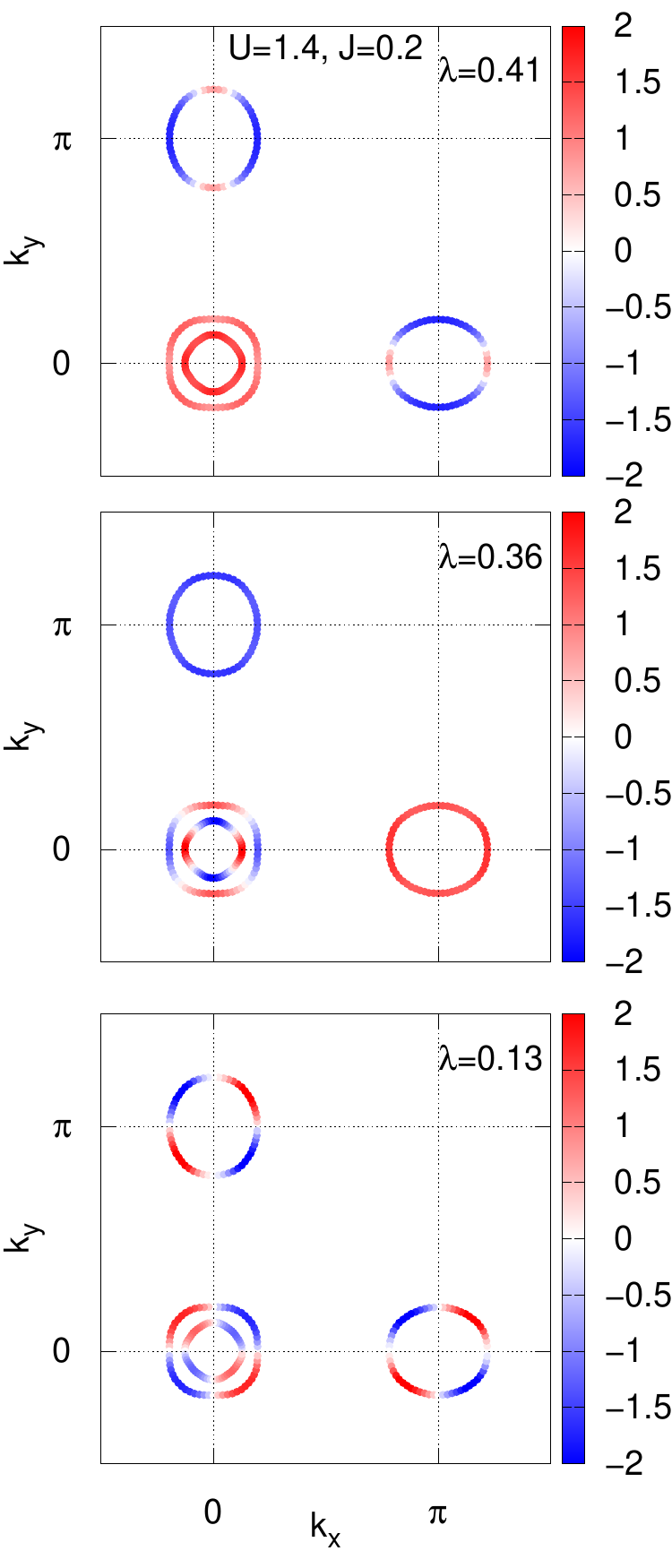}
 \caption{Gap functions $g(\k)$ in units of $\Delta_0 = 50$meV for the three leading eigenvalues $\lambda$ calculated for $U=1.4$~eV and the following values of Hund's exchange ($U'$ obeys the SRI): column (a) $J=0.1$~eV, column (b) $J=0.15$~eV, and column (c) $J=0.2$~eV.}
\label{fig:gk_tot171819}
\end{center}
\end{figure*}

\begin{figure*}
\begin{center}
 (a)\includegraphics[width=0.3\linewidth]{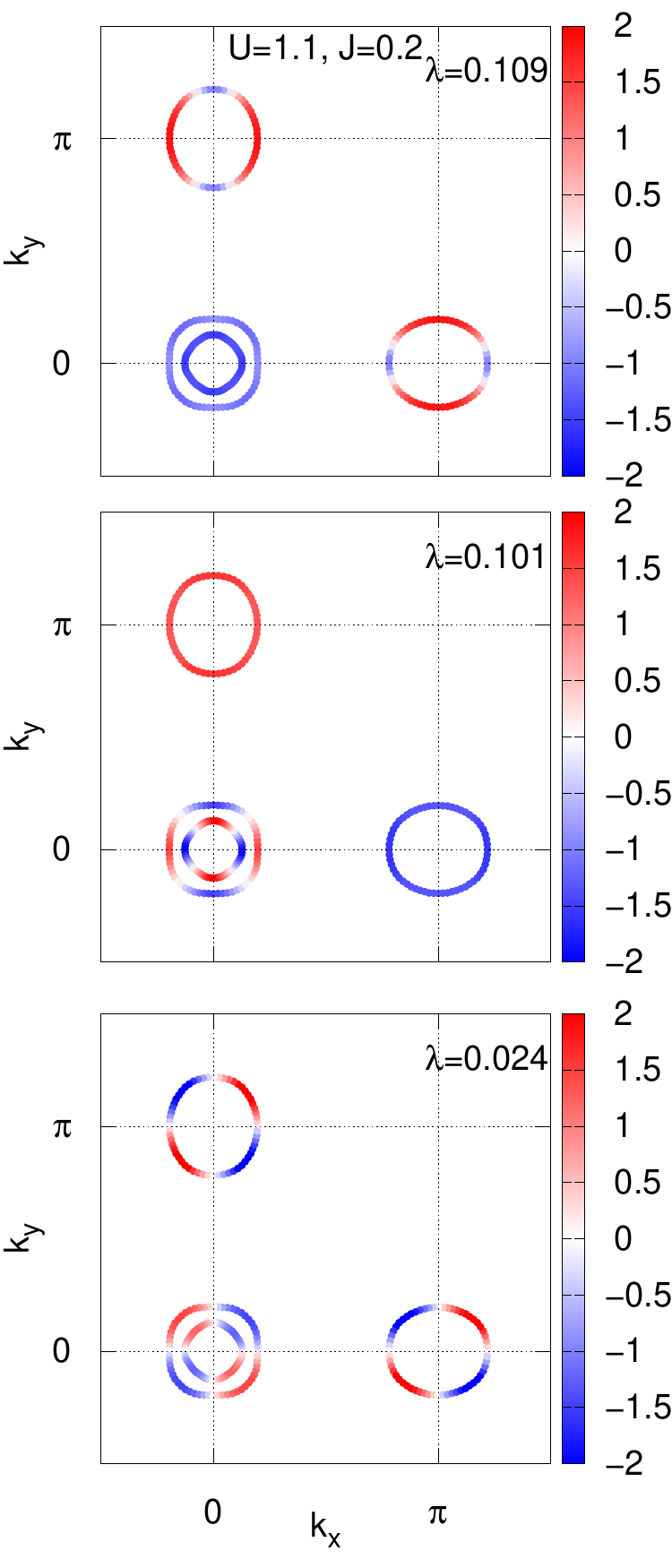}
 (b)\includegraphics[width=0.3\linewidth]{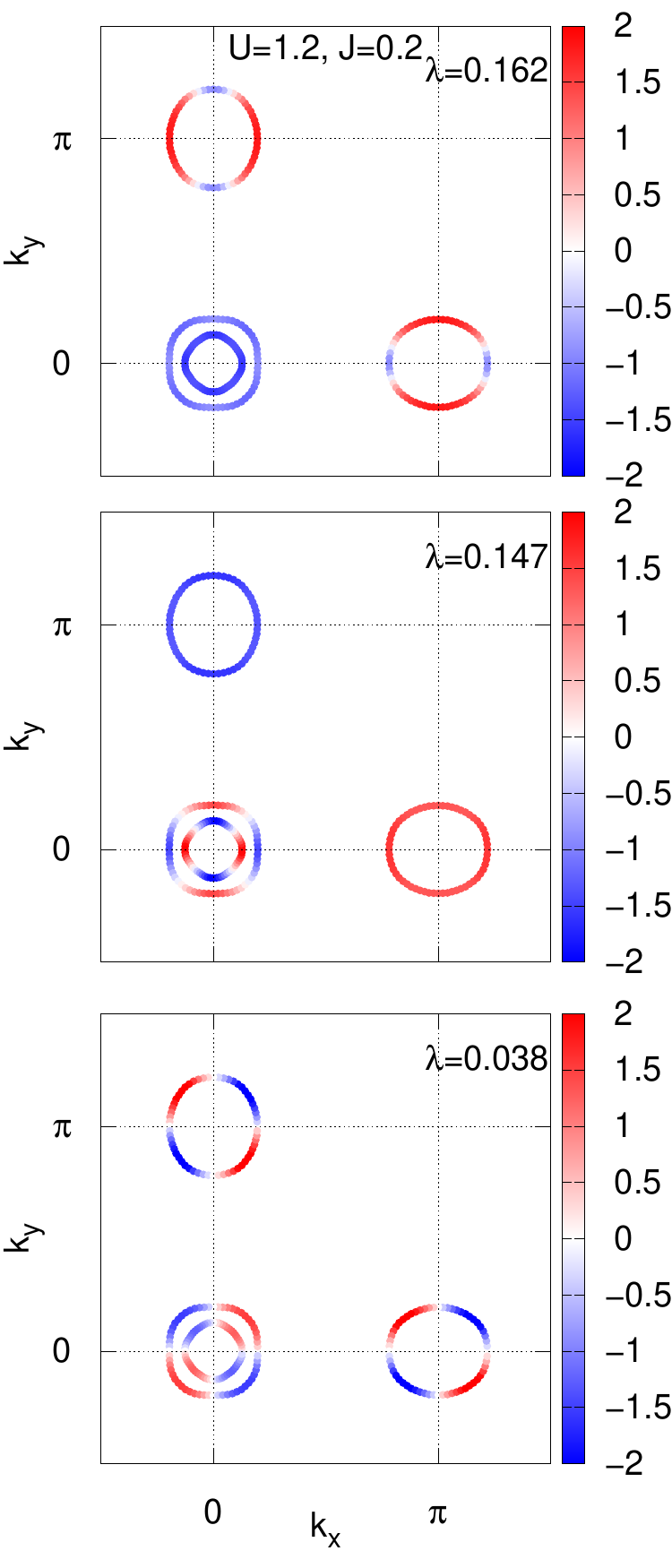}
 (c)\includegraphics[width=0.3\linewidth]{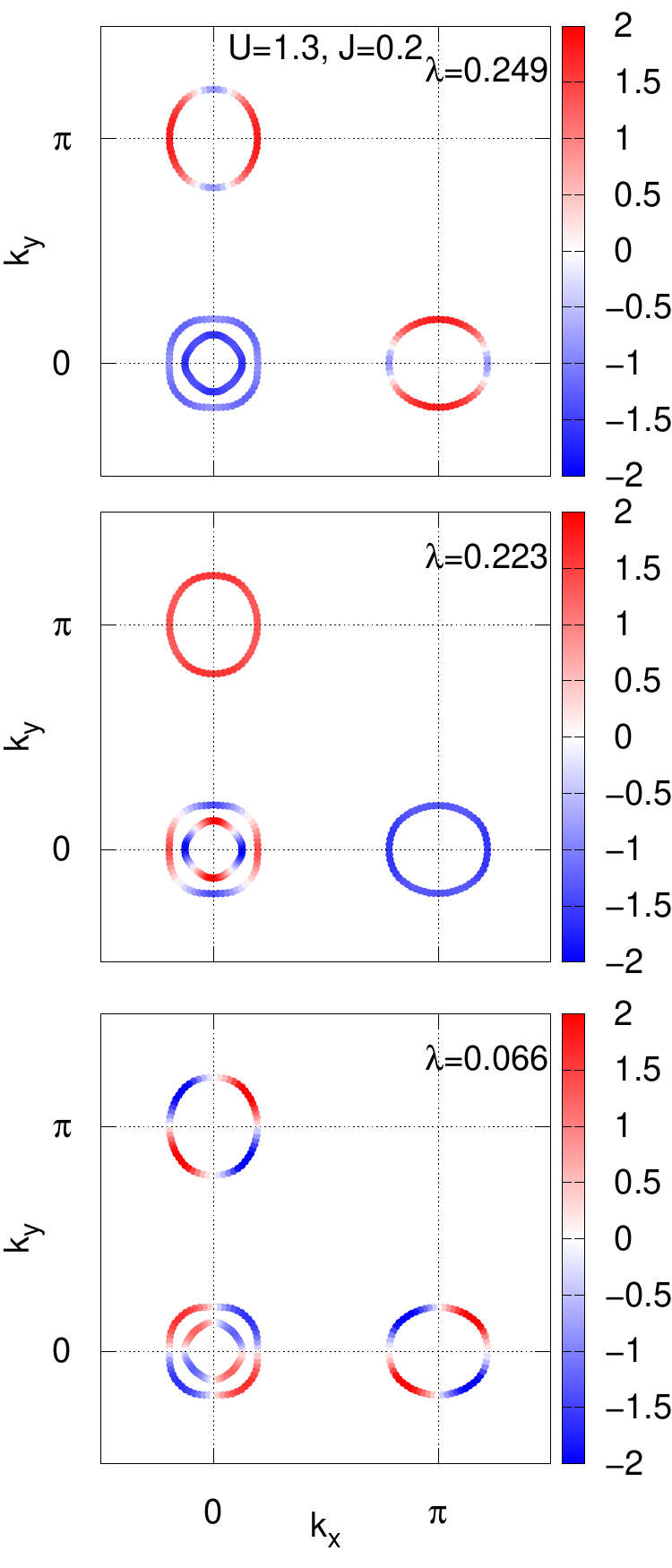}
 \caption{Gap functions $g(\k)$ in units of $\Delta_0 = 50$meV for the three leading eigenvalues $\lambda$ calculated for $J=0.2$~eV and the following values of Hubbard repulsion: column (a) $U=1.1$~eV, column (b) $U=1.2$~eV, and column (c) $U=1.3$~eV.}
\label{fig:gk_totJ02}
\end{center}
\end{figure*}

Spin fluctuation calculations were done for several sets of interaction parameters. In Fig.~\ref{fig:gk_tot171819}, gap functions for $U=1.4$eV and the three values of $J$ ($0.1$, $0.15$, and $0.2$~eV) are shown.
Gap functions shown in Fig.~\ref{fig:gk_totJ02} are calculated for the fixed value of $J=0.2$~eV and for the three values of $U$: $1.1$, $1.2$, and $1.3$~eV. Interorbital Hubbard repulsion $U'$ and pair hopping $J'$ were fixed by the spin-rotational invariance relation, $U'=U-2J$, $J'=J$. Note that the hierarchy of the gap symmetry and structure is the same for all presented cases. In particular, the leading state is of the $s_\pm$-type; subleading is the $d_{x^2-y^2}$-wave symmetry, and the next subleading is of the $d_{xy}$-type.

\end{document}